\documentclass[11pt]{article}
\usepackage{a4p}
\usepackage{amssymb}
\usepackage{epsfig}
\usepackage{cite}
\usepackage{pennames}
\usepackage{rotating}
\newcommand{\intLdt}  {10.36}
\newcommand{\intLdtfull}  {10.363}
\newcommand{\dLstat}  {0.045}
\newcommand{\dLsys}   {0.036}
\newcommand{\dLtot}   {0.06}
\newcommand{\rroots}  {172.12}  

\newcommand{\GENTxs}   {12.4}
\newcommand{\epem}{\ensuremath{\mathrm{e}^+\mathrm{e}^-}}

\newcommand{\tptm}{\ensuremath{\tau^+\tau^-}}
\newcommand{\lplm}{\ensuremath{\ell^+\ell^-}}
\newcommand{\Zz}{\ensuremath{{\mathrm{Z}^0}}}
\newcommand{\Hz}{\ensuremath{{\mathrm{H}^0}}}
\newcommand{\WW}{\ensuremath{\mathrm{W}^+\mathrm{W}^-}}
\newcommand{\Wp}{\ensuremath{\mathrm{W}^+}}
\newcommand{\Wpm}{\ensuremath{\mathrm{W}^\pm}}

\newcommand{\eeWW}{\ensuremath{\epem\rightarrow\WW}}
\newcommand{\qq}{\ensuremath{\mathrm{q\overline{q}}}}

\newcommand{\ff}{\ensuremath{\mathrm{f\overline{f}}}}
\newcommand{\lnu}{\ensuremath{\ell\overline{\nu}_{\ell}}}
\newcommand{\lpnu}{\ensuremath{\ell^+ \nu_{\ell}}}
\newcommand{\lmnu}{\ensuremath{{\ell^{\prime}}^-\overline{\nu}_{\ell^{\prime}}}}
\newcommand{\enu}{\ensuremath{\mathrm{e\overline{\nu}_{e}}}}
\newcommand{\mnu}{\ensuremath{\mu\overline{\nu}_{\mu}}}
\newcommand{\tnu}{\ensuremath{\tau\overline{\nu}_{\tau}}}

\newcommand{\taunu}{\ensuremath{\overline{\nu}_{\tau}}}
\newcommand{\ataunu}{\ensuremath{{\nu}_{\tau}}}

\newcommand{\qqqq}{\ensuremath{\qq\qq}}
\newcommand{\qqln}{\ensuremath{\qq\lnu}}
\newcommand{\WWqqln}{\ensuremath{\WW\rightarrow\qq\lnu}}
\newcommand{\WWqqqq}{\ensuremath{\WW\rightarrow\qq\qq}}
\newcommand{\WWqqen}{\ensuremath{\WW\rightarrow\qq\enu}}
\newcommand{\WWqqmn}{\ensuremath{\WW\rightarrow\qq\mnu}}
\newcommand{\WWqqtn}{\ensuremath{\WW\rightarrow\qq\tnu}}
\newcommand{\WWlnln}{\ensuremath{\WW\rightarrow\lpnu\lmnu}}
\newcommand{\lnln}{\ensuremath{\lpnu\lmnu}}
\newcommand{\Wenu}{\ensuremath{\epem \rightarrow \mathrm{W}\enu}}
\newcommand{\ZZqqqq}{\ensuremath{(\Zz/\gamma)(\Zz/\gamma)\rightarrow\qq\qq}}
\newcommand{\HZqqqq}{\ensuremath{\Hz\Zz\rightarrow\qq\qq}}

\newcommand{\Zee}{\ensuremath{\epem\rightarrow\Zz\epem}}
\newcommand{\Zqq}{\ensuremath{\Zz/\gamma\rightarrow\qq}}
\newcommand{\Ztt}{\ensuremath{\Zz/\gamma\rightarrow\tptm}}

\newcommand{\Mz}{\ensuremath{M_{\mathrm{Z}^0}}}
\newcommand{\Mw}{\ensuremath{M_{\mathrm{W}}}}
\newcommand{\Gw}{\ensuremath{\Gamma_{\mathrm{W}}}}

\newcommand{\LepII}{\mbox{LEP2}}
\newcommand{\LepI}{\mbox{LEP1}}

\newcommand{\Koralw}{\mbox{K{\sc oralw}}}
\newcommand{\KORALW}{\mbox{K{\sc oralw}}}
\newcommand{\KORALZ}{\mbox{K{\sc oralz}}}
\newcommand{\Excalibur}{\mbox{E{\sc xcalibur}}}
\newcommand{\EXCALIBUR}{\mbox{E{\sc xcalibur}}}
\newcommand{\grcff}{\mbox{grc4f}}
\newcommand{\Pythia}{\mbox{P{\sc ythia}}}
\newcommand{\PYTHIA}{\mbox{P{\sc ythia}}}
\newcommand{\Ariadne}{\mbox{A{\sc riadne}}}
\newcommand{\FERMISV}{\mbox{F{\sc ermisv}}}
\newcommand{\GENTLE}{\mbox{G{\sc entle}}}
\newcommand{\BHWIDE}{\mbox{B{\sc hwide}}}
\newcommand{\PHOJET}{\mbox{P{\sc hojet}}}
\newcommand{\TWOGEN}{\mbox{T{\sc wogen}}}
\newcommand{\Herwig}{\mbox{H{\sc erwig}}}
\newcommand{\HERWIG}{\mbox{H{\sc erwig}}}

\newcommand{\GeV}{\ensuremath{\mathrm{GeV}}}
\newcommand{\GeVc}{\ensuremath{\mathrm{GeV}}}
\newcommand{\GeVcc}{\ensuremath{\mathrm{GeV}}}

\newcommand{\dedx}{\ensuremath{\mathrm{d}E/\mathrm{d}x}}

\newcommand{\Rvis}{\ensuremath{R_{\mathrm{vis}}}}

\newcommand{\Yc}[1]{\ensuremath{y_{\mathrm{#1}}}}
\newcommand{\Jmom}{\ensuremath{J_{\mathrm{mom}}}}

\newcommand{\roots}{\ensuremath{\sqrt{s}}}
\newcommand{\rootsprime}{\ensuremath{\sqrt{s^\prime}}}
\newcommand{\Ecm}{\roots}
\newcommand{\Ebeam}{\ensuremath{E_{\mathrm{beam}}}}
\newcommand{\Ctmis}{\ensuremath{|\cos\theta_{\mathrm{mis}}|}}

\newcommand{\Ptsum}{\ensuremath{\sum p_T}}
\newcommand{\Zgamma}{\ensuremath{\Zz/\gamma}}

\newcommand {\ee}         {\ensuremath{\mathrm{e}^+ \mathrm{e}^-}}
\newcommand {\emu}        {\ensuremath{\mathrm{e}^{\pm} \mu^{\mp}}}
\newcommand {\et}         {\ensuremath{\mathrm{e}^{\pm} \tau^{\mp}}}
\newcommand {\mt}         {\ensuremath{\mu^{\pm} \tau^{\mp}}}
\newcommand {\mm}         {\ensuremath{\mu^+ \mu^-}}
\newcommand {\tautau}     {\ensuremath{\tau^+ \tau^-}}
\newcommand {\eeee}   {\ensuremath{\mathrm{e}^+ \mathrm{e}^-\rightarrow
\mathrm{e}^+ \mathrm{e}^-}}
\newcommand {\eemumu}   {\ensuremath{\mathrm{e}^+
            \mathrm{e}^-\rightarrow\mu^+ \mu^-}}
\newcommand {\eetautau}   {\ensuremath{\mathrm{e}^+
    \mathrm{e}^-\rightarrow\tau^+ \tau^-}}
\newcommand{\nunu}{\ensuremath{\nu\overline{\nu}}}
\newcommand {\eenunu}{\ensuremath{\epem\rightarrow\nunu\gamma\gamma}}

\newcommand{\Wmv}{\mbox{$\mathrm{W}\rightarrow\mnu$}}
\newcommand{\Wev}{\mbox{$\mathrm{W}\rightarrow\enu$}}
\newcommand{\Wtv}{\mbox{$\mathrm{W}\rightarrow\tnu$}}
\newcommand{\Wlv}{\mbox{$\mathrm{W}\rightarrow\lnu$}}
\newcommand{\Wqq}{\mbox{$\mathrm{W}\rightarrow\qq$}}
\newcommand{\Wten}{\mbox{$\mathrm{W}\rightarrow\tnu\rightarrow
(\enu\ataunu)\taunu$}}
\newcommand{\Wtmn}{\mbox{$\mathrm{W}\rightarrow\tnu\rightarrow
(\mnu\ataunu)\taunu$}}

\newcommand{\WWtelect}{\mbox{$\WW\rightarrow\qq\tnu\rightarrow
\qq(\enu\ataunu)\taunu$}}
\newcommand{\WWtonep}{\mbox{$\WW\rightarrow\qq\tnu\rightarrow
\qq(\pi^\pm n\pi^0\ataunu)\taunu$}}
\newcommand{\Wtonep}{\mbox{$\mathrm{W}\rightarrow\tnu\rightarrow
(\pi^\pm n\pi^0\ataunu)\taunu$}}
\newcommand{\Wtthreep}{\mbox{$\mathrm{W}\rightarrow\tnu\rightarrow 
(2\pi^\pm\pi^\mp\ataunu)\taunu$}}
\newcommand{\Tjet}{\mbox{$\theta_{\mathrm{jet}}$}}
\newcommand{\Etwo}{\mbox{$E_{\mathrm{200}}$}}

\newcommand{\ytwo}{\mbox{$y_{\mathrm{23}}$}}
\newcommand{\Elept}{\mbox{$E_{\mathrm{lept}}$}}

\newcommand{\Eisr}{\mbox{$E_{\mathrm{ISR}}$}}
\newcommand{\cslpmis} {\mbox{ $\cos\theta_{\mathrm{lpmis}}$}}

\newcommand{\Probe}{\mbox{$P({\mathrm{e}})$}}
\newcommand{\Probm}{\mbox{$P({\mu})$}}
\newcommand{\Psprime}{\mbox{$P(s^\prime)$}}
\newcommand{\qqen}{\ensuremath{\qq\enu}}
\newcommand{\qqee}{\mbox{\qq\epem}}
\newcommand{\qqff}{\mbox{\qq\ff}}
\newcommand{\eeff}{\mbox{\epem\ff}}
\newcommand{\qqmn}{\ensuremath{\qq\mnu}}
\newcommand{\qqtn}{\mbox{\qq\tnu}}
\newcommand{\Vij} {\mbox{$|\mathrm{V}_{ij}|$}}
\newcommand{\Vud} {\mbox{$|\mathrm{V}_{\mathrm{ud}}|$}}
\newcommand{\Vus} {\mbox{$|\mathrm{V}_{\mathrm{us}}|$}}
\newcommand{\Vcd} {\mbox{$|\mathrm{V}_{\mathrm{cd}}|$}}
\newcommand{\Vcb} {\mbox{$|\mathrm{V}_{\mathrm{cb}}|$}}
\newcommand{\Vub} {\mbox{$|\mathrm{V}_{\mathrm{ub}}|$}}
\newcommand{\Vcs} {\mbox{$|\mathrm{V}_{\mathrm{cs}}|$}}
\newcommand{\emnu}{\ensuremath{\mathrm{e^-\overline{\nu}_{e}}}}
\newcommand{\mmnu}{\ensuremath{\mu^-\overline{\nu}_{\mu}}}
\newcommand{\tmnu}{\ensuremath{\tau^-\overline{\nu}_{\tau}}}
\newcommand{\epnu}{\ensuremath{\mathrm{e^+{\nu_{e}}}}}
\newcommand{\mpnu}{\ensuremath{\mu^+{\nu}_{\mu}}}
\newcommand{\tpnu}{\ensuremath{\tau^+{\nu}_{\tau}}}
\newcommand{\semnu}{\ensuremath{\mathrm{e^{^-}\!\!\overline{\nu}_{e}}}}
\newcommand{\smmnu}{\ensuremath{\mu^{^-}\!\!\overline{\nu}_{\mu}}}
\newcommand{\stmnu}{\ensuremath{\tau^{^-}\!\!\overline{\nu}_{\tau}}}
\newcommand{\sepnu}{\ensuremath{\mathrm{e^{^+}\!\!{\nu_{e}}}}}
\newcommand{\smpnu}{\ensuremath{\mu^{^+}\!\!{\nu}_{\mu}}}
\newcommand{\stpnu}{\ensuremath{\tau^{^+}\!\!{\nu}_{\tau}}}
\newcommand{\enen}{\ensuremath{\epnu\emnu}}
\newcommand{\enmn}{\ensuremath{\epnu\mmnu}}
\newcommand{\entn}{\ensuremath{\epnu\tmnu}}
\newcommand{\mnmn}{\ensuremath{\mpnu\mmnu}}
\newcommand{\mntn}{\ensuremath{\mpnu\tmnu}}
\newcommand{\tntn}{\ensuremath{\tpnu\tmnu}}
\newcommand{\senen}{\ensuremath{\sepnu\semnu}}
\newcommand{\senmn}{\ensuremath{\sepnu\smmnu}}
\newcommand{\sentn}{\ensuremath{\sepnu\stmnu}}
\newcommand{\smnmn}{\ensuremath{\smpnu\smmnu}}
\newcommand{\smntn}{\ensuremath{\smpnu\stmnu}}
\newcommand{\stntn}{\ensuremath{\stpnu\stmnu}}
\newcommand{\WWenen}{\ensuremath{\WW\rightarrow\epnu\emnu}}
\newcommand{\WWenmn}{\ensuremath{\WW\rightarrow\epnu\mmnu}}
\newcommand{\WWentn}{\ensuremath{\WW\rightarrow\epnu\tmnu}}
\newcommand{\WWmnmn}{\ensuremath{\WW\rightarrow\mpnu\mmnu}}
\newcommand{\WWmntn}{\ensuremath{\WW\rightarrow\mpnu\tmnu}}
\newcommand{\WWtntn}{\ensuremath{\WW\rightarrow\tpnu\tmnu}}
\newcommand{\MW}{\ensuremath{M_{\mathrm{W}}}}
\newcommand{\CC}{\mbox{{\sc CC03}}}
\newcommand{\mrec}{\ensuremath{m_{\mathrm{rec}}}}

\def\etal{\mbox{{\it et al.}}}

\def\gappeq{\ensuremath{\mathrel{ \rlap{\raise.5ex\hbox{>}}
                      {\lower.5ex\hbox{\sim}}}}}
\def\lappeq{\ensuremath{\mathrel{ \rlap{\raise.5ex\hbox{<}}
                      {\lower.5ex\hbox{\sim}}}}}

\newcommand{\sigccthree}{\ensuremath{\sigma_{\mathrm{WW}}}}

\newcommand{\PLB}[3]  {Phys.\ Lett.\ \textbf{B#1} (#2) #3}
\newcommand{\ZPC}[3]  {Z.\ Phys.\ \textbf{C#1} (#2) #3}
\newcommand{\NIMA}[3] {Nucl.\ Instr.\ Meth.\ \textbf{A#1} (#2) #3}
\newcommand{\MPL}[3] {Mod.\ Phys.\ Lett.\ \textbf{A#1} (#2) #3}
\newcommand{\PPE}[1]  {CERN-PPE/{#1}}

\newcommand{\PRL}[3]  {Phys.\ Rev.\ Lett.\ \textbf{#1} (#2) #3}
\newcommand{\PRD}[3]  {Phys.\ Rev.\ \textbf{D#1} (#2) #3}
\newcommand{\NPB}[3]  {Nucl.\ Phys.\ \textbf{B#1} (#2) #3}

\newcommand{\CPC}[3]  {Comput.\ Phys.\ Commun.\ \textbf{#1} (#2) #3}
\newcommand{\Zzero}{\mbox{${\mathrm{Z}^0}$}}

\newcommand{\WWqqlnu}{\mbox{\WW$\rightarrow$ \qq\lnu }}
\newcommand{\WWqqenu}{\mbox{\WW$\rightarrow$\qq\enu}}
\newcommand{\WWqqmnu}{\mbox{\WW$\rightarrow$ \qq\mnu}}
\newcommand{\WWqqtnu}{\mbox{\WW$\rightarrow$ \qq\tnu}}

\def\opalackerstaff{OPAL Collaboration, K.\ Ackerstaff \etal}
\def\opalalexander{OPAL Collaboration, G.\ Alexander \etal}
\def\opalakers{OPAL Collaboration, R.\ Akers \etal}

\def\opalakrawy{OPAL Collaboration, M.Z.\ Akrawy \etal}
\newcommand{\QQQQ}{\ensuremath{\mathrm{4q}}}
\newcommand{\QQLV}{\ensuremath{\mathrm{qq}\ell\nu}}
\newcommand{\nch}{\ensuremath{n_{\mathrm{ch}}}}
\newcommand{\nchave}{\ensuremath{\langle n_{\mathrm{ch}}\rangle}}
\newcommand{\nchQQQQ}{\ensuremath{\langle n_{\mathrm{ch}}^{\QQQQ}\rangle}}
\newcommand{\nchQQLV}{\ensuremath{\langle n_{\mathrm{ch}}^{\QQLV}\rangle}}
\newcommand{\Dnch}{\ensuremath{\Delta\langle n_{\mathrm{ch}}\rangle}}
\newcommand{\xpQQQQ}{\ensuremath{\langle x_{p}^{\QQQQ}\rangle}}
\newcommand{\xpQQLV}{\ensuremath{\langle x_{p}^{\QQLV}\rangle}}
\newcommand{\xp}{\ensuremath{x_{p}}}
\newcommand{\thrQQQQ}{\ensuremath{\langle (1-T)^{\QQQQ} \rangle}}
\newcommand{\yQQQQ}{\ensuremath{\langle |y^{\QQQQ}| \rangle}}
\newcommand{\Dxp}{\ensuremath{\Delta \langle x_{p} \rangle}}
\newcommand{\Br}{\ensuremath{\mathrm{Br}}}
\begin{document}
\begin{titlepage}
 \begin{center}{\Large   EUROPEAN LABORATORY FOR PARTICLE PHYSICS
 }\end{center}\bigskip
\begin{flushright}
  CERN-PPE/97-116 \\  August 19, 1997
\end{flushright}
\bigskip\bigskip\bigskip
\begin{center}
 {\huge\bf \boldmath Measurement of the W Boson Mass and } \\
 \vspace{1mm}
 {\huge\bf \boldmath \WW\ Production and Decay Properties in } \\
 {\huge\bf \boldmath \epem\ Collisions at \roots=172~GeV}
\end{center}
\bigskip\bigskip
\begin{center}{\LARGE The OPAL Collaboration
    }\end{center}\bigskip\bigskip 
\bigskip\begin{center}{\large \bf
    Abstract }\end{center} This paper describes the measurement of the
W boson mass, \Mw, and decay width, \Gw, from the direct reconstruction of
the invariant mass of its decay products in W pair events collected at
a mean centre-of-mass energy of \roots=172.12~GeV with the OPAL
detector at LEP.  Measurements of the W pair production cross-section,
the W decay branching fractions and properties of the W decay final
states are also described.  A total of 120 
candidate
\WW\ events has been
selected for an integrated luminosity of 10.36~pb$^{-1}$.
The \WW\ production cross-section is measured to be
$\sigccthree=12.3\pm1.3 \mathrm{(stat.)}\pm0.3\mathrm{(syst.)}$~pb,
consistent with the Standard Model expectation.  The \WWqqln\ and
\WWqqqq\ final states are used to obtain a direct measurement of
$\Gw=1.30^{+0.62}_{-0.55}\mathrm{(stat.)}\pm0.18\mathrm{(syst.)}$~GeV.
Assuming the Standard Model relation between \Mw\ and \Gw, the W boson
mass is measured to be
$\Mw=80.32\pm0.30\mathrm{(stat.)}\pm0.09\mathrm{(syst.)}$~GeV.  The
event properties of the fully-hadronic decays of \WW\ events are
compared to those of the semi-leptonic decays.  At the current level
of precision there is no evidence for effects of colour reconnection
in the observables studied.  Combining data recorded by OPAL at
$\roots\sim161$--172~GeV, the W boson branching fraction to hadrons is
determined to be
$69.8^{+3.0}_{-3.2}\mathrm{(stat.)}\pm0.7\mathrm{(syst.)}$\%,
consistent with the prediction of the Standard Model.  The combined
mass measurement from direct reconstruction and from the \WW\ production
cross-sections measured at $\roots\sim161$ and $\roots\sim172$~GeV 
is $\Mw = 80.35 \pm 0.24\mathrm{(stat.)}\pm0.07\mathrm{(syst.)}$~\GeV.
 
\bigskip\bigskip
\begin{center}
  {\large To be submitted to Zeit. Phys. C}  
\end{center}
\begin{center}
\end{center}

\end{titlepage}
\begin{center}{\Large        The OPAL Collaboration
}\end{center}\bigskip
\begin{center}{
K.\thinspace Ackerstaff$^{  8}$,
G.\thinspace Alexander$^{ 23}$,
J.\thinspace Allison$^{ 16}$,
N.\thinspace Altekamp$^{  5}$,
K.J.\thinspace Anderson$^{  9}$,
S.\thinspace Anderson$^{ 12}$,
S.\thinspace Arcelli$^{  2}$,
S.\thinspace Asai$^{ 24}$,
D.\thinspace Axen$^{ 29}$,
G.\thinspace Azuelos$^{ 18,  a}$
% ...
}\end{center}\bigskip
\bigskip
$^{  1}$School of Physics and Space Research, University of Birmingham,
Birmingham B15 2TT, UK
\newline
$^{  2}$Dipartimento di Fisica dell' Universit\`a di Bologna and INFN,
I-40126 Bologna, Italy
\newline
% ...
$^{  5}$Cavendish Laboratory, Cambridge CB3 0HE, UK
\newline
% ...
$^{  8}$CERN, European Organisation for Particle Physics,
CH-1211 Geneva 23, Switzerland
\newline
$^{  9}$Enrico Fermi Institute and Department of Physics,
University of Chicago, Chicago IL 60637, USA
\newline
% ...
$^{ 12}$Indiana University, Department of Physics,
Swain Hall West 117, Bloomington IN 47405, USA
\newline
% ...
$^{ 16}$Department of Physics, Schuster Laboratory, The University,
Manchester M13 9PL, UK
\newline
% ...
$^{ 18}$Laboratoire de Physique Nucl\'eaire, Universit\'e de Montr\'eal,
Montr\'eal, Quebec H3C 3J7, Canada
\newline
% ...
$^{ 23}$Department of Physics and Astronomy, Tel Aviv University,
Tel Aviv 69978, Israel
\newline
$^{ 24}$International Centre for Elementary Particle Physics and
Department of Physics, University of Tokyo, Tokyo 113, and
Kobe University, Kobe 657, Japan
\newline
% ...
$^{ 29}$University of British Columbia, Department of Physics,
Vancouver BC V6T 1Z1, Canada
\newline
% ...
$^{  a}$ and at TRIUMF, Vancouver, Canada V6T 2A3
\newline
% ...
\newline
\bigskip

\newpage
\section{Introduction}
\label{intro} 

In 1996, the LEP collider at CERN entered a new phase of operation, \LepII,
with the first \Pep\Pem\ collisions above the \PWp\PWm\ production
threshold.  Approximately 10~pb$^{-1}$ of integrated
luminosity was delivered to each of the four LEP experiments at a
centre-of-mass energy $\roots\sim 161$~GeV.  Subsequently, 
a further $\sim$10~pb$^{-1}$ was delivered at a higher energy of $\roots\sim
172$~GeV.  This paper describes measurements of the
production and decay properties of
\PWp\PWm\ events in this higher energy run.

One of the
principal goals of the \LepII\ programme
is the measurement of the mass of the W boson, \Mw.  
Comparison between this
direct measurement and the value of \Mw\ determined indirectly from
precise electroweak analyses at $\roots\sim 91$~GeV (\LepI) and
elsewhere will eventually provide an important new test of the
Standard Model (SM) of electroweak interactions.  Direct measurements of
\Mw\ from hadron colliders currently yield
$80.33\pm0.15$~GeV~\cite{bib:pdg,bib:mwpdg,bib:mwd0}.  Ultimately it
is believed~\cite{bib:LEP2YR} that \LepII\ can achieve a precision on
the W mass of approximately 30--40~MeV.  The mass dependence of the
\WW\ production cross-section at threshold, $\roots \sim161$~GeV, 
was used to extract the
first measurements of \Mw\ from \LepII\ 
data~\cite{bib:opalmw1,bib:D-mw161,bib:L-mw161,bib:A-mw161}.  At higher
centre-of-mass energies measurement of the \PWp\PWm\ cross-section
itself provides an interesting test of the non-Abelian gauge structure
of the Standard Model, which
predicts substantial destructive interference 
between the different \PWp\PWm\ 
production diagrams,
 thus avoiding unitarity violation at higher energies.

This paper describes the measurement, above \WW\ production threshold, 
of \Mw\ and of the W decay width, \Gw,
by direct reconstruction of the invariant mass spectrum of the 
W decay products. 
The mass and width measurements use events in
the channels\footnote{Throughout this paper, a reference to \Wp\ or
its decay products implicitly includes the charge conjugate states.}
\WWqqqq\ and \WWqqln\ ($\ell$=e, $\mu$ or $\tau$).  For each event,
the mass of the W is reconstructed from its decay products, and \Mw\ 
and \Gw\ are determined by a fit to the resulting distribution.

The structure of the paper is as follows.  Section~\ref{sec-detector}
contains a brief overview of the OPAL detector 
and the Monte Carlo models used.  
Section~\ref{selection} describes the selection methods
used for the identification of samples of \WWlnln, \WWqqln\ and \WWqqqq\ 
events at $\roots\simeq172$~\GeV.  In Section~\ref{sec-results}, these
selections are used to determine the \PWp\PWm\ production cross-section and
the branching fractions 
of the W boson into various final states.  These results
may be interpreted in terms of a measurement of the
Cabibbo-Kobayashi-Maskawa (CKM) mixing parameter $|V_{\mathrm{cs}}|$.  
In Section~\ref{Wmass} the \WWqqqq\ and \WWqqln\ samples are used for a
determination of the W mass and width.  Finally,
Section~\ref{Wprop} discusses the properties of the hadronic W decays, which
are potentially sensitive to interesting QCD final-state interactions
between the products of the two W decays.

\section{The OPAL Detector, Data and Monte Carlo Models}
 \label{sec-detector}
 

 \subsection{Detector}
 A detailed description of the OPAL detector has been presented elsewhere
 \cite{bib:detector} and therefore only the features
 relevant to this analysis are summarised here.  Charged particle
 trajectories are reconstructed using the cylindrical central tracking
 detectors, which consist of a silicon microvertex detector, a high
 precision vertex detector, a large volume jet chamber and thin
 $z$-chambers.  The silicon microvertex detector consists of two layers of
 silicon strip detectors, allowing at least one hit per charged track in the
 angular\footnote{The OPAL right-handed coordinate system is defined such
   that the origin is at the geometric centre of the jet chamber, $z$ is
   parallel to, and has positive sense along, the e$^-$ beam direction, $r$
   is the coordinate normal to $z$, $\theta$ is the polar angle with respect
   to +$z$ and $\phi$ is the azimuthal angle around $z$.}  region
 $|\cos\theta|<0.93$.  It is surrounded by a vertex drift chamber.
 Outside this lies the jet chamber, about 400~cm in
 length and 185~cm in radius, which provides up to 159 space points per
 track, and measures the ionisation energy loss of charged particles,
 \dedx~\cite{bib:dEdx}. 
  The $z$-chambers, which improve considerably the measurement of
 charged tracks in $\theta$, are situated immediately beyond and co-axial
 with the jet chamber.
 Track finding is nearly 100\% efficient within the angular region $|\cos
 \theta |<0.97$.  The entire central detector is contained within a solenoid
 which provides an axial magnetic field of 0.435~T\@.  
 
 The electromagnetic calorimeter (ECAL) measures the energy of electrons and
 photons and provides a partial energy measurement for hadrons. It consists
 of a cylindrical ensemble of 9440 lead glass blocks arranged such that the
 inter-block gaps point slightly away from the origin, and of two endcaps,
 each having 1132 lead glass blocks aligned parallel to the beam axis. The
 barrel encompasses the angular region $|\cos\theta|<0.82$ whilst the
 endcaps cover the region $0.81<|\cos\theta|<0.98$. 
 
 Calorimeters close to the beam axis measure the luminosity using small
 angle Bhabha scattering events and complete the geometrical acceptance down
 to 34~mrad from the beam axis.  These include the forward detectors which
 are lead-scintillator sandwich calorimeters and, at smaller angles, silicon
 tungsten calorimeters~\cite{bib:SW} located on both sides of the
 interaction point.

 The iron return yoke of the magnet lies outside the electromagnetic
 calorimeter, and is instrumented with streamer tubes as a hadron
 calorimeter (HCAL).  Muon detectors are situated outside the hadron
 calorimeter.  
 Muons with momenta above 3~\GeVc\ usually penetrate
 to the muon detectors.  In addition, up to nine hits may be recorded for
 minimum ionising particles traversing the hadron calorimeter, further
 enhancing muon identification.

\subsection{Data and Monte Carlo}

\label{Monte Carlo}
 
The basic data selection, luminosity measurement, Monte Carlo (MC)
models and detector simulation are identical to those described in
\cite{bib:opalmw1}.  The accepted integrated luminosity, evaluated
using small angle Bhabha scattering events observed in the silicon
tungsten forward calorimeter, is $\intLdtfull\pm\dLstat
\mathrm{(stat.)} \pm\dLsys \mathrm{(syst.)}$~pb$^{-1}$ \cite{bib:LEP2SMpaper}, 
of which
approximately 1 pb$^{-1}$ was collected at 170.3 GeV and 9.3 pb$^{-1}$
at 172.3 GeV.
The luminosity weighted mean 
centre-of-mass energy for the data sample is $\roots
= \rroots \pm 0.06$~\GeV~\cite{bib:LEPenergy}.
 
 The \GENTLE\cite{bib:GENTLE} semi-analytic program is used to calculate the
 Standard Model \WW\ cross-section which is used throughout this paper to
 determine the expected number of \WW\ events.  The use of \GENTLE\ is
 motivated by the fact that it provides a more complete calculation than
 current Monte Carlo generators\cite{bib:LEP2YR}.  
 The calculated cross-section is \GENTxs~pb at
 $\roots = \rroots$~\GeV\ using the current world-average W boson mass of
 $\Mw=80.33$~\GeVcc\ \cite{bib:pdg,bib:mwpdg}.
 
 In the analyses described below, a number of Monte Carlo models were
 used to provide estimates of efficiencies and backgrounds as well as
 the shapes of the W mass distributions. The majority of the
 Monte Carlo samples were
 generated at \roots\ = 171~GeV with $\Mw=80.33$~\GeV.  Unless stated
 otherwise,
 all Monte
 Carlo samples were generated with a full simulation of the OPAL
 detector \cite{bib:GOPAL}. A number of Monte Carlo studies were 
 performed without detector simulation, referred to as
 generator level.
  
 The separation between signal and
 background processes is complicated by the interference between the
 \WW\ production diagrams (class\footnote{In this paper,
   the W pair production diagrams, {\em i.e.} 
   $t$-channel $\nu_{\mathrm{e}}$ exchange and $s$-channel \Zgamma\ exchange, 
   are referred to as ``\CC'',
   following the notation of \cite{bib:LEP2YR}.} \CC) and other
 four-fermion graphs.  For example, the process \Zqq\ where a \Wpm\ is
 radiated off one of the quarks can interfere with \WWqqln\ and
 \WWqqqq\ final states.
 Monte Carlo samples of \WW\ events, restricted to the CC03 diagrams, were
 obtained with the \KORALW \cite{bib:KORALW}, 
 \EXCALIBUR \cite{bib:EXCALIBUR}, \grcff \cite{bib:GRC4F}, \PYTHIA
 \cite{bib:PYTHIA} and \HERWIG \cite{bib:HERWIG} generators. \KORALW\ was
 used to determine the efficiencies for \WW\ events for the selections
 presented in this paper. This sample was generated at \roots\ = 171~GeV
 using a value for \Mw\ of 80.33~GeV. A number of \PYTHIA\ Monte Carlo
 samples generated with different values of \roots\ and \Mw\ were used
 to investigate the sensitivities of the analyses to these parameters.
 \EXCALIBUR\ was used to investigate the sensitivity to \Gw.
 
 The main background process, \Zqq, was simulated using \PYTHIA, with
 \HERWIG\ used as an alternative to study possible systematic effects.
 Other backgrounds involving two fermions in the final state were studied
 using \KORALZ\cite{bib:KORALZ} for \eemumu, \eetautau\ and \eenunu, and
 \BHWIDE\cite{bib:BHWIDE} for \eeee.  Backgrounds from processes with four
 fermions in the final state were evaluated using \grcff, \EXCALIBUR\ and
 \FERMISV\cite{bib:FERMISV}.  Backgrounds from two-photon processes were
 evaluated using \PYTHIA, \HERWIG, \PHOJET\cite{bib:PHOJET},
 \TWOGEN\cite{bib:TWOGEN} and the Vermaseren program\cite{bib:VERMASEREN}.
 At least two independent Monte Carlo estimates were available for each
 category of two-photon and four-fermion background.
  To study the influence of interference effects in the four-fermion final
 states the \grcff\ and \EXCALIBUR\ Monte Carlo generators were used. In
 both cases samples were generated using the full set of interfering four
 fermion diagrams. These four-fermion samples were compared to samples
 obtained with the same generator using only the CC03 set of W 
 pair production diagrams.

 Throughout this paper it is assumed that the production cross-section 
 at $\roots \sim$172~GeV for
 the Standard Model Higgs boson, \Hz, is negligible. This assumption is
 valid for a Higgs mass above 80~GeV. Below this mass, the cross-section
 becomes significant and the \WW\ event selections have high efficiencies to
 select \Hz\ events, particularly for the process \HZqqqq\, which would
 result in a non-negligible background.

\section{Event Selection}
\label{selection} 

Event selections have been developed to  identify efficiently 
all Standard Model 
\WW\ final states with low acceptance for background processes. 
The \WW\  event selection consists of three distinct parts to select
fully leptonic decays, \WWlnln, semi-leptonic decays, \WWqqln, and 
fully hadronic decays, \WWqqqq.
To ensure that the selections are mutually exclusive,
only events failing the  \WWlnln\ selection are considered as possible
\WWqqln\ candidates and only events failing both the \WWlnln\ and \WWqqln\
selections are considered as possible \WWqqqq\ candidates. 
The  fully leptonic and semi-leptonic selections
are separated into the individual lepton types giving ten selected 
final states:  \enen, \enmn, \entn, \mnmn, \mntn, \tntn, \qqen, \qqmn, 
\qqtn\ and \qqqq. No attempt has been made to identify the flavour
composition of the hadronic final states. 

The hadronic and semi-leptonic event samples are used to
determine the decay width and the mass of the W boson and also to
study hadronic event properties of W boson decays.
All final states are used in the measurement of the W pair
production
cross-sections and decay branching fractions. 
The classification into separate
leptonic final states allows the measurement of the W branching ratios
with or without the assumption of charged current lepton universality.

An overview of the event selections is given below. Emphasis is placed on
the performance of the selections and, in particular, the systematic 
uncertainties in the selection efficiencies and accepted background
cross-sections. More detailed descriptions 
of the \WWqqln\ and \WWqqqq\ event selections can be found in the Appendices.

\subsection{Treatment of Systematic Uncertainties}
 
The \KORALW\ Monte Carlo program 
(\CC\ diagrams only) was used to estimate the
efficiencies of each of the selections.  Two types of systematic
uncertainties in the selection efficiencies have been considered:
generator uncertainties and differences between data and the
Monte Carlo simulation, including detector simulation.  The generator
uncertainties were estimated by comparing the efficiencies for
four different Monte Carlo generators, \KORALW, \grcff, \EXCALIBUR\ 
and \PYTHIA.  The uncertainties arising from the beam
energy dependence of the selection efficiencies were assessed by
comparing \PYTHIA\ samples generated at different beam energies.  The
dominant uncertainties due to the beam energy dependence arise because
the Monte Carlo samples used to evaluate efficiencies
were generated at a centre-of-mass energy of 171~GeV rather than
$\rroots$~GeV, the mean energy
at which the data were recorded.  Similarly, the
propagation of uncertainties on \MW\ to the selection efficiencies was
estimated using samples of \PYTHIA\ events (including detector
simulation) generated with different values of \MW.

For each selection, the four-fermion backgrounds were determined using
the \grcff\ Monte Carlo program. The systematic uncertainties on the
four-fermion backgrounds were estimated from the difference between
the accepted cross-sections predicted by the \grcff\ and \EXCALIBUR\ 
Monte Carlo programs.  
For both the \WWqqln\ and \WWqqqq\ selections, the
dominant systematic uncertainty on the background was due to the
modelling of the \Zqq\ process.


\subsection{\boldmath \WWlnln\ Events}
 \label{sec-lnln}
\subsubsection{Selection} 
 Approximately 11\% of W pair events are expected to
 decay through the fully leptonic channel.
 These events may be observed as an acoplanar pair
 of charged leptons with missing momentum. The selection is sensitive to the
 six possible classes of observed leptons, $\ee$, $\mm$, $\tautau$, $\emu$,
 $\et$, $\mt$, which are
 expected to be produced in the ratio 1:1:1:2:2:2.  The main
 backgrounds are $\epem\rightarrow \Zz\Zz$, \Zee,
 \Wenu, $\epem\rightarrow\tptm$ and $\epem \rightarrow \epem \lplm$.
 
 The experimental signature for \WWlnln\ events
 is also the signature for a number of non-Standard Model processes. 
 The selection used to identify
 \WWlnln\ events is based on the general charged lepton pair 
 selection used in the search for anomalous production of 
 lepton pair events with significant missing transverse 
 momentum\cite{bib:ACOPLL}. 
 A high efficiency for selecting fully leptonic events 
 is obtained by forming the logical ``or''
 of two distinct 
 analyses\footnote{Referred to as Selection I and 
 Selection II in Reference \cite{bib:ACOPLL}. 
 For Selection I the additional \WW\ selection criteria are applied.}.
 The first analysis performs
 a general selection of events conforming to the acoplanar di-lepton
 topology. The second consists of
 several selections, each designed to select a particular di-lepton class.
 This analysis  was optimised to identify events consistent with
 being \WWlnln.
 Both analyses  
 require evidence that a pair of charged leptons has
 been produced in association with an 
 invisible system that carries away
 significant transverse momentum. 
 The first analysis simply requires leptonic events 
 with significant missing transverse momentum, 
 whereas the second analysis exploits  
 the kinematics of \WWlnln\ events and the flavour mixture
 of the expected backgrounds.
 This leads to a complementary acceptance. 

 From \Koralw\ Monte Carlo studies, 
 80\% of the expected efficiency for \WWlnln\ 
 events is common to
 the two analyses, 6\% is exclusive to the first analysis, 
 and 14\% to the second. Both analyses have 
 similar efficiencies for \WWlnln\ 
 events with two stable leptons, while the second analysis performs
 significantly better for events with one or two taus.

\subsubsection{\boldmath \WWlnln\ Classification}

Selected \WWlnln\ candidates 
are classified as one of the six possible di-lepton
combinations
according to the electron and muon identification 
results, the scaled energy of the leptons and
the charged track and cluster multiplicity associated with 
each lepton.
Jets are defined using the cone algorithm\cite{bib:conealg}. 
Those not identified as either electrons or muons are classified as taus.
A small fraction of selected events are expected to be reconstructed
with only one jet. 
These are events where one of the charged leptons is outside
the geometric acceptance of the tracking chambers.
For these one-jet events, the unobserved lepton is taken to be a muon.
For two-jet events, corresponding to most of the 
selected events, the classification is further elaborated 
to improve the assignment of secondary electrons and muons
from tau decays to the tau class.
Identified electrons or muons with energy scaled by the beam energy
less than 0.3 are classified as taus. 
In order to recuperate inefficiencies in the 
electron and muon identification algorithm,
jets failing the electron and muon identification, but with 
scaled energy greater than 0.5 are reclassified 
as electrons if E/p exceeds 0.5 and as muons if E/p is less than 0.5.
Lastly, jets with three or more charged tracks or three or more 
associated electromagnetic clusters (consistent with 
three-prong tau decays) are classified as taus.

 \subsubsection{Results and Systematic Errors} 

 A total of eight events is selected by the combined event selection
 at $\sqrt{s}=172$~GeV
 as \WWlnln\ candidates.
 Seven of the events are common to both analyses, while one 
 event is exclusively selected by the first analysis.
 
\begin{figure}
 \centerline{\epsfig{file=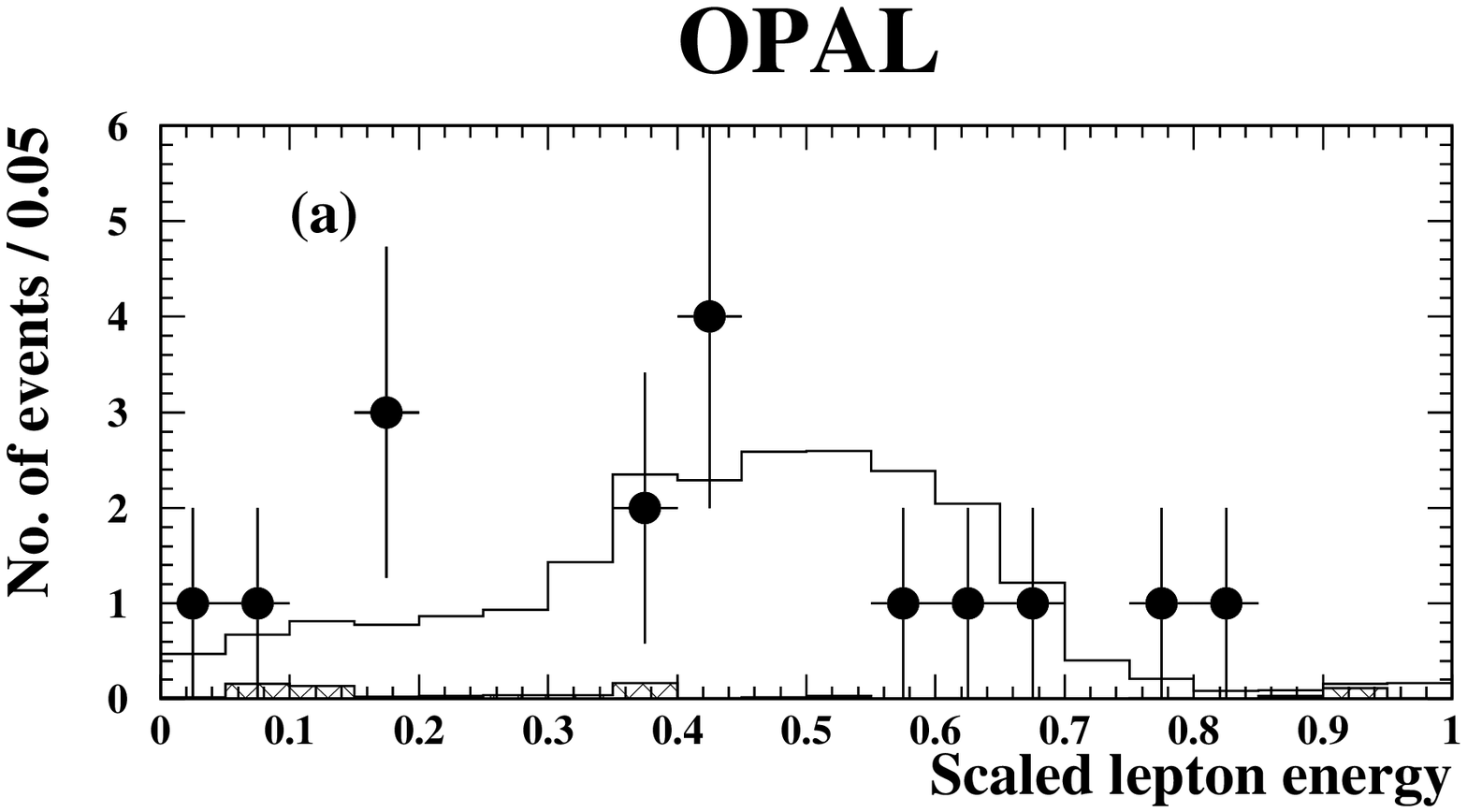,width=15cm}}
 \centerline{\epsfig{file=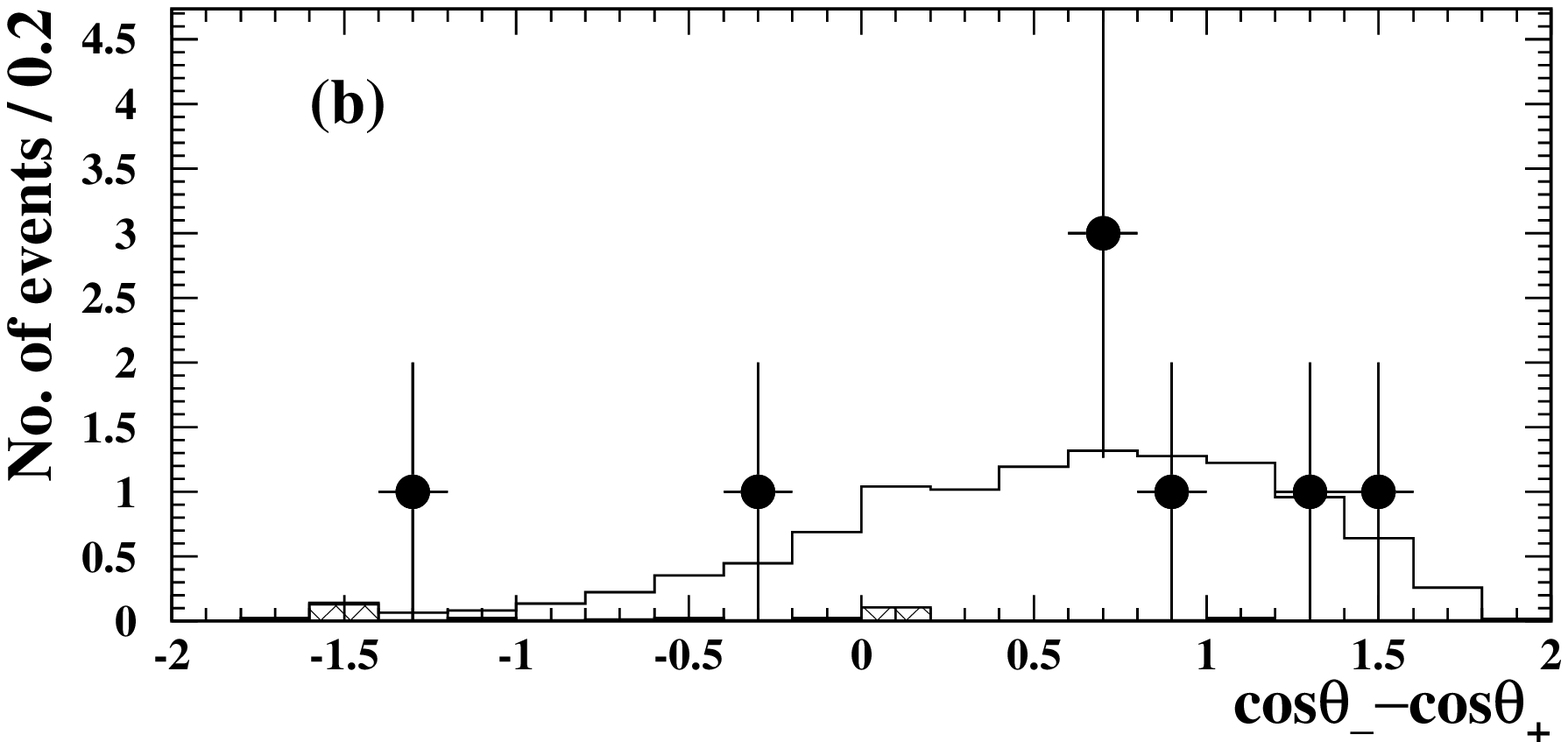,width=15cm}}
 \caption{Distributions of kinematic variables for selected 
   \WWlnln\ events, showing: (a) the lepton energy divided by the beam
   energy, and (b) $\cos\theta_{-}-\cos\theta_{+}$.  
   The data are shown as points with error
   bars. The Monte Carlo prediction for the sum of \WW\ and all other
   Standard Model processes is shown as the open histogram, while the
   non-\WW\ processes are represented by a doubly hatched histogram. The
   Monte Carlo samples have been normalised to the collected integrated
   luminosity of $\intLdt$~pb$^{-1}$.}
 \label{fig-lnln1}
\end{figure}

 Representative kinematic distributions for the selected events
 are given in Figure~\ref{fig-lnln1}, together with the Monte Carlo
 expectations.  Figure~\ref{fig-lnln1}(a) shows the energy of each
 charged lepton scaled by the beam energy. Figure~\ref{fig-lnln1}(b)
 shows  $\cos\theta_{-}-\cos\theta_{+}$, where 
 $\theta_{-}$ and $\theta_{+}$ are the polar angles of
 negatively charged and positively charged leptons respectively.
 The selected events favour positive values of
 $\cos\theta_{-}-\cos\theta_{+}$, 
 as expected for \WW\ production. These kinematic
 distributions are consistent with the Standard Model expectations.
 
 The Monte Carlo selection efficiencies for
 each di-lepton combination  are shown in Table~\ref{tab-lnln-eff}.
 The
 overall efficiency for selecting \WWlnln\ events is determined to be
 $(78.3\pm0.4\pm2.5)\%$, where the errors are statistical and
 systematic, respectively. Since events can be rejected on basis of having
 a significant energy deposit in the forward detectors,
 a correction factor of 0.99 with a systematic uncertainty of 0.01 has 
 been applied
 to the selection efficiencies and the background estimate to account for 
 detector occupancy due to off-momentum beam particles.  
 Systematic uncertainties on the efficiency were estimated 
 based on comparisons of the
 efficiencies for different Monte Carlo 
 models (3.0\% for $\enen$, 4.5\% for $\tntn$ and 2.0\% for
 the other fully leptonic final states). The Monte Carlo models have  
 different implementations of both
 initial and final state radiation effects and the modelling of
 tau decays. In addition, systematic errors
 were assigned to account for
 data/Monte Carlo agreement (0.8\%) and the knowledge of
 the trigger efficiency (0.4\%). 

 The expected background cross-sections from
 Standard Model processes are given in Table~\ref{tab-lnln1}. 
 The systematic
 errors are based on comparisons of different Monte Carlo generators. 
 Table~\ref{tab-sel_summary} lists the expected and observed numbers of events
 in each final state. The 8 fully leptonic events observed in the data
 are consistent with the Standard Model expectation.


\subsection{\boldmath \WWqqln\ Events}
 \label{sec-qqln}
 \subsubsection{Selection}
Semi-leptonic final states, \WWqqln, are expected to
comprise 44\% of \WW\ decays.
\WWqqen\ and \WWqqmn\ events are characterised by two 
well-separated hadronic jets, a high momentum lepton and 
missing momentum due to the unobserved neutrino. The signature for 
\WWqqtn\ events is two well separated jets from
the hadronic W decay and one low multiplicity jet typically consisting of
one or three tracks. The expected missing momentum is less well defined 
due to the additional neutrino(s) from the decay of the tau.

The \WWqqln\ event selection, described in Appendix
\ref{app-qqln},  consists of three separate selections, one
for each type of semi-leptonic decay. The \WWqqtn\ selection
is applied only to events which fail the \WWqqen\ and
\WWqqmn\ selections.
Each selection proceeds in four stages:
\begin{itemize}
 \item {\bf Identification of the Candidate Lepton:}
       The track with
       the highest probability of being 
       a lepton from either the decay \Wev\ or
       \Wmv\ is identified. The \WWqqtn\ selection 
       uses the track (or tracks) most consistent with being from a 
       tau decay from \Wtv .    
     \item {\bf Preselection:} Cuts are applied to the data to
       reduce the background from \Zqq\ events.
     \item {\bf Relative Likelihood Selection:} Relative likelihood
       selections, based on kinematic variables and the lepton candidate,
       are used to distinguish signal events (\WWqqen, \WWqqmn\ and \WWqqtn)
       from \Zqq\ background events.  Events passing the likelihood
       selection are considered to be \WWqqln\ candidates.
     \item {\bf Event Categorisation:} Having suppressed the background
       using the relative likelihood selection, a second relative likelihood
       is applied to the \WWqqen\ and \WWqqmn\ 
       candidates in order to categorise them as
       either \WWqqen, \WWqqmn\ or \WWqqtn\ events. 
\end{itemize}

\subsubsection{Results and Systematic Errors}
\label{sec-qqln-res}

Table \ref{qqln-xtalk} shows the efficiencies of the selections
for \WWqqln\ events after categorisation into the different channels.
These efficiences 
include corrections which account for observed differences between the data 
and the Monte Carlo simulation. 
The uncertainties include both systematic and statistical contributions.

%
Table~\ref{qqln-sys}
lists the sources of the uncertainties evaluated for the selection
efficiencies.  
The respective efficiencies of the \WWqqen, \WWqqmn\ and \WWqqtn\ 
selections determined from \PYTHIA, \KORALW, \EXCALIBUR\ and \grcff\ are
found to be consistent within errors.  Efficiency corrections and systematic
errors arising from discrepancies between data and Monte Carlo simulation
were determined by studying data and Monte Carlo ``mixed events'' formed by
superimposing LEP1 hadronic \Zz\ decay events and single hemispheres from LEP1 
events identified as \Zz\ decays to charged lepton pairs.
Overall corrections to the efficiencies of $0.987\pm0.006$, $0.999\pm0.004$
and $1.025\pm0.007$ were found for the \WWqqen, \WWqqmn\ and \WWqqtn\ 
selections respectively. The differences between data and Monte Carlo result
in a migration of events from the \WWqqen\ selection to the \WWqqtn\ 
selection. The main difference arises from the simulation of the electron
identification variables used in the relative likelihoods.  The
efficiencies summarised in
Table \ref{qqln-xtalk} include these corrections.

Table~\ref{qqln-back} shows the corrected background cross-sections and
total uncertainties for the three selections.  The systematic errors on the
expected background cross-sections are dominated by differences between data
and Monte Carlo for the two-fermion backgrounds and by differences between
generators in the case of the four-fermion backgrounds.

The \WWqqln\ selection efficiency for \Zqq\ events is small, $\sim0.1\%$. 
The background estimate
from \Herwig\ is consistent with that from \PYTHIA. Since the efficiency is
very low, the Monte Carlo estimate of the background level 
from this source is likely to be sensitive
to the simulation of the tails of distributions. 
For this reason, the \Zqq\ background is estimated from
the data. The background was found to be $1.2\pm0.5$ times
that predicted by the Monte Carlo.  This was determined by fitting the
observed likelihood distribution in the region 0.25--0.75, 
for all data recorded away from the \Zz\ 
peak (130--140~GeV, 161~GeV and 172~GeV ), with signal and
background components where the shapes of the respective likelihood 
distributions are taken from Monte Carlo. 
The quoted error includes a systematic
component arising from the variation of the fit region. 

Four-fermion backgrounds were estimated using the \grcff\ generator.
Differences between the four-fermion background predictions of the \grcff\ 
and \EXCALIBUR\ generators were used to assign systematic uncertainties. 
The W boson pair production cross-section for \qqen\ 
was compared with that from the full four-fermion treatment and 
the difference in the accepted cross-sections taken as the effective 
background. This is important  since  the background
from \Wenu\ can 
interfere with \WWqqen. Generator level ({\em i.e.} 
without detector simulation)
studies  
using \EXCALIBUR\ indicated that the  contributions
from non-\CC\ diagrams for the \qqmn\ and
\qqtn\ final states are negligible for the experimental acceptance.
Two-photon backgrounds are included in the \eeff\ four-fermion background.
For the two-photon background and 
each class of four-fermion background at least two independent Monte Carlo
determinations were used and differences between the predicted background
cross-sections taken as systematic uncertainties.

\begin{figure}
 \centerline{\epsfig{file=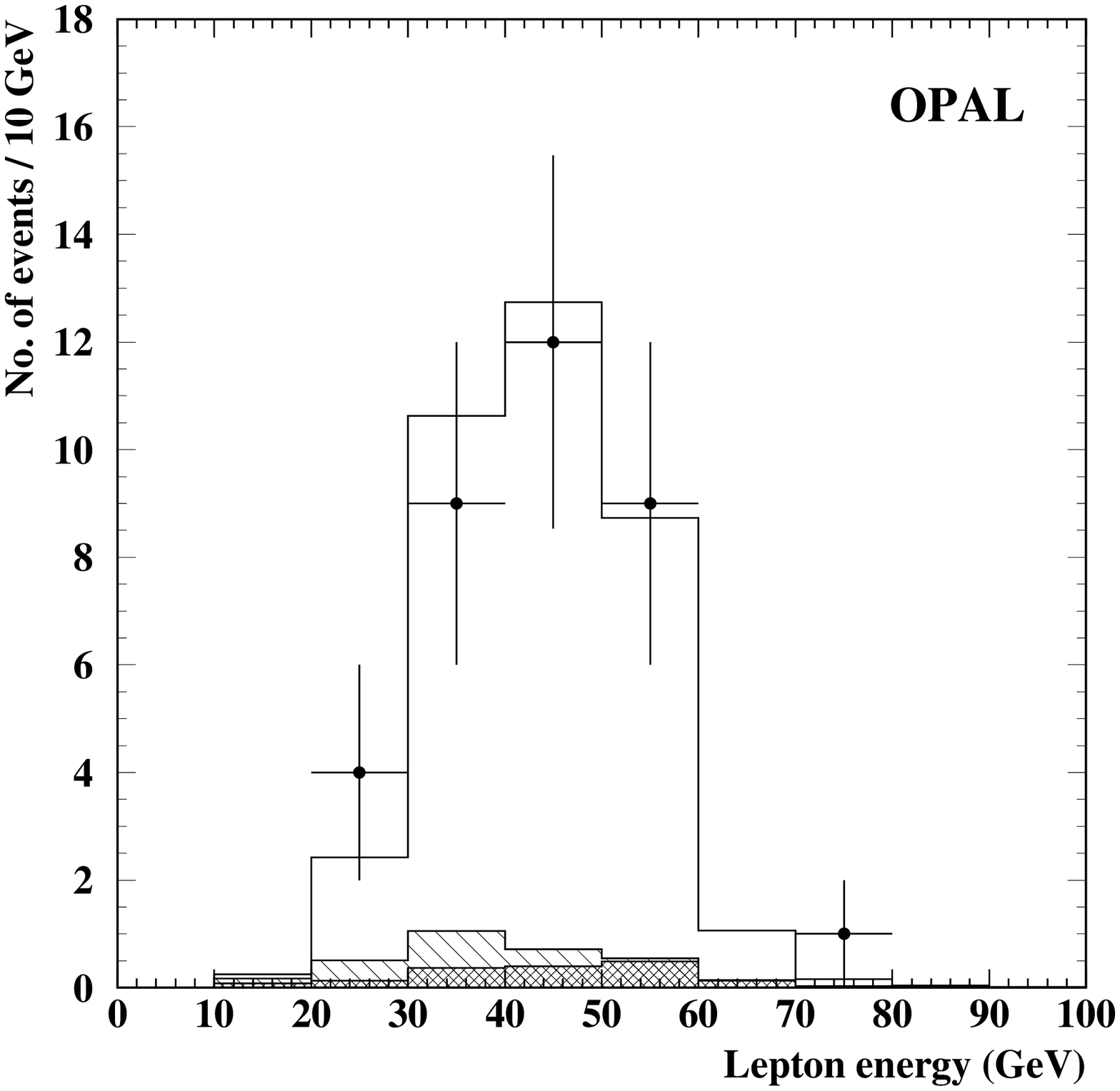,%
width=15 cm}}
 \caption{The energy of the lepton for events selected as \WWqqen\ or 
   \WWqqmn\ after categorisation.  For electrons this is determined using
   the ECAL energy and for muons using the track momentum.  The open
   histogram shows the Monte Carlo prediction (\PYTHIA) for an integrated
   luminosity of $\intLdt$~pb$^{-1}$ and total \WW\ cross-section of
   \GENTxs~pb.  The contribution from \WWqqtn\ decays is shown as the single
   hatched histogram and the contribution from background processes as the
   doubly hatched histogram. The data are shown as the points with error
   bars.}
 \label{fig_qqln_3}
\end{figure}

\begin{figure}
 \centerline{\epsfig{file=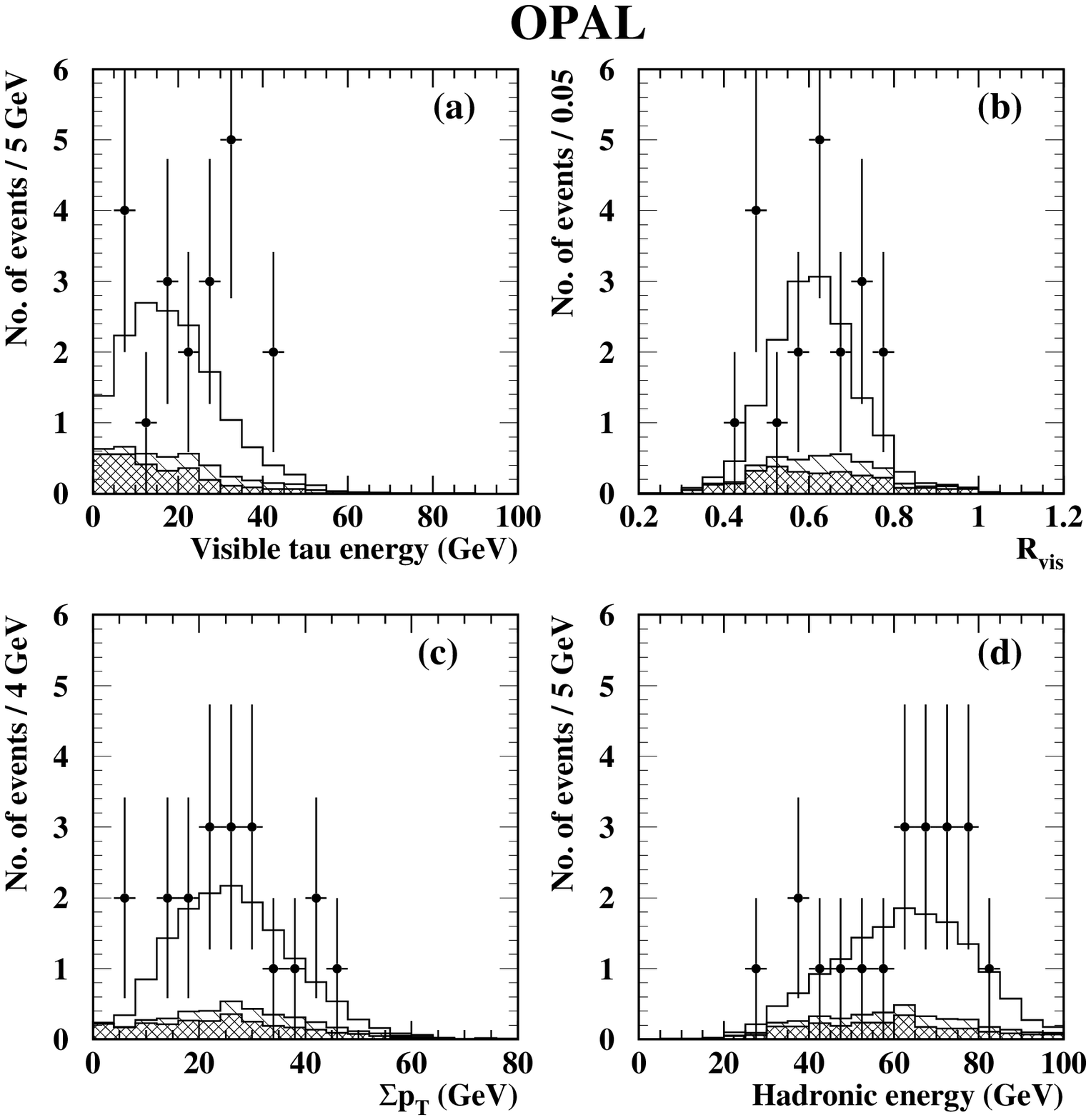,%
width=15 cm}}
 \caption{Distributions of kinematic variables for selected 
   \WWqqtn\ events, showing: (a) estimated visible energy of the tau decay,
   excluding the energy of the neutrino(s), (b) visible energy of the event
   divided by $\protect\roots$ \Rvis,
   (c) net transverse momentum \Ptsum\ of the
   event, and (d) uncorrected energy of the two hadronic jets in the event,
   calculated using tracks and unassociated ECAL clusters, when the jet
   containing the tau candidate is excluded (this variable is not used in
   the likelihood).The open histogram shows the Monte Carlo prediction
   (\PYTHIA) for an integrated luminosity of $\intLdt$~pb$^{-1}$ and total
   \WW\ cross-section of \GENTxs~pb. The contribution from \WWqqen\ and
   \WWqqmn\ decays is shown by the single hatched histogram and the
   contribution from background sources as the doubly hatched histogram. The
   data are shown as the points with error bars.}
 \label{fig_qqln_4}
\end{figure}

In the 172~GeV data sample 19 \WWqqen\ events, 16 \WWqqmn\ events and 20
\WWqqtn\ events were observed in agreement with the expectation shown in
Table \ref{tab-sel_summary}. Figure~\ref{fig_qqln_3} shows the distribution
of lepton energy for accepted events categorised as \WWqqen\ and \WWqqmn.
Figure~\ref{fig_qqln_4} shows four example distributions for selected
\WWqqtn\ events. The observed distributions for the data are in good
agreement with the Monte Carlo expectations.


\subsection{\boldmath \WWqqqq\ Events}
\label{sec-qqqq}
 \subsubsection{Selection}

 Fully hadronic decays, \WWqqqq, which are expected to 
 comprise 46\% of the total \WW\ cross-section, are
 characterised by four energetic, hadronic jets and little missing energy.
 A selection, described in Appendix \ref{app-qqqq},
 consisting of preselection cuts and a  
 likelihood analysis
 is used to separate \WWqqqq\ events from the background.
 The preselection removes events which are likely to be from the
 process \Zqq. 
 A relative likelihood for an event being from the process 
 \WWqqqq\ rather than from the dominant \Zgamma\ background is then
 estimated using seven kinematic variables. 
 For the selection used in the determination of the mass of the W boson
 and the studies of \WW\ event properties
 a cut is placed on the value of the relative likelihood.
 The relative likelihood
 selection is designed to maximise the product of efficiency and purity
 while limiting possible distortion of the W mass spectrum.  

 \subsubsection{Event Weights}

 A modified
 version of the above relative
 likelihood is used to give a weight, $w_i$, to each event 
 reflecting the probability that the event originates from a \WWqqqq\ 
 decay.
 These event weights, described in more detail in 
 Appendix \ref{sec-event-weights}, 
 are used to reduce the statistical uncertainty on the
 cross-section measurement. 
 Given the expected average event weights for
 both \WWqqqq\ and background events, 
 $\overline{w_s}$ and  $\overline{w_b}$, 
 the measured \WWqqqq\ cross-section, for an integrated luminosity, $L$,
 can be expressed as
\begin{eqnarray*}
        {{\sum_i w_i-
   L\sigma^{\mathrm{pre}}_{\mathrm{bgd}}\overline{w_b}}\over{
     L \epsilon^{\mathrm{pre}}_{\mathrm{sig}}
         \overline{w_s}}},
\end{eqnarray*}
where the summation over $i$ corresponds to summing the weights for
the observed events which pass the preselection,
$\epsilon^{\mathrm{pre}}_{\mathrm{sig}}$ is the efficiency of the 
preselection for \WWqqqq\ and $\sigma^{\mathrm{pre}}_{\mathrm{bgd}}$ is the 
accepted background cross-section after the preselection.
The use of event weights results in a 5\% reduction of the expected 
statistical uncertainty on the measured \WWqqqq\ cross-section. 

\subsubsection{Results and Systematic Errors}
\label{sec-qqqq-res}

The efficiency of the likelihood selection for \WWqqqq\ events is estimated
from the \KORALW\ Monte Carlo simulation 
to be $(79.8\pm0.2\pm1.2)$\%, where the errors are statistical
and systematic respectively.  
The total expected
background cross-section, $\sigma_{\mathrm{bgd}}$, and the contributions
from different processes are given in Table~\ref{qqqq-back}. 
Also shown is the weighted background cross-section
which is the expected sum of the event weights for background events passing
the preselection.  For the event weight method, the other
parameters necessary to
determine the cross-section are the preselection efficiency for \WWqqqq\ 
events, $\epsilon^{\mathrm{pre}}_{\mathrm{sig}}=(90.3\pm$0.1)\%, 
and the average event weight for
\WWqqqq\ events which pass the preselection, $\overline{w_s}=0.778\pm0.001$, 
where the errors are statistical. 

The systematic uncertainties evaluated for the \WWqqqq\ selection efficiency
are given in Table~\ref{qqqq-sys}. This table shows uncertainties on the
efficiencies and backgrounds for
both the likelihood cut selection and for the event weight method.
The main systematic uncertainty arises
from the QCD modelling of both the \WWqqqq\ process and the dominant
background from \Zqq.  The likelihood function has been re-evaluated using
an alternative QCD Monte Carlo model (\HERWIG) and by varying the 
parameters $\sigma_q, b, \Lambda_{QCD}$ and $Q_0$ of the \PYTHIA\ Monte Carlo
by $\pm$ one standard deviation
about their tuned values\cite{bib:qqqqQCD}.  The Monte
Carlo distributions used for the likelihood probabilities were rescaled so
that the means of the distributions agreed with those observed in the 172
GeV data and the analysis repeated.  The numbers of LEP1 data and Monte
Carlo events passing the selection, with a modified preselection, were
compared. The selection was compared for Monte Carlo events generated at
different beam energies and with different W masses.  Finally the effect of
rebinning the distributions in the likelihood probability calculation was
investigated. The differences observed in each case are taken as systematic
errors.

The estimated signal efficiency and expected background cross-section are
used to determine the expected numbers of signal and background events in
Table~\ref{qqqq-events}. The observed numbers for the preselection,
the likelihood cut and the summed event weight are consistent with the 
expectations. The statistical
error on the observed summed event weight is calculated
as the square root of the sum of the squares of the weights 
for the events passing the preselection.

 \section{\boldmath \WW\ Cross-Section and the W Decay
 Branching Fractions}
 \label{sec-results}
 \subsection{\boldmath 172~\GeV\ Results}

 The observed 
 numbers of selected \WW\ events have been used to measure the
 \WW\ production cross-section 
 and the W decay branching fractions to leptons and
 hadrons.  The measured cross-section corresponds to 
 that of W pair production from the \CC\ diagrams. 
 Additional diagrams can result in the same 
 four-fermion final states as produced in the decays of \WW\ and can
 therefore interfere.
 When calculating the expected non-\CC\ backgrounds for the various
 event selections the effects of this interference have been included.
 The expected four-fermion backgrounds, quoted throughout this 
 paper, include both contributions from non-\CC\ diagrams and the effects of 
 interference.  The four-fermion backgrounds
 for each final state  are calculated from the difference
 between the accepted four-fermion cross-section including all diagrams,  
 and the accepted \CC\ cross-section. For this determination
 the \grcff\ Monte Carlo was used. The
 associated systematic uncertainty is estimated by comparing the
 predictions of \grcff\ and \Excalibur. 
 At the current level of statistical precision
 such interference effects are small, but not negligible,
 for the experimental acceptance.
 In making the measurement, Standard Model expectations
 for four-fermion processes are used.
 
 This division into \CC\ \WW\ events and background events is shown
 in Table~\ref{tab-sel_summary}, for a W mass of
 $80.33$~\GeVcc~\cite{bib:pdg,bib:mwpdg} and a centre-of-mass energy
 of $\rroots$~GeV.  The data are consistent with the Monte Carlo
 expectation.  The systematic uncertainties on the expected numbers of
 signal events include contributions from the current errors of
 $\pm$0.15~\GeVcc~\cite{bib:pdg,bib:mwpdg} on \Mw\ and $\pm$0.03~GeV
 on the beam energy~\cite{bib:LEPenergy} (0.9\% and 0.2\%, respectively).  A
 \WW\ cross-section of \GENTxs~pb predicted by \GENTLE\ was used, with
 a theoretical uncertainty of $\pm2\%$. 
  Uncertainties in the selection efficiencies, accepted
 background cross-sections and luminosity have been given in
 Sections~\ref{Monte Carlo}--\ref{sec-qqqq}.

 The \WW\ cross-section and branching fractions are 
 measured using the information from the
 ten separate channels.  For each of the ten 
 final states, $i$, the probability of obtaining the number of
 observed events is calculated as a function of the \WW\ cross-section,
 \sigccthree, and the W branching fractions.
 A likelihood
 $\mathcal{L}$ is formed from the product of these Poisson probabilities,
 $\mathcal{P}_i$, of observing $N_i$ events for a Monte Carlo prediction
 of $\mu_i$ events:
\begin{eqnarray*}
  \mathcal{L} = \prod_i \mathcal{P}_i(N_i,\mu_i) & = &
  \prod_i \frac{\mu_i^{N_i}e^{-\mu_i}}{N_i!}.
\label{eqn-likelihood}
\end{eqnarray*}
For the \WWqqqq\ selection the likelihood is calculated using the
summed event weights for the 99 preselected events
with a Gaussian error of $\pm6.6$ corresponding to the 
square root of the sum of the squared event weights.  
The expected number of events in each of the ten channels,
$\mu_i$, can be expressed in terms of the luminosity, the total (CC03)
cross-section at \roots\ of 172.12~GeV, $\sigccthree(172~\GeV)$, the W
boson branching fractions, $\Br(\mathrm{W}\rightarrow X)$, the
background cross-section in each channel and the efficiency matrix for
the \WW\ selections, $\epsilon_{ij}$. The entries in the matrix,
$\epsilon_{ij}$, shown in Table~\ref{tab:eff_matrix}, are the
efficiencies of the event selections $i$ for a \WW\ decay of type $j$.
The off-diagonal elements of
the matrix determine the acceptances of each selection for other \WW\ 
decays.

Three different maximum likelihood fits have been performed. In the first case
\sigccthree(172~GeV), $\Br(\Wev)$, $\Br(\Wmv)$ and $\Br(\Wtv)$ are
extracted under the assumption that
\begin{eqnarray*}
     \Br(\Wev)+\Br(\Wmv)+\Br(\Wtv)+\Br(\Wqq) & = & 1,
\end{eqnarray*}
giving:
\begin{eqnarray*}
 \Br(\Wev)    & = & 0.107^{+0.025}_{-0.022} \pm 0.004, \\ 
 \Br(\Wmv)    & = & 0.076^{+0.021}_{-0.019} \pm 0.003, \\ 
 \Br(\Wtv)    & = & 0.128^{+0.032}_{-0.029} \pm 0.005, \\ 
 \sigccthree(172~\GeV) & = & 12.5  \pm 1.3 \pm 0.4~\mathrm{pb},
\end{eqnarray*}
where the first uncertainty is statistical
and the second systematic. The systematic error includes
contributions from the uncertainties in the efficiency, background
cross-section and luminosity. The largest systematic error arises from
the uncertainty in the backgrounds in the \WW\ selections.  
The correlations between the above measurements are less than 30\%.
In the second fit the additional constraint of charged current lepton
universality is imposed\footnote{For the current level 
of experimental precision, the effect of \Br(\Wtv) being $\sim0.1\%$ 
lower\cite{bib:LEP2YR}
than \Br(\Wev) and \Br(\Wmv) has been neglected.}:
\begin{eqnarray*}
     \Br(\Wev)=\Br(\Wmv)=\Br(\Wtv)=\Br(\Wlv),  
\end{eqnarray*}
giving
\begin{eqnarray*}
 \Br(\Wlv)    & = & 0.102^{+0.012}_{-0.011}\pm0.003, \\ 
 \sigccthree(172~\GeV) & = & 12.3  \pm 1.3 \pm 0.4 ~\mathrm{pb}.
\end{eqnarray*}
This value for $\Br(\Wlv)$ implies a value for the hadronic
branching fraction, $\Br(\Wqq)$, of
\begin{eqnarray*}
 \Br(\Wqq) & = & 0.694^{+0.033}_{-0.035}\pm0.008.
\end{eqnarray*}
In the final fit, the branching fractions are fixed to their Standard
Model values and the \CC\ cross-section is determined to be
\begin{displaymath}
 \sigccthree(172~\GeV) = 12.3 \pm 1.3 \pm 0.3 ~\mathrm{pb}.
\end{displaymath}
The \CC\ cross-section at 172.12~GeV depends, 
albeit weakly, on the W mass.  Therefore the
measured cross-section, \sigccthree(172~\GeV), 
can be used to obtain a measurement of \Mw:
\begin{displaymath}
 \Mw= 80.5 ^{+1.4}_{-2.2} \mbox{} ^{+0.5}_{-0.6}~\GeV,
\end{displaymath}
where the first uncertainty is statistical and the second is
systematic. The latter uncertainty includes a 2\% theoretical
component.
%
The result of the fit using the Standard Model branching fractions is
consistent with that obtained by taking the total number of observed
events divided by the luminosity, subtracting the total expected
background cross-section and dividing by the overall effective
selection efficiency of $(77.4\pm1.0)\%$. This gives \sigccthree\ =
$12.3^{+1.4}_{-1.3}~\mathrm{pb}$, where the error is the combined
statistical and systematic uncertainty.

\begin{figure}[tbhp]
 \epsfxsize=\textwidth
 \epsffile{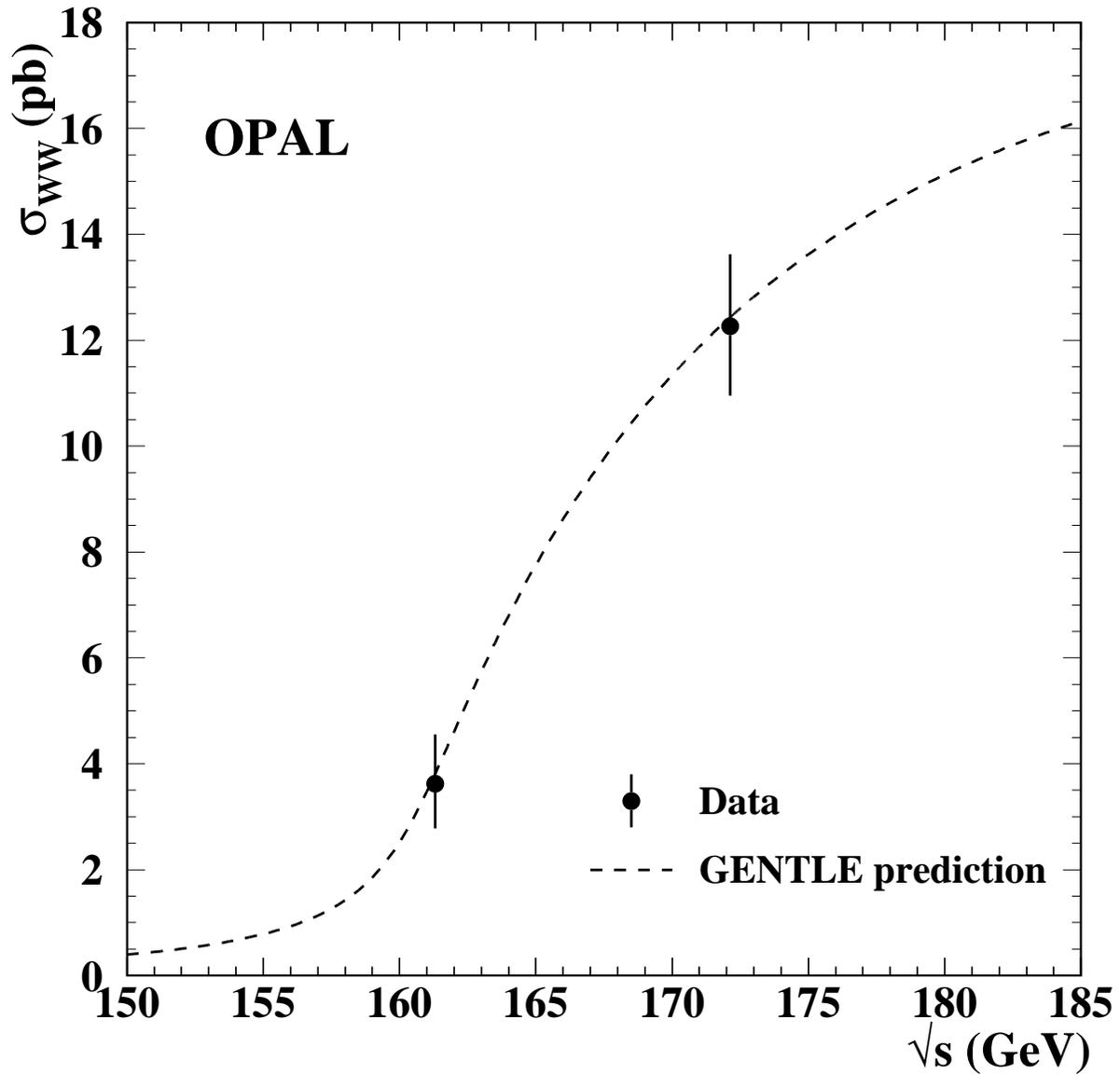}
   \caption{The dependence of \sigccthree\ on
     $\protect\roots$, as predicted by \GENTLE\ for $\Mw = 80.33$~GeV. The
     \WW\ cross-sections measured at $\protect\roots=\rroots$~GeV (this
     paper) and at $\protect\roots=161.3$~GeV \protect\cite{bib:opalmw1}, are
     shown.  The error bars include statistical and systematic
     contributions.}
 \label{fig-sigmaww} 
\end{figure}

The measured \WW\ production cross-section at $\roots=\rroots$~GeV is
shown in Figure~\ref{fig-sigmaww}, together with the recent OPAL
measurement of \sigccthree\ at $\roots=161.3$~GeV \cite{bib:opalmw1}.
The energy dependence of the \WW\ production cross-section, as
predicted by the \GENTLE\ program for a representative W mass of
$80.33$~\GeVcc~\cite{bib:pdg,bib:mwpdg}, is also given in the figure.
The data are seen to be consistent with the Standard Model
expectation.

\subsection{Combination with 161 GeV Results}

The branching fraction results from the 172.12~GeV data can be combined
with the OPAL data recorded at 161.3~GeV\cite{bib:opalmw1}.
Simultaneous fits to the \CC\ cross-section at 161.3~GeV,
\sigccthree(161~\GeV), the \CC\ cross-section at 172.12~GeV
and the W boson branching fractions are
performed in the same way as presented above for the 172.12~GeV data.
The event selections, efficiencies, backgrounds and observed numbers
of events for the 161.3~GeV data have been described 
previously\cite{bib:opalmw1}. The results of this combination are
summarised in Table~\ref{tab:xsecbr_results}.

Within the framework of the Standard Model, the W boson branching
fractions depend on the six elements of the CKM mixing matrix, \Vij,
which do not involve the top quark \cite{bib:LEP2YR}
\begin{eqnarray*}
     {{\Br(\Wqq)}\over{(1-\Br(\Wqq))}} & = & 
 \left( 1+\frac{\alpha_s(\Mw)}{\pi} \right)
\sum_{i={\mathrm{u,c}}; \, j={\mathrm{d,s,b}}} \Vij^2,
\end{eqnarray*}
where $\alpha_s(\Mw)$ is taken to be $0.120\pm0.005$. The effect of
finite quark masses, $m_q$, on the W branching ratios are of order
$(m_{\mathrm{q}}/\Mw)^2$ \cite{bib:LEP2YR}, and are therefore neglected.  The
result, given in
Table~\ref{tab:xsecbr_results}, for the branching fraction $\Br(\Wqq)$ 
obtained from the fit assuming lepton universality gives:
\begin{eqnarray*}
 \sum_{i={\mathrm{u,c}};\, j={\mathrm{d,s,b}}}
  \Vij^2 & = & 2.22^{+0.32}_{-0.34}\pm0.07.
\end{eqnarray*}
This is consistent with a value of 2 which is expected from unitarity.
Using the experimental knowledge\cite{bib:pdg} of the sum, 
$\Vud^2+\Vus^2+\Vub^2+\Vcd^2+\Vcb^2 = 1.05\pm0.01$, the above result can
be interpreted as a measure of \Vcs, the least well determined of these 
elements: 
\begin{eqnarray*}
   \Vcs & = & 1.08 ^{+0.15}_{-0.16} \pm 0.03.
\end{eqnarray*}
This result is consistent in value with and has a comparable uncertainty
to other determinations of \Vcs\ which do not invoke unitarity
\cite{bib:pdg,bib:L-xs172}.

\section{Measurement of the Mass and the Decay Width of the W Boson}
\label{Wmass}

The W boson mass, \Mw, and decay width, \Gw, are determined from fits to the
reconstructed invariant mass spectrum of W pair candidate events. A
kinematic fit is employed to improve the mass resolution for the \WWqqqq\ 
and \WWqqlnu\ channels.  The fully leptonic final state is underconstrained
as it contains at least two neutrinos and is therefore not used.  Two
different methods are used to fit the reconstructed mass distribution.  In
the first, \Mw\ is determined by performing an unbinned likelihood fit which
uses a Breit-Wigner function to describe the signal shape. Monte Carlo
events are used to correct the fit result for biases introduced by the event
selection, detector resolution and effects of initial state radiation (ISR).
In the second method, a reweighting technique is employed to produce Monte
Carlo mass spectra, including detector simulation, corresponding to any
given mass and width.  A binned likelihood fit is used to determine \Mw\ and
\Gw\ by comparing the shape of the reconstructed mass distribution from the
data to that from the reweighted Monte Carlo spectra.  This method takes
into account all resolution, acceptance and ISR effects and is free of bias
as long as these effects are simulated correctly.  In addition, the
extension from a one-parameter fit determining \Mw\ to a two-parameter fit
determining both \Mw\ and \Gw\ is straightforward. Therefore, the
reweighting method (RW) is used to derive the central results of this paper
for \Mw\ and \Gw.  The simpler Breit-Wigner fit (BW) provides a valuable
cross-check of the RW results and systematics.  By appropriately choosing
the fit normalisations, both methods are made independent of cross-section
information and are sensitive only to the shape of the reconstructed mass
spectrum.

\subsection{Invariant Mass Reconstruction}

For each selected \WW\ event, the masses of the two W bosons 
can be determined from measured invariant masses of the decay products.  
Incorporating the constraints of energy and momentum conservation 
into a kinematic fit significantly improves the invariant
mass resolution.  The resolution of the kinematic fit is further improved by
neglecting the finite W width event-by-event, and constraining the masses of
the two W boson candidates to be equal, thus yielding a single reconstructed
invariant mass per event, $\mrec$. For \WW\ events, $\mrec$ is strongly
correlated to the average mass of the two W bosons in the event\footnote{At
  the generator level, the distribution of the average mass of the two W
  bosons is described by the same Breit-Wigner function as the distribution
  of the two separate W masses, multiplied by a phase space factor.}.
Incorporating the measured jet masses into the kinematic fit, rather than
treating jets as massless, gives an improved mass resolution and yields a
better agreement between the reconstructed and average masses.  Cuts on the
fit probability remove poorly reconstructed events and reduce background.
In addition, there is an ambiguity in the choice of the correct jet-jet
pairing in \WWqqqq\ events which leads to a combinatorial background.

\subsubsection{The \boldmath \WWqqqq\ Channel}
In each selected \WWqqqq\ event, four jets are defined using the Durham
algorithm \cite{bib:durham}.  These jets are used as input to a
five-constraint kinematic fit requiring energy and momentum conservation and
equality of the two W boson masses.  For each event, three kinematic fits
are performed, corresponding to the three possible jet-jet pairings, and
placed in descending order of the resulting fit probabilities
$P_1$, $P_2$, $P_3$.  In Monte Carlo studies, the highest fit probability,
$P_1$, corresponds to the correct jet-jet pairing in about 68\% of \WWqqqq\ 
events.  In approximately 25\% of events the second highest probability fit,
$P_2$, corresponds to the correct combination.
The lowest probability fit is dominated by incorrect combinations and is not
used.  Both fit methods require $P_1 >0.01$. In the BW method, the second
fit is also included in the reconstructed mass distribution if $P_2 >0.01$
and if $P_2 / P_1 > 1/3$.  This prescription is adopted to optimise the
ratio of signal to background.
In the RW method, for events in which both $P_1$ and $P_2$ exceed $0.01$, a
two-dimensional distribution, $\left(M_{1},M_{2}\right)$ is used, where
$M_i$ is the reconstructed mass corresponding to the combination having fit
probability $P_i$.  For those events in which only $P_1$ exceeds $0.01$, a
one-dimensional spectrum is used.  The numbers of selected events used in
the mass analyses are listed in Table~\ref{tab:evno} for the fit ranges
discussed in Sections~\ref{ssec:BWfit} and~\ref{ssec:RWfit}.

\subsubsection{The \boldmath \WWqqlnu\ Channel}
For the selected \WWqqlnu\ candidate events ($\ell=\Pe$ or $\Pgm$), the
reconstructed mass is estimated by defining two jets in the hadronic system
and then performing a kinematic fit, including the jets and lepton, using
the same constraints as for the \WWqqqq\ channel.  Since the three-momentum
of the neutrino is not known, this results in a two-constraint fit.  Events
with fit probability greater than 0.001 are retained for the analysis.  The
probability cut is lower than for the \WWqqqq\ channel because the
semi-leptonic channels have less background.  
In the case of the \WWqqtnu\ 
channel, the energy of the visible tau decay products is often a poor
estimate of the tau lepton energy because of the presence of 
additional neutrinos in the tau decay. 
To account for this in a kinematic fit, 
the direction of the tau lepton is taken to be that of the identified
tau jet and the energy of the tau lepton is set to half the beam
energy, with an error large enough to cover the kinematically allowed range.
However, Monte Carlo studies show that the mass resolution of this fit
is not significantly better than that obtained by dividing the
invariant mass of the hadronic system by the ratio of the visible energy of
the hadronic system to the beam energy.  Furthermore, the fit introduces
more bias and therefore both the BW and the RW method use the reconstructed
mass as determined by the scaling method for the \WWqqtnu\ channel.
To reduce background and remove poorly reconstructed events,
the BW method requires that the kinematic fit probability be greater than
$0.01$ for \WWqqtnu\ events.  The numbers of selected events in these
channels are listed in Table~\ref{tab:evno}.

\subsection{Breit-Wigner Fit to the Reconstructed Invariant Mass Spectrum}
\label{ssec:BWfit}
The first method used to determine the W mass involves fitting an
analytic function to the distribution of the masses obtained from the
kinematic fit. Because of
biases introduced by event selection, detector resolution, ISR, and 
phase-space effects, the method has been calibrated by performing similar 
fits to Monte Carlo samples generated with known W masses.

\subsubsection{Signal Shape}
A variety of analytic forms to describe the signal and background shapes
were investigated.   From Monte Carlo studies the signal
shape is found to be well described up to $\mrec \sim 84~\GeV$ by a function
based on a relativistic Breit-Wigner function with fixed width,
\begin{eqnarray*}
  S(\mrec) & =  &
    \frac{\mrec^{2}\Gamma^{2}}
{(\mrec^{2}-m_{0}^{2})^{2} + \mrec^{2}\Gamma^{2}},
\label{eq_rbw}
\end{eqnarray*}
where $m_{0}$ and $\Gamma$ are characteristic parameters of the signal peak.
 The fits are limited to the
range 40--84~\GeV, where the lower boundary is determined by considerations
regarding the background normalisation and is discussed below.  The width,
$\Gamma$, should be regarded as a parameter which embodies both the width of
the W boson and experimental effects and is fixed to the value predicted by
the Monte Carlo, 3.8~GeV.

\subsubsection{Background Shape}
Background arises mainly in the \WWqqqq\ channel, both from \Zqq\ events,
and from incorrect combinations in \WWqqqq\ events.  Monte Carlo studies
show that a quadratic form in $\mrec$ describes the combinatorial background
well up to $\mrec \sim 84~\GeV$.  The \Zqq\ background has a different
shape. It was found that this 
can be described by the function $A p^{\alpha} e^{-\beta p}$,
where $p=\sqrt{E^2_{\mathrm{beam}}-\mrec^2}$ is the momentum with which a W
pair of mass $\mrec$ would have been produced and where $\alpha$ and $\beta$
are positive parameters. The shapes and relative normalisation of the
background contributions are determined from Monte Carlo samples, but their
overall normalisation is allowed to vary in the fit.  Extending the allowed
fit range down to 40~\GeV\ improves the determination of the background
fraction in the peak region, as this low mass region is dominated by
background.

\subsubsection{Results of the Breit-Wigner Fit Method}
\begin{figure}
  \epsfxsize=\textwidth
  \epsfbox[0 0 567 680]{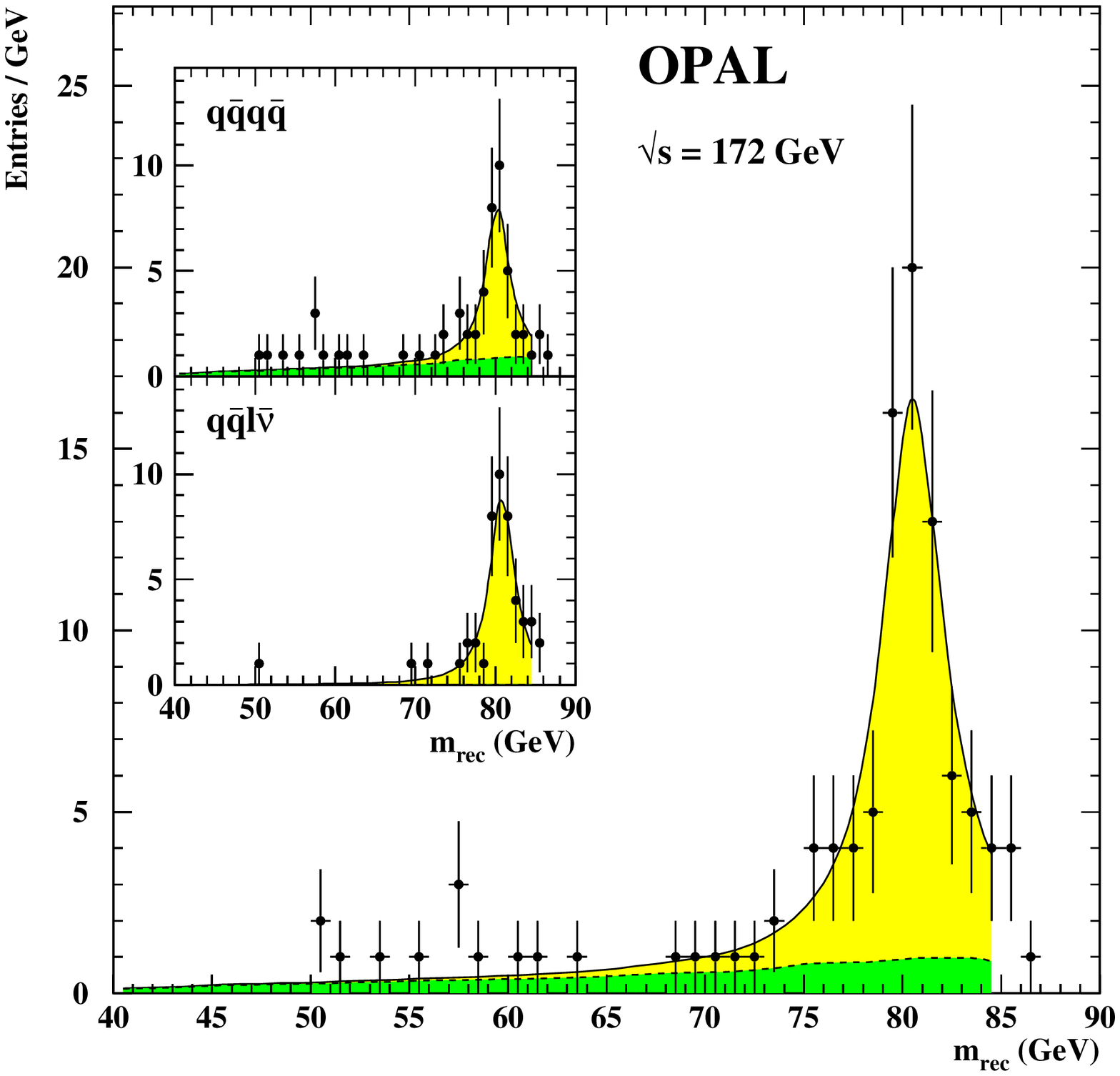}
  \caption
  {Reconstructed mass distributions for the data.  The main plot shows the
    \WWqqqq\ and \WWqqlnu\ samples combined, and the insets present them
    separately.  The solid curves display the results of unbinned maximum
    likelihood fits to a relativistic Breit-Wigner signal plus background in
    the range 40-84~GeV as described in the text.  The background function
    alone is shown by the dark shaded region.}
\label{fig:data}
\end{figure}

To extract \Mw\ from the reconstructed mass distribution, an unbinned
maximum likelihood fit to a relativistic Breit-Wigner plus background
function is used.
The fit is performed
using the reconstructed mass distribution from all channels combined.  For
comparison, it is also performed on the \WWqqqq\ and \WWqqlnu\ channels
separately\footnote{The widths are fixed to different values in the two
  channels.}.  The fit results are given in Table~\ref{tab:simplefit}, and
compared with the data in Figure~\ref{fig:data}.  The expected statistical
error has been studied using independent subsamples of Monte Carlo events
corresponding to the same integrated luminosity as the data.  These studies
demonstrate that the error returned by the fit accurately reflects the
r.m.s. spread of the $m_0$ distribution.

\begin{figure}
  \epsfxsize=\textwidth
  \epsfbox[0 0 567 680]{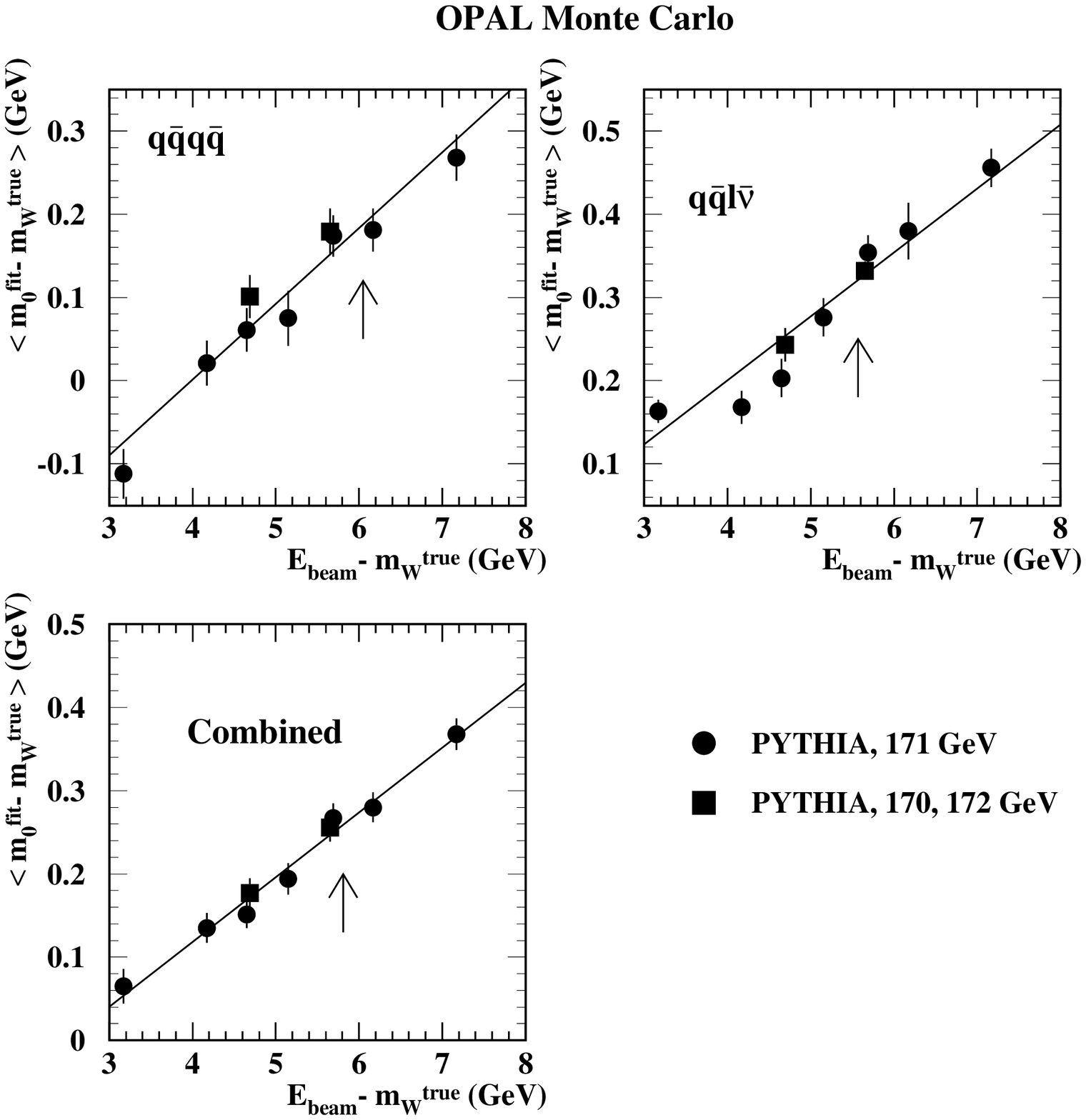}
  \caption
  {Mean of the difference between fitted and true mass from fits to many
    Monte Carlo subsamples, for \Pythia\ Monte Carlo events generated with
    various input W masses and at different beam energies, plotted as a
    function of the difference between beam energy and true mass.  The
    \WWqqqq\ and \WWqqlnu\ samples are shown separately, as well as both
    combined.  The lines represent fits to the points, with parameters given
    in the text.  The arrows indicate the actual corrections applied.  }
\label{fig:bias}
\end{figure}

The correction for bias in the BW method is
estimated using \Pythia\ Monte Carlo events corresponding to various input
values of \Mw\ and the beam energy, $E_{\mathrm{beam}}$.  Small samples of
Monte Carlo signal and background events corresponding to the integrated
luminosity of the data are processed through the same event selection, mass
reconstruction, and fitting routines as the data.  The mean correction is
determined by fitting a Gaussian to the resulting distribution of
$(m_0^{\mathrm{fit}}-m_{\mathrm{W}}^{\mathrm{true}})$ where
$m_{\mathrm{W}}^{\mathrm{true}}$ is the W boson mass with which the Monte
Carlo sample is generated.  This correction is found to depend linearly on
$(E_{\mathrm{beam}} - m_{\mathrm{W}}^{\mathrm{true}})$, as shown in
Figure~\ref{fig:bias}. A straight line fit yields: 
\begin{eqnarray*}
   \WWqqqq: \;\;\;\;
   &  m_0^{\mathrm{fit}} - m_{\mathrm{W}}^{\mathrm{true}} 
 = & -0.363+0.0911 (E_{\mathrm{beam}} - m_{\mathrm{W}}^{\mathrm{true}})\\
   \WWqqlnu: \;\;\;\;
   &  m_0^{\mathrm{fit}} - m_{\mathrm{W}}^{\mathrm{true}}
 = & -0.107+0.0768 (E_{\mathrm{beam}} - m_{\mathrm{W}}^{\mathrm{true}})\\
   \mathrm{Combined:} \;\;\;\;
   &  m_0^{\mathrm{fit}} - m_{\mathrm{W}}^{\mathrm{true}} 
 = & -0.193+0.0779 (E_{\mathrm{beam}} - m_{\mathrm{W}}^{\mathrm{true}}) 
\end{eqnarray*}
with the masses and energies expressed in GeV.  These relations are used to
correct the fitted mass for all biases. The fitted masses, corrections and
corrected masses are given in Table~\ref{tab:simplefit}.

\subsection{Fit to the Reconstructed Mass Spectrum using a Reweighting Method}
\label{ssec:RWfit}
In the second method, the W boson mass and width are extracted by directly
comparing the reconstructed mass distribution of the data to mass spectra
estimated from fully simulated Monte Carlo events corresponding to various
values of \Mw\ and \Gw.  A likelihood fit is used to extract \Mw\ and \Gw\ 
by determining which Monte Carlo spectrum gives the best description
of the data.  In order to obtain the Monte Carlo spectrum for arbitrary
values of \Mw\ and \Gw, a Monte Carlo reweighting technique is employed.

\subsubsection{The Monte Carlo Reweighting Technique}
For a sample of \CC\ W pair Monte Carlo events produced with a given 
mass and width, ($\Mw^{\mathrm{MC}}$, $\Gw^{\mathrm{MC}}$), the detector level
mass spectrum corresponding to a different mass and width, 
($\Mw^{\mathrm{new}}$, $\Gw^{\mathrm{new}}$), is estimated by assigning to
each event a reweighting factor with which it enters the new spectrum. This 
factor is the ratio of the probability that the event would be produced 
assuming the new values of $(\Mw^{\mathrm{new}},\Gw^{\mathrm{new}})$ 
to the probability that this same event would be 
produced for the input values $(\Mw^{\mathrm{MC}},\Gw^{\mathrm{MC}})$.   
These production 
probabilities are given by the W pair production cross-section~\cite{muta}
\begin{eqnarray*}
\label{proba1}
  P(\Mw,\Gw;m_1,m_2) \propto  
  {\cal BW}(\Mw,\Gw;m_1) {\cal BW}(\Mw,\Gw;m_2) \sigma_0(m_1,m_2,s),
\end{eqnarray*}
where $m_1$ and $m_2$ are the generator level masses of the two W bosons 
produced in the event and $\sigma_0$ is the Born level cross-section for 
producing a pair of W bosons with masses $\left(m_1,m_2\right)$ at a 
centre-of-mass energy \roots.  The values $m_1$ and $m_2$ are distributed
according to the running-width, relativistic Breit-Wigner function, 
\begin{eqnarray*}
  \label{bw}
  {\cal BW}(\Mw,\Gw;m) =  \frac{1}{\pi} 
    \frac{ \frac{m^2}{\Mw} \Gw}{(m^2-\Mw^2)^2 + \frac{m^4}{\Mw^2}\Gw^2}.
\end{eqnarray*}
The reweighting factor for the $i$th event, $f_i$, is then given by
\begin{eqnarray*}
  \label{rewfactor}
  f_i =
  \frac{ {\cal BW}(\Mw^{\mathrm{new}},\Gw^{\mathrm{new}};m_1^i)  {\cal
      BW}(\Mw^{\mathrm{new}},\Gw^{\mathrm{new}};m_2^i) }{{\cal
      BW}(\Mw^{\mathrm{MC}},\Gw^{\mathrm{MC}};m_1^i) {\cal 
      BW}(\Mw^{\mathrm{MC}},\Gw^{\mathrm{MC}};m_2^i) }.
\end{eqnarray*}
The Born level cross-section cancels in the ratio since it depends only on 
$m_1$, $m_2$ and \roots.

An estimate of the detector level reconstructed mass spectrum
for any $(\Mw^{\mathrm{new}},\Gw^{\mathrm{new}})$ can be obtained from one
such sample alone with a given input
$(\Mw^{\mathrm{MC}},\Gw^{\mathrm{MC}})$.  In general, deviations from unity
in the reweighting factors increase as $\Mw^{\mathrm{new}}$ and
$\Gw^{\mathrm{new}}$ differ from $\Mw^{\mathrm{MC}}$ and
$\Gw^{\mathrm{MC}}$, respectively.  For very large $f_i$, the reweighted
spectrum becomes sensitive to statistical fluctuations in the input sample.
To reduce the sensitivity to large reweighting factors, 
a total of 850\,000 events from
all \WW\ Monte Carlo samples with full detector simulation
are used (\Pythia, \Herwig,
\Excalibur, \Koralw\ and \grcff\ Monte Carlo generators).
These samples cover a range of 
input $(\Mw^{\mathrm{MC}},\Gw^{\mathrm{MC}})$ values, spanning
$\Mw = 78.33$~\GeV\ to $\Mw=82.33$~\GeV\ and
widths of $1.7~\GeV$ to $2.5~\GeV$.  The inclusion
of all of these samples ensures that the reweighted spectra are reasonably
smooth over a large range of ($\Mw^{\mathrm{new}}$, $\Gw^{\mathrm{new}}$)
values.  

\subsubsection{The Signal and Background Shapes}
\begin{figure}[htb]
  \epsfxsize=\textwidth
  \epsfbox[0 0 567 680]{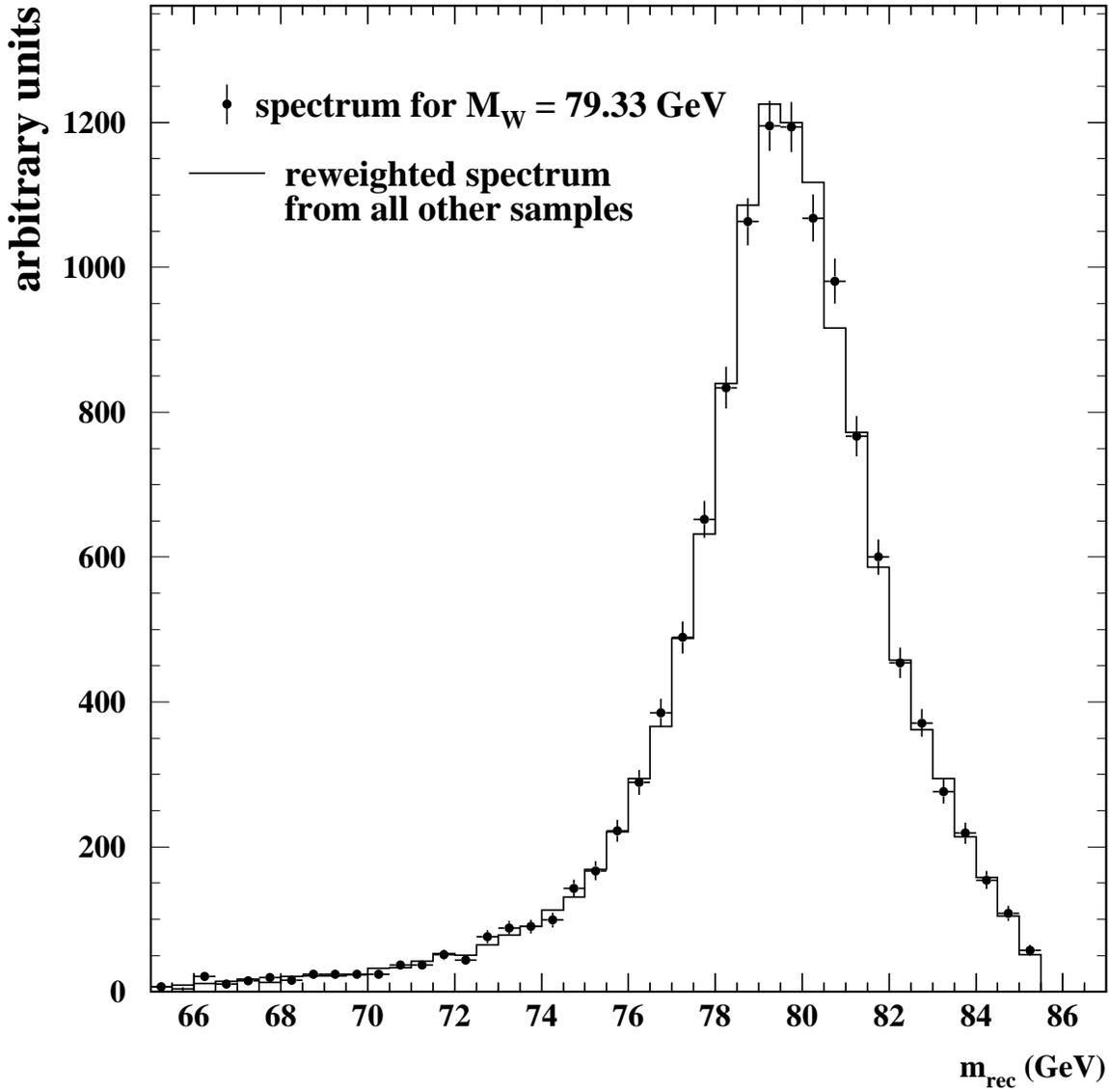}
  \caption
  {Illustration of the reweighting procedure for the reconstructed mass
    spectrum of the \WWqqln\ signal events.  The spectrum of a sample
    produced with $M_W = 79.33~\GeV$ is compared with the reweighted
    spectrum obtained from samples with \PW\ boson masses between $M_W =
    78.33~\GeV$ and $M_W = 82.33~\GeV$. Only samples produced with the same
    Monte Carlo generator (\PYTHIA) are used for this comparison and the
    $M_W = 79.33~\GeV$ sample has not been included in the reweighting
    procedure. The Monte Carlo statistics for the final fit including all
    samples, will be larger by more than a factor of 3.}
 \label{fig:testrew}
\end{figure}

The normalised reweighted mass spectra of the individual Monte Carlo samples
are combined into a single reweighted spectrum by taking the weighted
average, bin-by-bin.  As an example of the performance of this procedure,
Figure~\ref{fig:testrew} compares the reweighted mass distribution for the
\WWqqln\ channel for $(\Mw^{\mathrm{new}}, \Gw^{\mathrm{new}})=(79.33,
2.047)~\GeV$ to the spectrum from the fully-simulated \Pythia\ MC sample
generated with $(\Mw^{\mathrm{MC}}, \Gw^{\mathrm{MC}}) =(79.33,
2.047)~\GeV$. Good agreement is found.

The background mass spectra are taken from Monte Carlo and are assumed to be
independent of \Mw\ and \Gw.  The background distributions are normalised to
the expected number of background events.  The reweighted signal spectra are
then normalised such that the total number of signal plus background events
corresponds to the observed number of events. This is done separately for
each channel. For this purpose, the \WWqqqq\ events in which both the first
and second highest probability fits are used are treated as a separate
channel from those \WWqqqq\ events in which only the highest probability fit
is used.

Most of the Monte Carlo events have been produced for a beam energy of 
$E^{\mathrm{MC}}_{\mathrm{beam}} = 85.50~\GeV$, whereas the luminosity weighted
average beam energy of the data is 
$E^{\mathrm{data}}_{\mathrm{beam}} = 86.06~\GeV$.   In order to correct for 
this, the mass $\Mw^{\mathrm{new}}$  used for reweighting is modified to
\begin{eqnarray*}
  \Mw^{\mathrm{new}} \rightarrow \Mw^{\mathrm{new}} &-&
(E^{\mathrm{data}}_{\mathrm{beam}} - E^{\mathrm{MC}}_{\mathrm{beam}}),
\end{eqnarray*}
while shifting the resulting spectrum by
$(E^{\mathrm{data}}_{\mathrm{beam}} - E^{\mathrm{MC}}_{\mathrm{beam}})$.
Since not all the fully simulated samples are produced with the same 
centre-of-mass energy, this is done for each sample separately before combining
the reweighted  spectra. A corresponding shift is made to the background 
spectra.

\subsubsection{Results of the Reweighting Method}

A binned likelihood is formed from the product of the Poisson 
probabilities for
obtaining the numbers of events observed in the data in each bin of mass 
(and width) using the Monte Carlo expectation for $(\Mw,\Gw)$.
The log-likelihood curve is determined 
separately for each of the different channels.  Since the channels are 
statistically independent, the results are combined by  adding together 
these likelihood curves.  The resulting log-likelihood distribution as a 
function of the mass and the width allows the determination of \Mw\ and \Gw\ 
and their associated statistical uncertainties.

To reduce further the sensitivity to statistical fluctuations in the 
determination of the expected number of events, 
only events with a reconstructed mass 
in the range $65~\GeV < \mrec < \Ebeam$ are included in the fit.
For the two-dimensional mass spectrum of the \WWqqqq\ channel, at least one of 
the two reconstructed masses is required to be greater than 70~\GeV\ and the 
sum of the two masses must exceed 120~\GeV.  The rather limited mass range is
further motivated by the fact that events with small reconstructed masses have
very little sensitivity to \Mw\ and \Gw. 

\begin{figure}[htb]
  \epsfxsize=\textwidth
  \epsfbox[0 0 567 680]{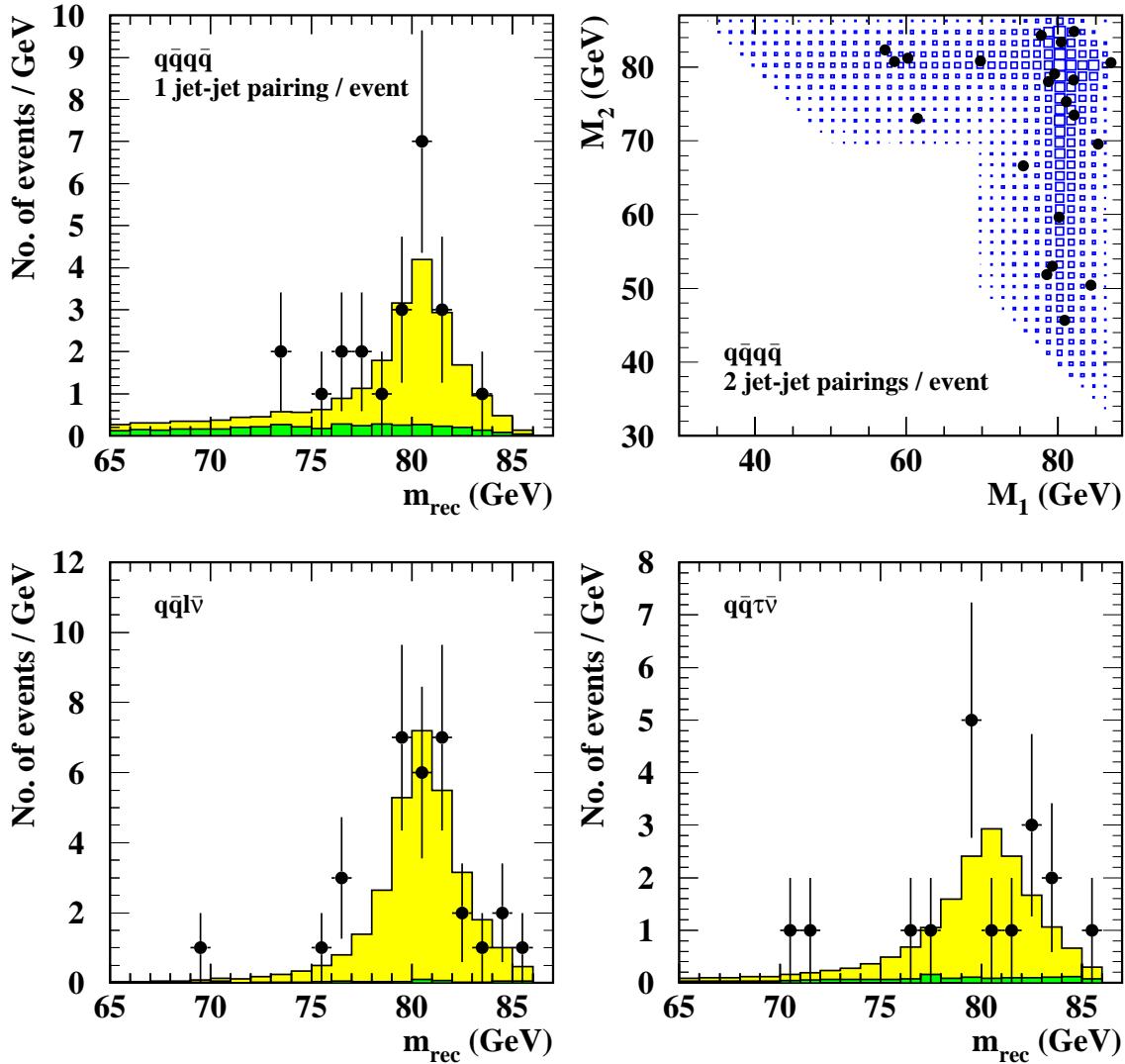}
  \caption
  {Fits to the mass spectra of the different channels using the reweighting
    method.  The dots represent the data, the lightly shaded histograms the
    result from the fit for (\Mw,\Gw)  
    and the darkly shaded ones the background contributions to
    it. For the \WWqqqq\ candidates in which two jet-jet pairings per event
    are used, the underlying histogram represents the fit result.
    For clarity, the bin size used for the one-dimensional histograms 
    is twice that used for the fit.}

 \label{fig:rew_spectra}
\end{figure}

\begin{figure}[htb]
  \epsfxsize=\textwidth
  \epsfbox[0 0 567 680]{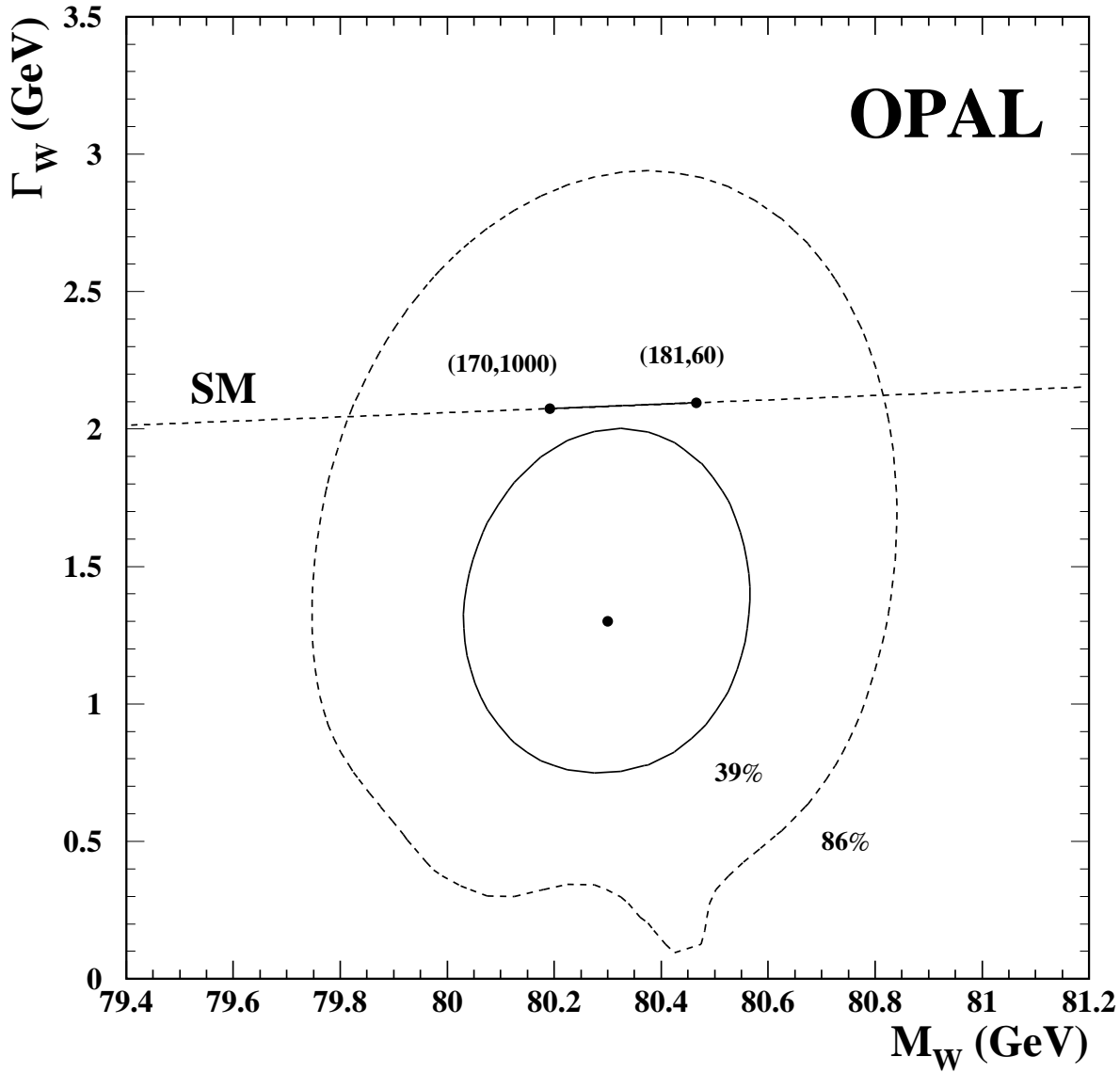}
  \caption
  {The 39\% and 86\% contour levels of the two-parameter fit using the
    reweighting method. The projections of these contours onto the axes give
    the one and two standard deviation errors quoted in the text.  The
    straight dashed line gives the dependence of the width on the mass
    according to the Standard Model. The one-parameter fit for the mass
    alone is constrained to this line.  The solid line is the Standard Model
    prediction for variations of
    (\mbox{$M_{\mathrm{top}}$},\mbox{$M_{\mathrm{Higgs}}$}), both given in
    \GeV. For small fitted values of \Gw, the 86\% contour is very sensitive
    to the particular distribution of masses observed in the data.}
 \label{fig:errcontour}
\end{figure}

The results of the fit to \Mw\ and \Gw\ using \WWqqqq\ and \WWqqln\ events
are:
\begin{eqnarray*}
 \Mw & = & 80.30 \pm 0.27 \pm 0.09 ~\GeV, \\
 \Gw & = &  1.30^{+0.62}_{-0.55} \pm 0.18 ~\GeV,
\end{eqnarray*}
where the errors are statistical and systematic, respectively.  The
correlation coefficient between \Mw\ and \Gw\ is 0.06.  The systematic
errors are discussed in Section~\ref{sec:mwsyserr}.  The mass spectra of the
individual channels together with the fitted reweighted spectra are shown in
Figure~\ref{fig:rew_spectra}.  The likelihood contours for the statistical
errors are displayed in Figure~\ref{fig:errcontour} together with the
Standard Model prediction.  From the shape of the contours and the
asymmetric errors on the width, it can be seen that the errors on \Gw\ for
the present statistics are non-Gaussian.  The two standard deviation errors
on \Gw\ are $^{+1.64}_{-1.21}~\GeV$.

A one-parameter fit for the mass is 
performed by constraining the width using the Standard Model relation
\cite{bib:LEP2YR}
\begin{eqnarray*}
\Gamma_W  & = & (2.0817~\GeV)  { \Mw^3 \over (80.26~\GeV)^3 },
\end{eqnarray*}
giving:
\begin{eqnarray*}
   \WWqqqq: \;\;\;\;           \Mw & = & 80.08 \pm 0.44 \pm 0.14 ~\GeV,\\
   \WWqqlnu: \;\;\;\;          \Mw & = & 80.53 \pm 0.41 \pm 0.10 ~\GeV,\\
   \mathrm{Combined:} \;\;\;\; \Mw & = & 80.32 \pm 0.30 \pm 0.09 ~\GeV.\\
\end{eqnarray*}
\vspace{-9mm}

The error on the fitted mass is correlated with the width of the
reconstructed mass distribution.  Consequently, the statistical error on the
mass resulting from the one-parameter fit is larger than that resulting from
the two-parameter fit.
Since
this measurement is principally intended to determine \Mw\ as a Standard
Model parameter, the result of the one-parameter fit is taken as the central
result of this paper for the mass of the W boson as determined from the
direct reconstruction method. The value of \Mw\ obtained 
from the BW fit is consistent with that
obtained using the RW method, thus providing a useful
cross check of the RW analysis.

\subsection{Systematic Uncertainties}
\label{sec:mwsyserr}
The main sources of systematic uncertainty on \Mw\ and \Gw\ are 
studied by changing some feature of the analysis.
The effect of this can typically be assessed in two ways -- either by
repeating the fit to the reconstructed mass spectrum and observing the
change in the fit results, or by determining the average of the
event-by-event change in the reconstructed mass.  The latter method tends to
have greater statistical sensitivity, but does not readily yield a
determination of the error on \Gw.  For this reason, the former method was
generally preferred in the RW analysis, and the latter in the BW analysis.
Despite these differences,
the overall systematic errors calculated for the two 
analyses are in good agreement. The following
descriptions concentrate on the procedure used to estimate the uncertainty
in the RW fit method.  The estimated errors are summarised in
Table~\ref{tab:sys}.
\begin{description}
\item[Beam Energy:] Uncertainties in the LEP beam energy affect the
  reconstructed mass spectrum through the energy constraints imposed by the
  kinematic fit.  The precision on the beam energy is estimated to be $\pm
  30$~MeV~\cite{bib:LEPenergy}.  The influence of this uncertainty on the
  fit results has been estimated by changing the beam energy in independent
  samples of signal and background Monte Carlo before performing the
  kinematic fits.
 In addition, the effect of a possible asymmetry 
 between the \Pep\ and \Pem\ energies
 is found to be negligible.
\item[Initial State Radiation:] The effect of initial state radiation is to
  increase systematically the reconstructed invariant mass,
  and consequently  the resulting fitted mass, 
  due to biases introduced through the energy
  conservation constraint of the kinematic fit.  This bias is taken into
  account in both the RW and BW fit methods to the extent that ISR is
  accurately modelled in the Monte Carlo.  The \Koralw\ generator is used to
  estimate the systematic error associated with the incomplete modelling of
  ISR for the RW fits. Distributions of the mean\footnote{At the generator
    level, the average mass of two W bosons is used as an approximation to
    the results of the five-constraint kinematic fit at the detector level.}
  \Mw\ per event are compared in two samples, one including only first
  order corrections and one
  including the full second order ISR corrections.  From fits to these
  generator level distributions systematic uncertainties of 20 MeV for the
  mass and 60 MeV for the width are assigned.
\item[Hadronisation:] Sensitivity of the results to the choice of the
  hadronisation model was studied in two ways.
  In the first approach, a single sample of \WW\ 
  four-fermion final states was
  hadronised separately by both \Pythia\ and \Herwig\, and the fit results
  compared. By using the same four-fermion final states the comparison is
  sensitive only to differences in hadronisation.
  Alternatively, the mean
  difference between the reconstructed mass and the generator level average
  mass is determined for the \Pythia\ and \Herwig\ \WW\ samples separately.
  No significant difference is observed in either comparison and
  the statistical precisions of the tests are taken as the systematic
  errors.  Varying the hadronisation model for the background is found to
  have a negligible effect.
\item[Four-fermion Interference Effects:] The Monte Carlo samples used to
  calibrate the BW method and in the reweighting procedure of the RW method
  include only the CC03 diagrams. In order to test the sensitivity of the
  results to the interference between \PWp\PWm\ diagrams and other
  four-fermion processes, the fit results of a sample generated including
  the full set of interfering four-fermion diagrams are compared to one
  restricted to the \CC\ \PWp\PWm\ diagrams alone.  The comparison is made 
  using both the \grcff\ generator and the \Excalibur\ generator.  
   A small mass shift was found when using
  \Excalibur, while no such indication was found using \grcff.  
  The assigned uncertainties accommodate the results from these two models.
\item[Detector Effects:] The effects of detector mis-calibrations and
  deficiencies in the Monte Carlo simulation of the data have been
  investigated by varying the inputs to the kinematic fit over reasonable
  ranges.  The range is determined in each case from a detailed comparison
  of data and Monte Carlo using approximately $1.2\:\mathrm{pb}^{-1}$ of
  data recorded at $\roots \sim M_Z$ during 1996.
 The following effects are considered:
\begin{itemize}
\item The jet energy scale is determined to be accurate to within 1\%.
  The effect on both data and Monte Carlo reconstructed masses is found
  to be negligible.
\item The calibration of the lepton momentum or energy scale is accurate to
  within 1\%.  The corresponding changes for the semi-leptonic (e/$\mu$)
  channels are taken as systematic errors.
\item The results of the kinematic fit depend on the covariance matrix of
  the input parameters, which have been determined using Monte Carlo events.
  Studies of \Zzero\ decays indicate that the jet energy errors in the data
  are understood to within 3\%.  No indication of mis-modelling of the
  errors associated with the jet angles or of the corresponding lepton
  parameters was found. As an estimate of the associated uncertainty, each
  parameter of the covariance matrix was varied within its estimated
  uncertainty.
\item The algorithm used to correct jet energies and angles combines
  information from tracks, and electromagnetic and hadron calorimeter
  clusters.  Three possible variants in the procedure are studied.  Firstly,
  the mass reconstruction is repeated with the hadron calorimeter removed
  from the analysis. Secondly, an alternative parametrisation of the jet
  errors is employed. Finally, a different correction algorithm for the
  combination of track and calorimetric information is used.  The standard
  correction procedure combines all tracks and clusters into jets and then
  applies an average correction based on parametrisations of the total
  energy in each detector component contributing to the jet. The alternative
  algorithm first associates tracks with clusters then subtracts the
  associated charged particle energy from the individual clusters prior to
  jet finding.  In all of these checks, the mean and r.m.s.\ of the shifts
  in the reconstructed masses observed in the data are compatible with the
  expectations from Monte Carlo.  The statistical precision of each
  comparison is taken as the associated uncertainty, and the largest of
  these is included in the total systematic error.
\end{itemize}
The errors from all of these checks are added in quadrature to give the total 
systematic error due to uncertainties associated with detector mis-calibrations
and mis-modelling. This is the dominant experimental systematic uncertainty.

\item[Fitting Procedure and Background Treatment, 
      RW Fit Method:]\mbox{}\newline
Using 800 fully simulated Monte Carlo subsamples with
input values of $\left(\Mw,\Gw\right)=$ \newline
$\left(80.33~\GeV,2.085~\GeV\right)$, 
each corresponding to the same integrated luminosity as the data sample, it was
verified that both the one-parameter and the two-parameter fits have no
observed bias at the precision studied.  In addition, 2000 Monte Carlo test
samples, each corresponding to the same integrated luminosity as the data,
were used to study the reliability of the fit errors, the validity of the
contour levels, and correlations between the various fitted parameters and
their associated errors.  Test samples were 
constructed by randomly sampling
a parametrisation of the 
reconstructed mass spectrum (including detector simulation)
in each channel.

The possibility of residual bias is further investigated as follows:
\begin{itemize}
\item To check for any biases in the reweighting procedure that could depend
  on the input mass or width, fits are performed
  on \Pythia\ Monte Carlo samples with input \PW\ 
  masses between $78.33~\GeV$ and $82.33~\GeV$.  Only the \Pythia\ samples
  are used in the reweighting procedure, excluding the sample being fitted.
  An equivalent test is done with the different width samples.  No
  significant shifts between the fitted and input values are found in the
  \WWqqln\ channels and the statistical precision of this test is taken as a
  systematic error.  In the \WWqqqq\ channel there is an indication of a small
  systematic effect which is believed to be an artifact of finite Monte
  Carlo statistics of the samples employed in the reweighting procedure.
  For values of \Mw\ in the range of interest, $80.33 \pm 0.50$~\GeV, this
  bias is $\pm 20$~MeV, which is taken as the systematic error.
\item The effect of the finite Monte Carlo statistics in the background
  samples is estimated by comparing the results of repeated fits to a Monte
  Carlo sample using only fractions of the background sample.
\item Variations in the normalisation of the predicted 
  backgrounds, based upon uncertainties in the background cross-sections and
  efficiencies discussed in Section~\ref{selection}, were studied.
  The combined background is varied by 20\%
  in the \WWqqqq\ channel, and 40\% in the \WWqqln\ channels. The
  observed shifts in the fit results are taken as systematic uncertainties.
\item The procedure to correct for the difference between the actual
  beam energy and the beam energy with which the Monte Carlo samples are
  generated was checked by fitting the value for \Mw\ in a
  \Pythia\ sample generated with $\roots =
  172~\GeV$ using only the \Pythia\ samples generated with $\roots =
  171~\GeV$ for the reweighting and background estimation. No significant
  bias is observed. The statistical accuracy of this test is taken as a
  systematic error.
\end{itemize}
The errors from all of these checks are added in quadrature to give the
total systematic error due to uncertainties associated with the fitting
procedure and background treatment.

\item[Fitting Procedure and Background Treatment, 
      BW Fit Method:]\mbox{}\newline
The effects of using different analytic functions to describe the signal and
background shapes, of fixing various fitted parameters, and of varying the fit
range are investigated.  In each case, the shift in the fit mass observed in
the data is compared with the expected shift estimated using 400 Monte Carlo
subsamples, each corresponding to the same integrated luminosity as the
data.  The following variations are considered:
\begin{itemize}
\item 
A simple non-relativistic Breit-Wigner is used for the signal.
\item The relativistic Breit-Wigner is multiplied by a factor $p^\alpha$,
  where $\alpha$ is a free parameter, and $p$ is the centre-of-mass momentum
  of a pair of particles of mass $\mrec$.\footnote{The
    mass distribution would be expected to be modified by a phase-space factor
    $\propto p$. However, this fails to describe the reconstructed Monte
    Carlo distribution,
    while a factor $p^{\alpha}$ gives
    a reasonable fit up to the kinematic limit.}  In this case, the
  data are fitted up to the kinematic limit in $\mrec$ with $\alpha$ fixed
  to the value determined from Monte Carlo.
\item
The width parameter, $\Gamma$, is left as a free parameter in the fit.
\item
The background is represented by a quadratic in~$\mrec$, with parameters
determined in the fit to the data.
\item
The background normalisation is fixed to that expected from Monte Carlo.
\item
The relative normalisation of the combinatorial and \Zqq\ backgrounds is 
varied by $\pm 50 \%$.
\item
The masses of the combinatorial and \Zqq\ backgrounds are displaced by 1~GeV.
\end{itemize}
In each case, the shift in fitted mass observed in the data is consistent
with that estimated from the Monte Carlo.  The r.m.s. deviation of the
differences of the observed shift between the data and Monte Carlo is taken as a
systematic error.

\item[Colour Reconnection Effects and Bose-Einstein Correlations:] As
  discussed in \cite{bib:LEP2YR} and references therein, a
  significant bias to the apparent W mass measured in the \WWqqqq\ channel
  could arise from the effects of colour reconnection and Bose-Einstein
  correlations between the decay products of the \PWp\ and \PWm.  Neither of
  these effects are included in the various hadronisation models used.
  There is no consensus as to the magnitude of such effects, though in some
  models they produce a shift in the measured \Mw\ by as much as
  +100~MeV~\cite{bib:LEP2YR}.  In some models, colour reconnection
  also affects other properties of \WWqqqq\ events, for example, the charged
  particle multiplicity. As will be discussed in Section~\ref{Wprop}, with
  the sensitivity afforded by the present statistics, no indication of such
  reconnection effects is observed. In the absence of an experimental limit
  to the biases associated with these phenomena, a systematic
  uncertainty of $\pm 100$~MeV is assigned to the \WWqqqq\ channel due to
  these effects.  As the \WWqqqq\ channel comprises approximately half the
  total signal sample an uncertainty of $\pm 50$~MeV is assigned in the
  combined analysis.  One possible test of such effects is to compare \Mw\ 
  measured in the \WWqqqq\ channel with that determined from \WWqqlnu.
  Within the present experimental errors the results are compatible.
  Monte Carlo studies on colour reconnection effects indicate that possible
  effects on the width are typically of the same size as on the mass.  Thus,
  a corresponding error of $\pm 50$~MeV is assigned to the width as
  determined from a combined fit to all channels.
\end{description}

\subsection{Combination with Cross-section Measurements}

The measurements of \Mw\ from direct reconstruction and from the measurement
of the \PWp\PWm\ production cross-section at \Ecm=\rroots~GeV, as presented
in this paper, may be combined with the value obtained from the
\PWp\PWm\ production cross-section at
\Ecm=161.3~GeV~\cite{bib:opalmw1}, 
\begin{eqnarray*}
 \Mw = 80.40^{+0.44\;+0.09}_{-0.41\;-0.10}\pm 0.03 \; \GeV,
\end{eqnarray*}
where the errors are the statistical,
systematic and beam energy uncertainties, respectively. 
The uncertainty of
the LEP beam energy has been updated since Reference~\cite{bib:opalmw1} as
described in Reference~\cite{bib:LEPenergy}.  The systematic effects of the
three measurements are quite different, and therefore they are combined
assuming that they are uncorrelated, apart from the uncertainty associated with
the LEP beam energy, which is taken to be fully correlated.  The result
obtained is:
\begin{displaymath}
   \Mw = 80.35 \pm 0.24 \pm 0.06  \pm 0.03 \pm 0.03\; \GeV, 
\end{displaymath} 
where the errors are the statistical, systematic, reconnection effects
and beam energy uncertainties, respectively.

\section{Properties of W Pair Events}
\label{Wprop}
 
Hadronic data in \epem\ collisions can be characterised by event shape
distributions and inclusive observables such as charged particle
multiplicities and momentum spectra. In \eeWW\ events, although relatively
little data have been collected, it is useful to study such characteristics.
In addition to tests of Monte Carlo models, measurement of the properties of
the hadronic sector of \WW\ decays allows the question of colour
reconnection to be addressed experimentally.  At present there is general
consensus that observable effects of interactions between the colour
singlets during the perturbative phase are expected to be small.  In
contrast, significant interference in the hadronisation process is
considered to be a real possibility.  With the current knowledge of
non-perturbative QCD, such interference can be estimated only in the context
of specific models
\cite{bib:GPZ,bib:PYTHIA,bib:SK,bib:GH,bib:ARIADNE,bib:HERWIG,bib:EG1}.

It has been suggested \cite{bib:SK,bib:GH} that simple observable
quantities, such as the charged multiplicity in restricted rapidity
intervals, may be sensitive to the effects of colour reconnection.
Such effects were predicted to be undetectable with data samples
corresponding to 10~pb$^{-1}$. More recently \cite{bib:EG2} it was
suggested that the effect on the inclusive charged multiplicity itself
may be larger than previously considered and that the mean hadronic
multiplicity in \WWqqqq\ events, \nchQQQQ, may be as much as 10\%
smaller than twice the hadronic multiplicity in \WWqqln\ events,
\nchQQLV. The visible effects of such phenomena are expected to
manifest themselves most clearly in low momentum regions. Therefore
studies of the fragmentation function, i.e.\ the distribution of
the scaled momentum, $\xp=p/\Ebeam$, are also relevant.

Most studies of sensitivity to colour reconnection have been estimated
within the context of a given model, comparing ``reconnection'' to ``no
reconnection'' scenarios; in general, both the size and sign of any changes
are strongly dependent upon the model considered.  At the expense of a
reduction in statistical sensitivity, such model dependence can be avoided
by comparing directly the properties of the hadronic part of \WWqqln\ events
with \WWqqqq\ events.  In the current study, the inclusive charged
multiplicity and the fragmentation function are measured and compared for
\WWqqqq\ and \WWqqln\ events.  The quantities $\Dnch=\nchQQQQ-2\nchQQLV$ and
$\Dxp=\xpQQQQ-\xpQQLV$ are also examined.  By way of characterising the
global properties of \WWqqqq\ events themselves, mean values of the thrust
distribution, \thrQQQQ, and the rapidity distribution relative to the thrust
axis, \yQQQQ, are also measured in this channel.  Mean values are used for
these comparisons due to the relatively small size of the current data
sample.

The models of colour reconnection implemented\footnote{No retuning of any
  model was performed in generating events in the various reconnection
  scenarios.}  in the event generators \Pythia, \Herwig\ and \Ariadne\ 
\cite{bib:ARIADNE} were
used to assess the sensitivities of the quantities above to such effects.
Observables such as \thrQQQQ\ and \yQQQQ, considered in earlier studies of
colour reconnection, had a predicted sensitivity that was sufficiently small
as to be unobservable at present.  The models \Ariadne\ 
and a `colour octet' variant\footnote{Merging of partons to form clusters
  was performed on a nearest neighbour basis, as a partial emulation of the
  model of Reference \cite{bib:EG1}.} 
  of \Herwig\ \cite{bib:BWOXFORD} predict shifts in
\nchQQQQ\ and \xpQQQQ\ similar in size to the
statistical uncertainty observed on these quantities in the current data.

\subsection{Correction Procedure}


The distributions of \nch, \xp, $1-T$ and $y$ are corrected for
background contamination using a bin-by-bin subtraction of the
expected background, based on Monte Carlo estimates.  Corrections are
then applied for finite acceptance and the effects of detector
resolution, after which mean values of the distributions are
calculated.  Each observable is evaluated using two samples of Monte
Carlo events.  The first includes full simulation of the OPAL detector
and contains only those events which pass the cuts applied to the data
(detector level).  The second does not include initial state radiation
or detector effects and allows all particles with lifetimes shorter
than $3\times10^{-10}$~s to decay (hadron level).  Both samples are
generated at the same $\sqrt{s}$.  Distributions normalised to the
number of events at the detector and the hadron level are compared to
derive bin-by-bin correction factors which are used to correct the
observed distributions of \xp, $1-T$ and $y$.

This correction method is not appropriate for the multiplicity distribution,
since resolution and acceptance effects cause significant migration of
charged tracks between bins.  Instead, a matrix correction is used to
correct for detector resolution effects, followed by a bin-by-bin correction
which accounts for the effects due to acceptance cuts and residual initial
state radiation, as in previous OPAL multiplicity studies
\cite{bib:LEP1.5QCD}.

\begin{figure}[htbp]
 \centerline{\epsfig{file=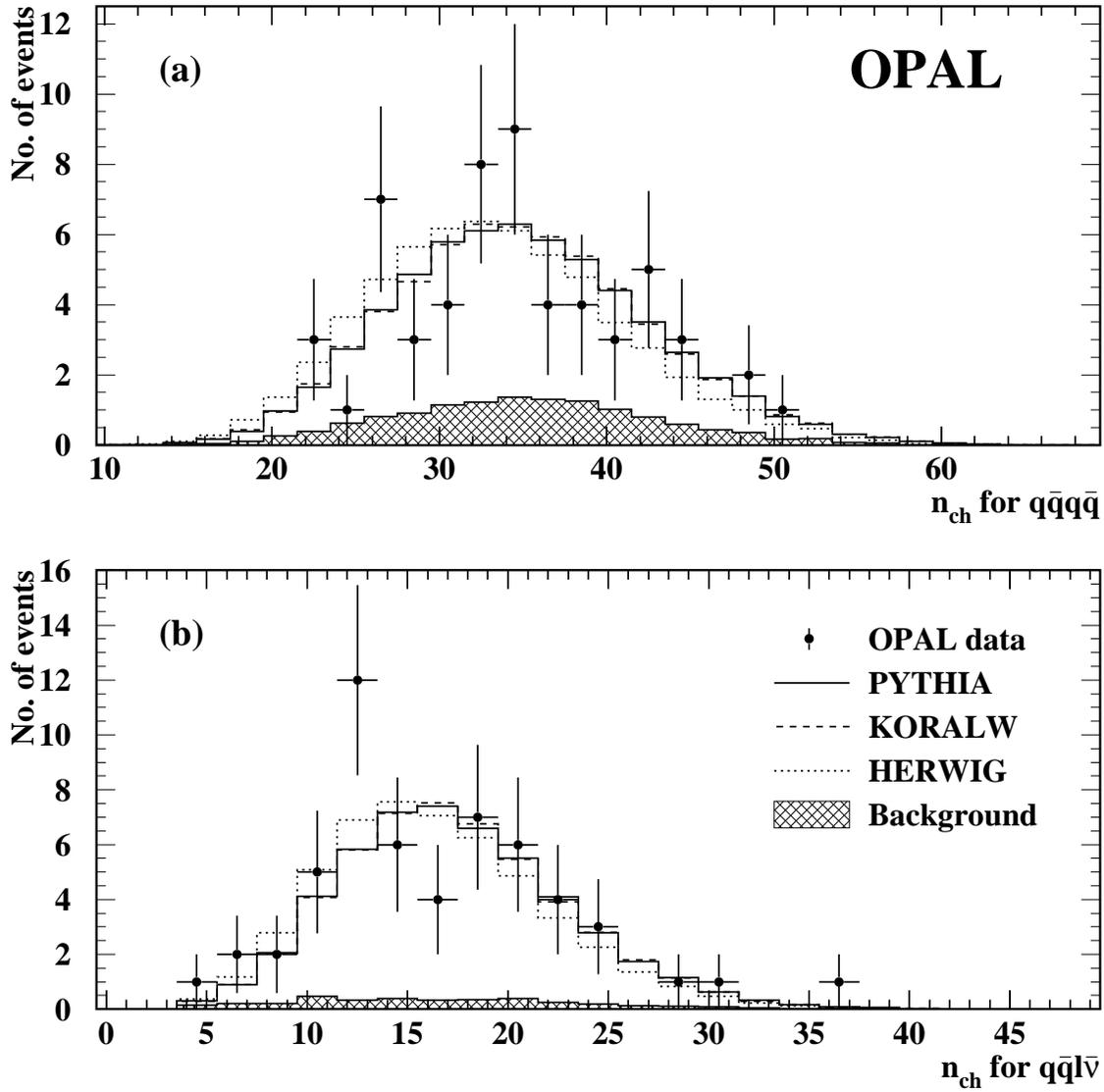,width=\textwidth}}
 \caption{Uncorrected charged multiplicity distributions for
   (a) \WWqqqq\ events and (b) the hadronic part of \WWqqln\ events.  The
   points indicate the data, the histograms show the total expected signal
   and background contribution for various signal models, and the hatched
   histogram shows the expected background.  The bins are two units wide.}
 \label{fig-wwprop-nch}
\end{figure}

\begin{figure}[htbp]
 \centerline{\epsfig{file=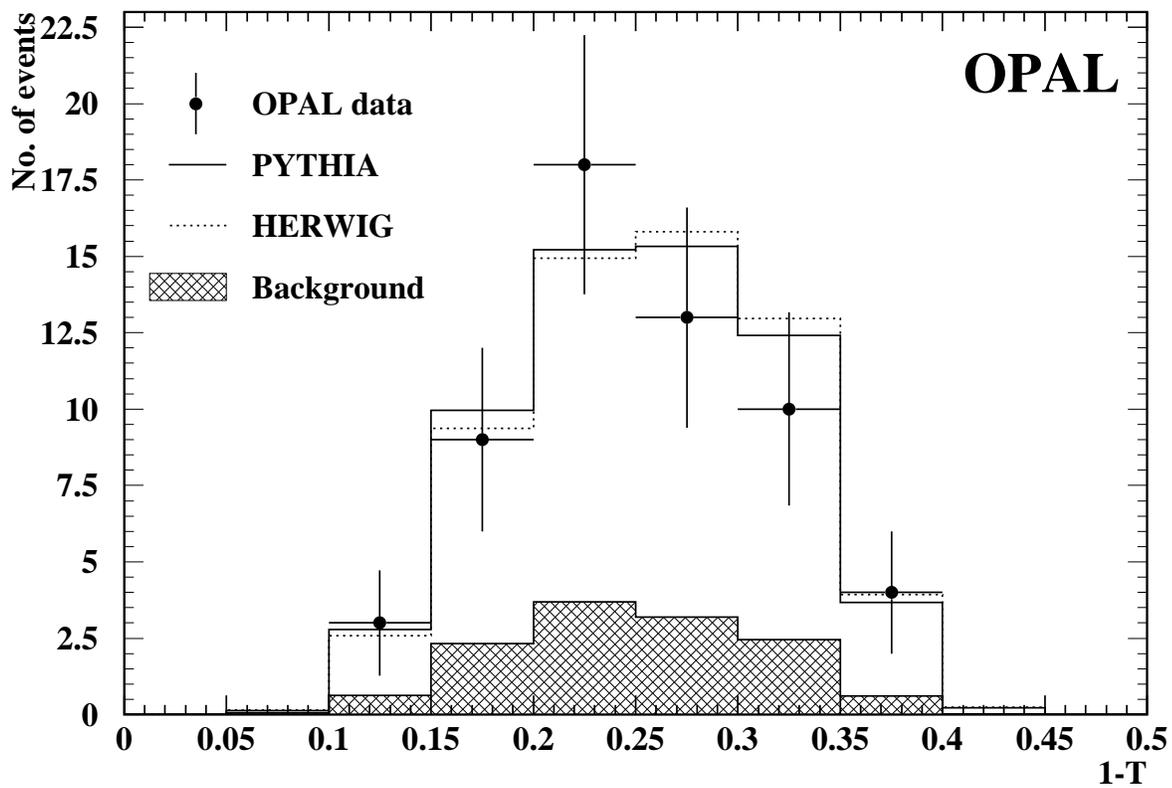,width=\textwidth}}
 \caption{Uncorrected thrust distribution for
   \WWqqqq\ events.  The points indicate the data, the histograms show the
   total expected signal and background contribution for various signal
   models, and the hatched histogram shows the expected background.}
 \label{fig-wwprop-thr}
\end{figure}

The uncorrected multiplicity and thrust distributions for the \WW\ candidate
events before background subtraction are illustrated in
Figures~\ref{fig-wwprop-nch} and \ref{fig-wwprop-thr}, together with the
predictions of Monte Carlo events including detector simulation. The
background prediction is the sum of all Standard Model processes, as
described by the models used in Section~\ref{selection}.  Good agreement is
seen between the data and predictions from the models. The simulated \WW\ 
events do not include colour reconnection effects.

\subsection{Systematic Uncertainties}

A number of systematic uncertainties have been considered in the analysis,
as summarised in Table~\ref{tab-wwprop-syst}.  Each systematic uncertainty
is taken as a symmetric error and the total uncertainty is defined by adding
the individual contributions in quadrature.  The dependence of the
correction procedure on the Monte Carlo model is evaluated by comparing
results obtained using \PYTHIA, \KORALW\ or \HERWIG\ as the \WW\ signal
samples.  In each case, the same model is used for the subtraction of the
small \WW\ background contamination in each channel: \WWqqqq\ or \WWlnln\ 
events selected as \WWqqln, for example.  Uncertainties arising from the
selection of charged tracks are estimated by varying the track selection
cuts and repeating the analysis.  The maximum allowed values of the
distances of closest approach to the interaction region in $r$-$\phi$ and
$z$ are varied from 2~cm to 5~cm and from 25~cm to 50~cm, respectively, and
the minimum number of hits on tracks is varied from 20 to 40. The
dependence on charged track quality cuts is the sum in quadrature of these
three effects.

To test the dependence of the results on the event selection, alternative
selections were considered.  The \WWqqln\ selection was replaced by a
cut-based selection similar to that used in~\cite{bib:opalmw1}, which has a
lower efficiency but comparable purity to the likelihood selection.  For the
\WWqqqq\ case, the standard likelihood selection is modified by adding a cut
on the jet resolution parameter \Yc{34} such that 10\% of the selected
signal Monte Carlo events are removed from the sample.  The entire analysis
was repeated with the alternative selections.

The background estimates presented in Section~\ref{selection} have overall
uncertainties of order 25\% for the \qq\qq\ and \qq\lnu\ channels.  These were
taken into account by scaling the background by $\pm25\%$ before subtraction
from the data.  An additional uncertainty was included to allow for the fact
that bin contents could become negative after background subtraction due to
low statistics in the data.

An uncertainty was evaluated for the Monte Carlo tuning or model dependence
of the \Zgamma\ background by shifting the multiplicity distribution by
$\pm1$ unit or comparing \PYTHIA\ with \HERWIG.  These gave similar shifts
in \nchave.  Other background contributions were also varied by considering
alternative two-photon Monte Carlo samples from \PYTHIA, \HERWIG\ or
\TWOGEN, by comparing the corrected values when treating \grcff\ \WW\ or
four-fermion events as data, or by neglecting the
$\Zgamma\rightarrow\tau^+\tau^-$ background.  The corresponding systematic
uncertainty was taken to be the sum in quadrature of the uncertainties from;
the four-fermion study, the effect of
neglecting the $\Zgamma\rightarrow\tau^+\tau^-$ background and the
largest shift from the use of any two-photon sample.

Since most of the Monte Carlo samples used in the study were generated
at $\roots=171$~GeV, the analysis was repeated with \WW\ and \Zgamma\ 
samples generated at 172~GeV.  The effect of varying the signal and
background cross-sections over the range expected at these
centre-of-mass energies was found to be negligible.

As a further systematic check, the mean values \nchQQQQ, \nchQQLV, \xpQQLV,
\xpQQQQ, \thrQQQQ\ and \yQQQQ\ were evaluated by applying a correction
factor to each of the uncorrected values.  This correction is the ratio
between the \PYTHIA\ prediction without detector simulation or initial state
radiation, to the corresponding prediction for the same observable when
these two effects are included.  The change in the corrected value is
included as an estimate of the systematic error due to the unfolding
process.  For \thrQQQQ, this is the largest single contribution to the
systematic uncertainty, as the event selection in the \WWqqqq\ channel
rejects events having two-jet like characteristics which are similar to the
dominant $\Zgamma\rightarrow\qq$ background.  Hence, part of the thrust
distribution is unmeasured in data.

\subsection{Results}

The mean values of the event properties are as follows,
where in each case the first uncertainty is statistical and the
second systematic.  
\begin{eqnarray*}
           \nchQQQQ & = & 38.3 \pm 1.1 \pm 0.6 \\
           \nchQQLV & = & 18.4 \pm 0.9 \pm 0.4 \\
           \Dnch    & = & +1.5 \pm 2.2 \pm 0.8 \\
           \xpQQQQ  & = & (3.22  \pm 0.13  \pm 0.08)\times 10^{-2} \\
           \xpQQLV  & = & (3.60  \pm 0.20  \pm 0.11)\times 10^{-2} \\
           \Dxp     & = & (-0.38 \pm 0.23  \pm 0.10)\times 10^{-2} \\
           \thrQQQQ & = & 0.227 \pm 0.036 \pm 0.021 \\
           \yQQQQ   & = & 1.033 \pm 0.042 \pm 0.025
\end{eqnarray*} 
The values of \thrQQQQ\ and \yQQQQ\ agree well with the predictions of
models, which are 0.219 and 1.031 for \PYTHIA, and 0.216 and 1.032 for
\HERWIG, respectively.  It is interesting to compare the mean charged
particle multiplicity in \WW\ events with that in $\Zgamma\rightarrow\qq$
events at $\roots=172$~\GeV, which has a value of approximately
26~\cite{bib:QCD172}.

\begin{figure}[htbp]
 \centerline{\epsfig{file=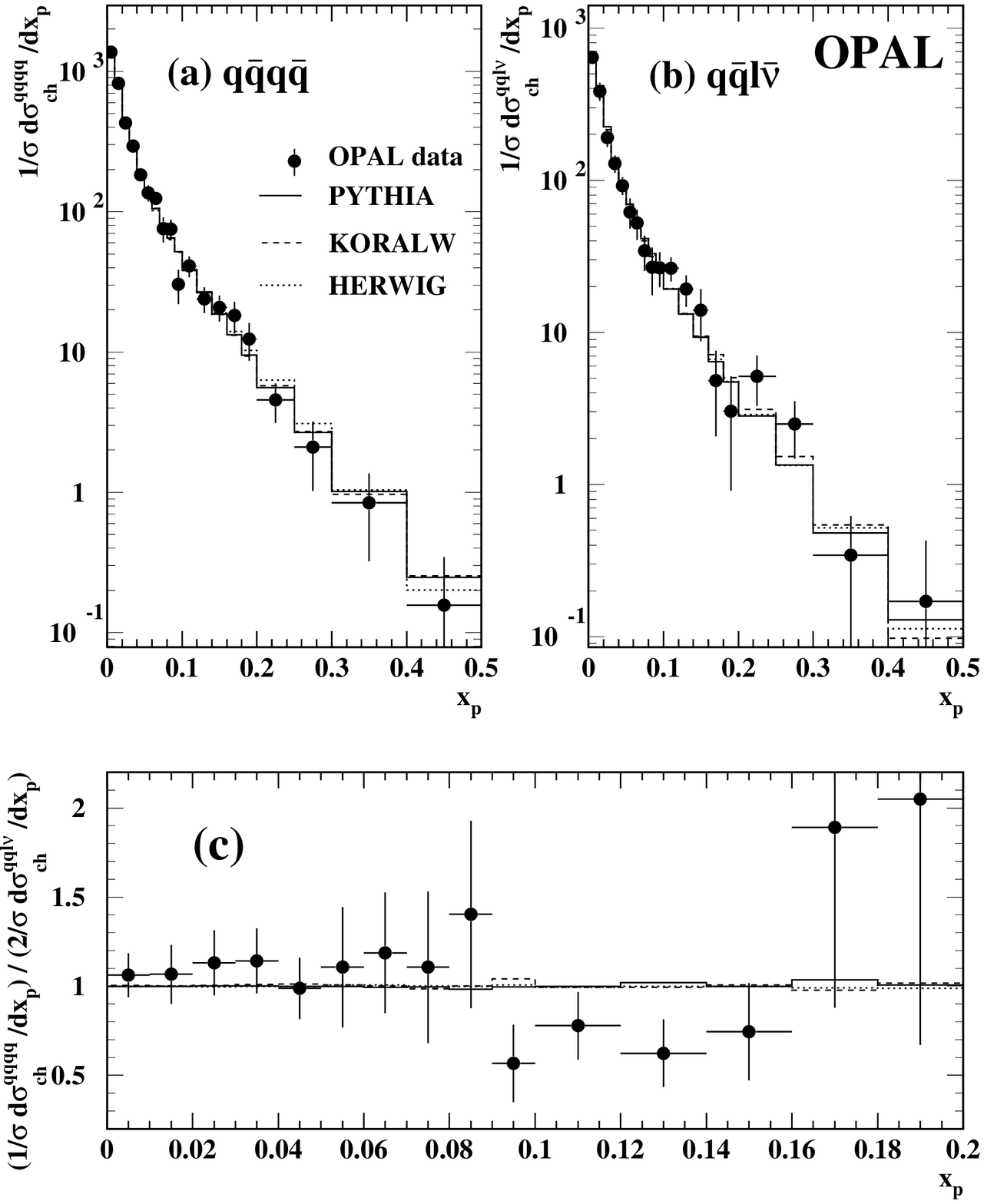,width=\textwidth}}
 \caption{Corrected \protect\xp\ distributions for
   (a) \WWqqqq\ events, (b) the hadronic part of \WWqqln\ events and (c) the
   ratio of the \WWqqqq\ distribution to twice the \WWqqln\ distribution.
   The points indicate the data, with statistical and systematic
   uncertainties added in quadrature, and the predictions of various Monte
   Carlo models (without colour reconnection) are shown as histograms.}
 \label{fig-wwprop-xp}
\end{figure}

The difference in mean charged multiplicities in hadronic W decays in
\qq\qq\ and \qq\lnu\ events, \Dnch, is found to be consistent with zero at
the current level of statistical precision.  Similarly, the measurements of
the mean scaled charged particle momenta are consistent in the two channels.
Figures~\ref{fig-wwprop-xp}(a) and (b) show the corrected fragmentation
functions for the \WWqqqq\ and \WWqqln\ channels, together with predictions
from the \PYTHIA\ and \HERWIG\ models.  The models are in good agreement
within statistical uncertainties in both cases.  An alternative measurement
of the mean charged multiplicity may be obtained from the integral of the
fragmentation function.  The values determined in this way are \nchQQQQ=38.2
and \nchQQLV=18.2.  Figure~\ref{fig-wwprop-xp}(c) shows the ratio of the
$x_{p}^{\QQQQ}$ distribution to twice the $x_{p}^{\QQLV}$
distribution for low particle momenta, $\xp<0.2$.  The ratio is slightly
greater than unity in the high statistics region, in agreement with the
positive value of \Dnch\ measured above.

In summary, the \WW\ event properties presented here are in good agreement
with expectations of standard QCD models.  From studies of reconnection
phenomena implemented with the \Herwig, \Ariadne\ and \Pythia\ models,
changes up to approximately one statistical standard deviation in the
current data may be expected in \nchQQQQ\ and \xpQQQQ. Shifts several times
larger have also been predicted for \nchQQQQ\ \cite{bib:EG2}.  
Defining \Dnch\ and \Dxp\ using data alone 
provides a model independent test of possible reconnection effects.
The maximum shifts in these variables predicted by the models 
considered are at the level 0.5--1.5 standard deviations for the current 
data set. There is no indication of the effects of colour
reconnection on these observables at the current level of statistical
precision.

\section{Summary}
%
%
W pair events produced in \epem collisions at 172.12~\GeV\ were
analysed.  A Monte Carlo based reweighting method was used to fit the reconstructed mass
spectra of the selected hadronic and semi-leptonic W pair events to obtain
the following values for the mass and decay width of the \PW\ boson:
\begin{eqnarray*}
 \Mw & = & 80.30 \pm 0.27 \pm 0.09 \; \GeV, \\
 \Gw & = &  1.30^{+0.62}_{-0.55} \pm 0.18 \; \GeV,
\end{eqnarray*}
where the first uncertainty is statistical and the second is
systematic.  The fitted width is approximately one standard deviation
below the Standard Model prediction for this value of \Mw.
Constraining \Gw\ to its Standard Model relation gives a measurement
of the mass of
\begin{eqnarray*}
 \Mw  =  80.32 \pm 0.30 \pm 0.09 \; \GeV.
\end{eqnarray*}
These results are consistent with a recent result\cite{bib:L-mass172}.
A Breit-Wigner fit to the reconstructed mass spectra was used to check
the reweighting method for the extraction of \Mw, and found to give a
consistent result.  Combining the above value for 
\Mw\ with the OPAL threshold
measurement \cite{bib:opalmw1} of the W boson mass and that derived from
the measurement of $\sigccthree(172~\GeV)$, gives
\begin{eqnarray*}
 \Mw = 80.35 \pm 0.24 \pm 0.06  \pm 0.03 \pm 0.03\; \GeV, 
\end{eqnarray*}
where the uncertainties are statistical, systematic, colour reconnection and
Bose-Einstein, and beam energy, respectively.

The production cross-section for \WW\ events at this energy 
was found to be 
\begin{displaymath}
 \sigccthree(172~\GeV) = 12.3 \pm 1.3 \pm 0.3 ~\mathrm{pb}.
\end{displaymath}
The decay branching fractions of the W
boson were measured assuming only Standard Model W decay modes. 
When combined with the previous OPAL results at
\roots=161.3~\GeV\ the following leptonic branching fractions for the \PW\ 
boson are obtained:
\begin{eqnarray*}
 \Br(\Wev)    & = & 0.098^{+0.022}_{-0.020} \pm 0.003, \\ 
 \Br(\Wmv)    & = & 0.073^{+0.019}_{-0.017} \pm 0.002, \\ 
 \Br(\Wtv)    & = & 0.140^{+0.030}_{-0.028} \pm 0.005.
\end{eqnarray*}
If charged current lepton universality is assumed then the
leptonic and hadronic branching fractions are determined to be
\begin{eqnarray*}
 \Br(\Wlv)         & = & 0.101^{+0.011}_{-0.010} \pm 0.002, \\ 
 \Br(\Wqq)         & = & 0.698^{+0.030}_{-0.032} \pm0.007,
\end{eqnarray*}
where the errors are 100\% anti-correlated. 
%
%
The hadronic branching fraction result gives 
a direct measurement of the sum of
squares of kinematically accessible elements of the CKM matrix,
\begin{eqnarray*}
  \sum_{i={\mathrm{u,c}};\, j={\mathrm{d,s,b}}}
  \Vij^2 & = & 2.22^{+0.32}_{-0.34}\pm0.07.
\end{eqnarray*}
 This can in turn be interpreted as a measurement of the element \Vcs,
\begin{eqnarray*}
  \Vcs & = & 1.08 ^{+0.15}_{-0.16} \pm 0.03,
\end{eqnarray*}
by using measured values of the other $\Vij^2$ parameters. The measurement of
\Vcs\ is consistent in value with and has a comparable uncertainty to
other determinations which do not invoke unitarity
\cite{bib:pdg}. The above cross section and branching fration measurements
are consistent with the Standard Model expectations
and with recent results\cite{bib:L-xs172}. 

%
The predicted effects of colour reconnection and Bose-Einstein phenomena in
the fully hadronic channel, and also their influence on \Mw, are model
dependent. A first investigation of these effects was performed in data
alone by comparing the distribution of the fragmentation function, \xp, and
mean values of the charged particle multiplicity and \xp\ for \WWqqqq\ and
the non-leptonic component of \WWqqln\ events. The mean values obtained in
the two channels are found to be consistent, with differences:
\begin{eqnarray*}
  \Dnch = \nchQQQQ-2\nchQQLV  & = & +1.5 \pm 2.2 \pm 0.8 \\
  \Dxp  = \xpQQQQ-\xpQQLV     & = & (-0.38 \pm 0.23  \pm 0.10)\times 10^{-2}.
\end{eqnarray*} 
In addition, mean values of rapidity and thrust are determined for \WWqqqq\ 
events.  All measurements are in agreement with predictions of standard QCD
models. At the current level of statistical precision no evidence for colour
reconnection effects was found in the observables studied.

\section{Acknowledgements}

We particularly wish to thank the SL Division for the efficient operation
of the LEP accelerator at all energies
and for
their continuing close cooperation with
our experimental group.  We thank our colleagues from CEA, DAPNIA/SPP,
CE-Saclay for their efforts over the years on the time-of-flight and trigger
systems which we continue to use.  In addition to the support staff at our own
institutions we are pleased to acknowledge the  \\
Department of Energy, USA, \\
National Science Foundation, USA, \\
Particle Physics and Astronomy Research Council, UK, \\
Natural Sciences and Engineering Research Council, Canada, \\
Israel Science Foundation, administered by the Israel
Academy of Science and Humanities, \\
Minerva Gesellschaft, \\
Benoziyo Center for High Energy Physics,\\
Japanese Ministry of Education, Science and Culture (the
Monbusho) and a grant under the Monbusho International
Science Research Program,\\
German Israeli Bi-national Science Foundation (GIF), \\
Bundesministerium f\"ur Bildung, Wissenschaft,
Forschung und Technologie, Germany, \\
National Research Council of Canada, \\
Hungarian Foundation for Scientific Research, OTKA T-016660, 
T023793 and OTKA F-023259.\\

\clearpage
\appendix

\section*{Appendices}
\label{app-selections} 


\section{\boldmath \WWqqln\ Event Selection}
 \label{app-qqln}
 
 The selection of each flavour of \WWqqln\ events is divided into four
 distinct stages as discussed in Section~\ref{sec-qqln}, namely
 identification of a charged lepton candidate, preselection, relative
 likelihood selection to separate signal from the remaining
 background, and event categorisation. Details of each of these stages
 are given below. Only events which fail the \WWlnln\ selection are 
 considered as possible \WWqqln\ candidates.   

\subsection{Identification of the Candidate Electron and Muon}
\label{sec-qqev-eid}

In order to maximise efficiency no explicit lepton identification is required.
Instead, the track in the event which is most consistent 
with being an electron (muon) from the decay \Wev\ (\Wmv\ ) is 
taken to be the candidate lepton. 
This track is identified using two types of variables: 
a) lepton identification variables, for example, number of
hits in the hadron calorimeter or specific energy loss in the 
central tracking chamber; and b)
variables representing the probability that the lepton arose 
from a W decay, for example,  energy and isolation. 
These variables are used to calculate, for each track, the probability 
that the track arose from a \Wev\ decay, \Probe, or from a \Wmv\ 
decay, \Probm. These probabilities are the products of 
probabilities from the individual variables, which are determined using 
probability density functions obtained from \WW\ Monte Carlo.     
The track with the highest value of \Probe\ is taken to be the
candidate electron in the  \WWqqen\ selection and 
the track with the highest value of \Probm\ is taken to be the
candidate muon in the \WWqqmn\ selection. Each event will yield one
candidate electron track and one candidate muon track.

\subsection{Definition of Variables}

Having selected the most likely electron and muon candidates,
variables are constructed which are used in the preselection and,
subsequently, in the likelihood selection. 
The following variables are used:
\begin{itemize}
 \item \Elept, the energy of the candidate lepton. For electrons this is
               calculated using the electromagnetic calorimeter
               energy;
               for muons the track momentum is used,
 \item \cslpmis, the cosine of the angle between the lepton track and the 
  missing momentum vector,
 \item \Ctmis, the cosine of the angle the missing momentum vector makes 
              with the beam axis, 
 \item \Etwo,  the energy in a cone of 200 mrad around the candidate lepton
               evaluated using tracks and ECAL clusters,
 \item \Probe\ or \Probm, the electron or muon identification   
               probability for the 
               candidate lepton track,
 \item \Rvis, the visible energy of the event scaled by \roots,
 \item \Yc{23}, where \Yc{{ij}} is the value of the jet
                resolution parameter (Durham scheme~\cite{bib:durham})
                at which an event is reclassified from $i$ jets to $j$ jets,
 \item \Ptsum,  the transverse momentum of the event relative to the beam 
                axis, calculated using tracks, ECAL clusters and HCAL clusters,
 \item \Psprime, the probability from a kinematic fit which estimates
                 the invariant mass of the event, \rootsprime,
 \item \Tjet,  the angle between the lepton candidate and the jet axis of
               the nearest hadronic jet.
\end{itemize}

\subsection{Preselection}
\label{sec-qqev-pres}

Preselection cuts are applied to remove events which are clearly
inconsistent with \WWqqln\ decays. In particular, the preselections
are designed to remove most two-photon events and a significant
fraction of the \Zqq\ background.  Slightly different preselection
cuts are applied in the \WWqqen\ and \WWqqmn\ selections. Firstly,
events
are required to have more than five charged tracks and more than seven
electromagnetic calorimeter clusters.  In addition, the main
preselection cuts are: \Elept $ > 10.0$ GeV, $0.30 < \Rvis < 1.05$,
the total energy in the forward luminosity monitors $<40$~GeV,
$\cslpmis<0.0$ and the energy of the highest energy isolated photon
in the event $< (\Eisr - 10)$~GeV, where \Eisr\ is the expected energy
of an initial state photon for radiative events with \rootsprime$\sim$\Mz. An
isolated photon is defined as an ECAL cluster which is not associated
to a track and which satisfies the isolation requirement of less than
2.5~GeV of energy in a 200~mrad cone around the cluster.  Finally,
loose cuts are made on the lepton identification probabilities and, in
the case of the \WWqqen\ selection, several cuts are made to reduce
the background from converting photons.  The preselection is
approximately 92\% efficient for \WWqqen\ and \WWqqmn\ events, where
about half the loss in efficiency arises from cases where the lepton
is outside the experimental acceptance for well reconstructed charged
tracks.  The preselection cuts remove approximately 90\% of the \Zqq\ 
background.
   
\subsection{Relative Likelihood Selection}
\label{sec-qqev-like}

\begin{figure}
 \centerline{\epsfig{file=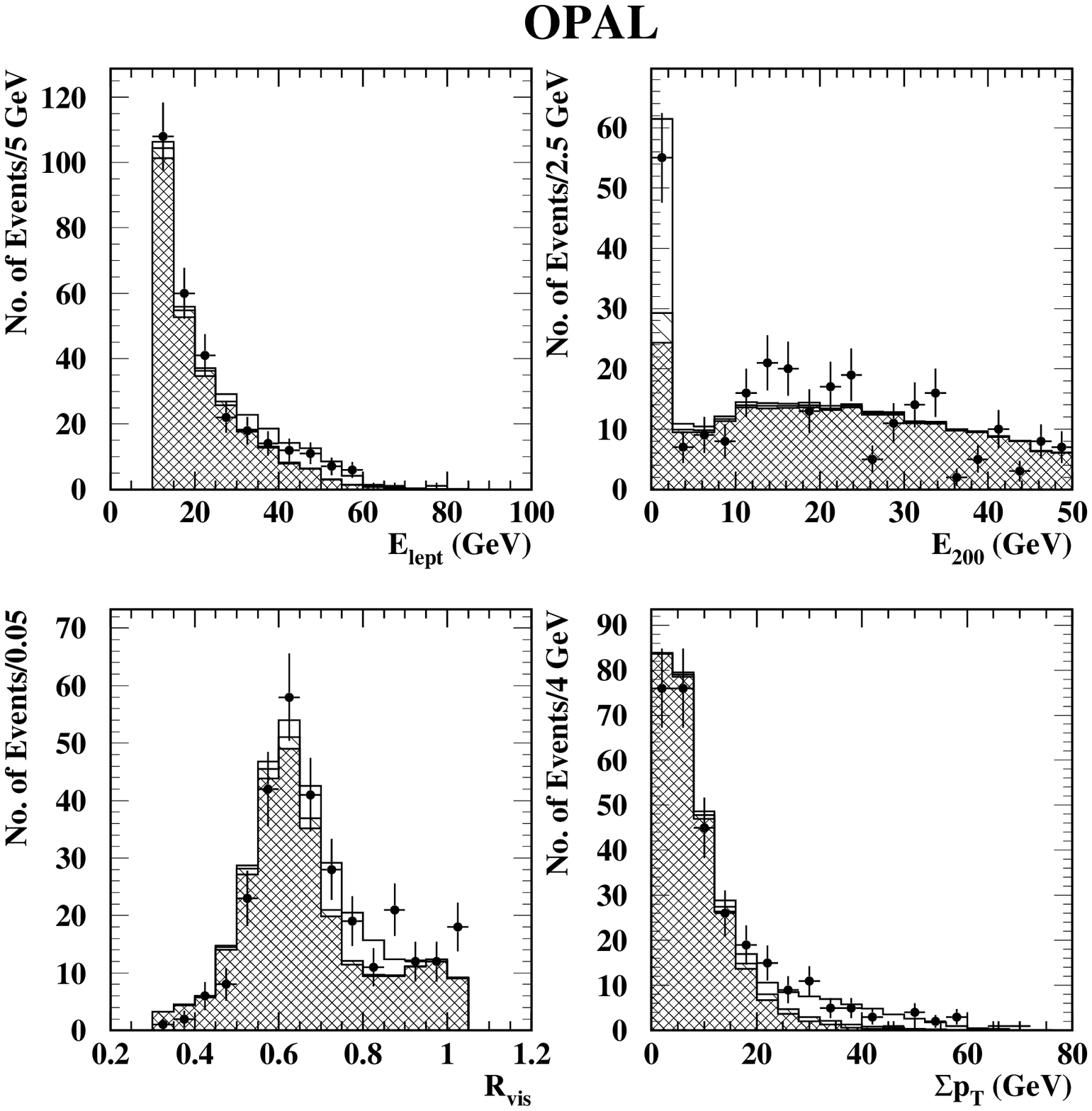,%
 width=15 cm}}
 \caption{ Distributions of some of the 
   variables used in the \WWqqen\ and \WWqqmn\ selections for events passing
   either the \WWqqen\ or the \WWqqmn\ preselection cuts.  The contribution
   from \WWqqtn\ decays is shown as the single hatched histogram and the
   contribution from background processes as the doubly hatched histogram.
   The data are shown as the points with error bars.}
 \label{fig_qqln_1}
\end{figure}

For events passing the electron 
preselection, a relative likelihood 
method is used to distinguish \WWqqen\ events from the 
background of \Zqq\ events. After the preselection cuts  
there is still a signal to background  ratio of less than 0.1.
The likelihoods are based on a set of variables, ${x_i}$, where
the observed values are compared to the expected distributions 
(obtained from Monte Carlo)
for  \WWqqen\ events from which the probabilities, $P_i(x_i)$, are obtained.
The likelihood, $L^{\qqen}$, is calculated 
as the product of these probabilities for the individual variables used in the 
analysis. 
The background likelihood, $L^{\qq}$, is obtained in the same manner 
using Monte Carlo distributions for \Zqq\ events. The relative likelihood,   
$\cal{L}^{\qqen}$, is calculated as:
\begin{eqnarray*}
  \cal{L}^{\qqen} &=& {{L^{\qqen}}\over{L^{\qqen}+f\times L^{\qq}}},
\end{eqnarray*}  
where the normalisation factor, $f$, is the ratio of preselected
background to signal cross-sections from Monte Carlo.  The variables
used in the \WWqqen\ likelihood are: \Elept, \Etwo, \Probe, \ytwo,
\Rvis, \Ctmis, \Ptsum, \cslpmis, \Psprime\ and \Tjet.  The \qqmn\ 
relative likelihood, $\cal{L}^{\qqmn}$, is obtained in the same manner
with a slightly modified set of variables; with \Probm\ replacing
\Probe\ and by not using \Tjet.  The inclusion of \Tjet\ in the \qqen\ 
likelihood was observed to improve the signal and background
separation whereas in the \qqmn\ selection this was not the case.
Figure \ref{fig_qqln_1} shows the distributions of a number of these
variables where each plot shows the combination of the distributions
for events passing the \WWqqen\ and \WWqqmn\ preselections.

\begin{figure}
 \centerline{\epsfig{file=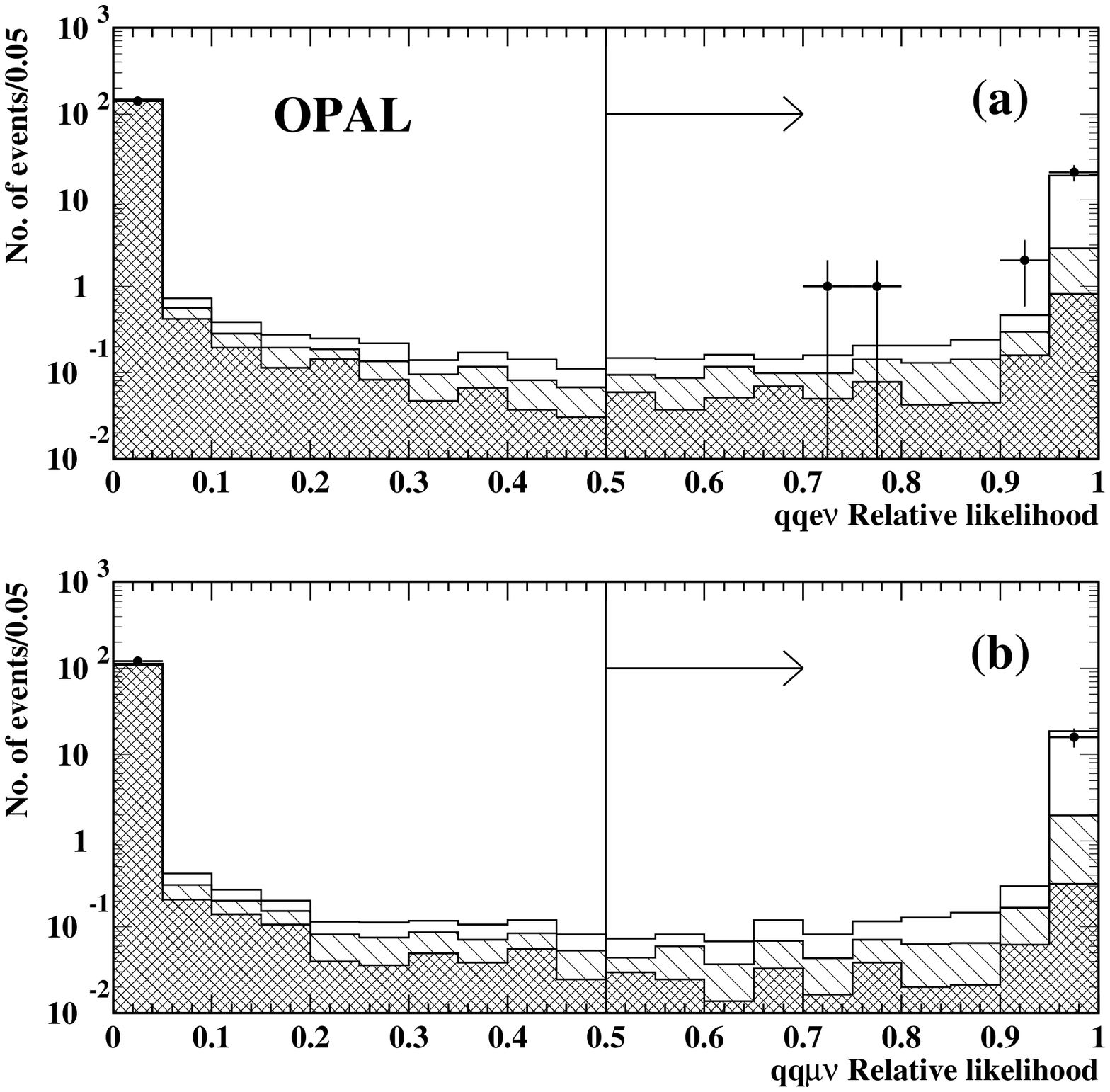,%
 width=15 cm}}
 \caption{ Values of the \WWqqen\ and
   \WWqqmn\ relative likelihoods, ${\cal{L}}^{\qqen}$ and
   ${\cal{L}}^{\qqmn}$, for events passing the \WWqqen\ and \WWqqmn\ 
   preselection cuts respectively.  The contribution from \WWqqtn\ decays is
   shown as the single hatched histogram and the contribution from
   background processes as the doubly hatched histogram. The peaks in the
   background distributions near one arise predominantly from four-fermion
   background.  The data are shown as the points with error bars. Events
   having relative likelihood values greater than 0.5 are selected, as
   indicated shown by the arrows.}
 \label{fig_qqln_2}
\end{figure}

Figure~\ref{fig_qqln_2}(a) shows the distribution of $\cal{L}^{\qqen}$
for events passing the \WWqqen\ preselection.  The value of the
relative likelihood peaks at around one for \WWqqen\ events and zero
for \Zqq\ events.  Figure~\ref{fig_qqln_2}(b) shows the equivalent
distribution of $\cal{L}^{\qqmn}$ 
for events passing the \WWqqmn\ preselection.  Events
with ${\cal{L}}^{\qqen} > 0.5$ or ${\cal{L}}^{\qqmn} > 0.5$ are
selected as \WWqqen\ or \WWqqmn\ candidates, respectively.  The
combination of preselection and likelihood selection rejects about
99.95\% of the \Zqq\ background and is approximately 90\% efficient
for \WWqqen\ and \WWqqmn\ events.

\subsection{Event Categorisation}
\label{sec-qqlv-categ}
Although the above likelihood selections were optimised to 
separate \WWqqen\ and \WWqqmn\ events from the \Zqq\ background, they also
select approximately 25\% of \WWqqtn\ decays.  
For this reason events passing the \qqen\ selection are re-classified as
either \WWqqen\ or \WWqqtn\ and events passing the
\qqmn\ selection are re-classified as
either \qqmn\ or \qqtn. 
This assignment enables separate cross-sections for the three
\WWqqln\ channels to be  evaluated cleanly.
The predominant \WWqqtn\ 
contamination in the \qqen\ selection arises from cases where the 
tau lepton decays to an electron or decays into a one prong hadronic final 
state. Therefore, for events identified as 
\WWqqen, two relative 
likelihoods, analogous to
those described in Section \ref{sec-qqev-like},
are constructed using the same variables as were used in the 
\WWqqen\ likelihood selection. The first relative likelihood
attempts to separate \WWtelect\ decays from \WWqqen\ decays 
and the second attempts to separate \WWtonep\ from \WWqqen. 
If either of these relative likelihoods
is greater than 0.5 the event is re-categorised as \WWqqtn.
A similar procedure is applied to events passing the \WWqqmn\ selection.
The result of the event categorisation is that 
all events passing the  \qqen\ and \qqmn\ selections are categorised
as either \WWqqen, \WWqqmn\ or  \WWqqtn.

\subsection{\boldmath \WWqqtn\ Event Selection}
\label{sec-qqtn-sel}

\begin{figure}
 \centerline{\epsfig{file=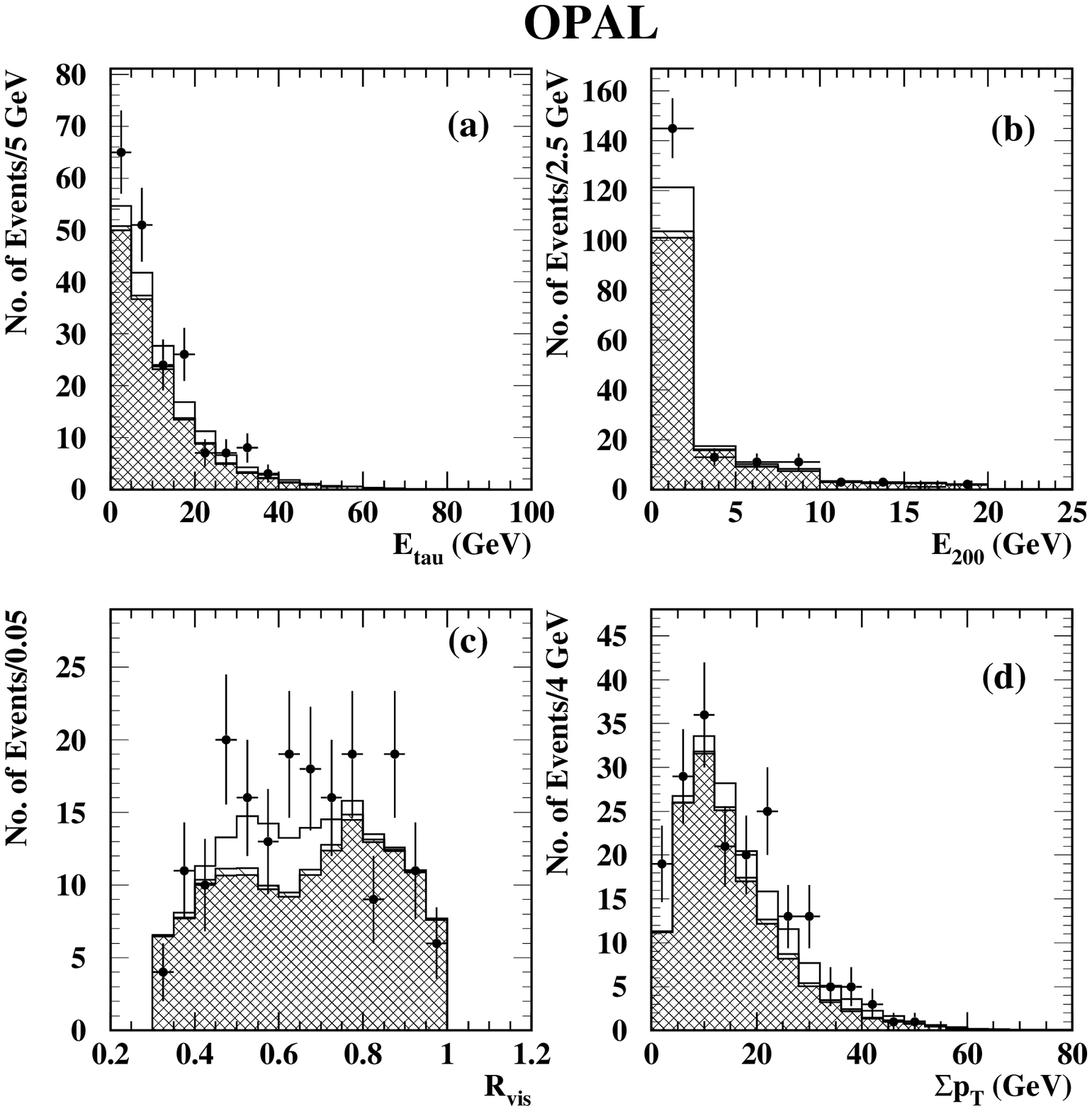,%
 width=15 cm}}
 \caption{ Distributions of some of the 
   variables used in the \WWqqtn\ selections for events passing the \WWqqtn\ 
   preselection cuts. Some events appear more than once in the plot since
   they can satisfy more than one of the \WWqqtn\ preselections.  The
   variables shown are; (a) the energy of the tau decay products, excluding
   the neutrino(s), (b) the energy within a 200 mrad cone about the tau
   candidate track(s) calculated using tracks only, (c) the scaled visible
   energy, after preselection cuts of $0.3<\Rvis<1.0$ and (d) the transverse
   momentum of the event.  The contribution from \WWqqen\ and \WWqqmn\ 
   decays is shown as the single hatched histogram and the contribution from
   background processes as the doubly hatched histogram.  The data are shown
   as the points with error bars.}
 \label{fig_wwqqtn}
\end{figure}

A relative likelihood selection designed to separate \WWqqtn\ from \Zqq\ 
background is applied to events which fail the \WWqqen\ and \WWqqmn\ 
selections described in Sections \ref{sec-qqev-eid}--\ref{sec-qqev-like}.
Approximately 25\% of all events finally selected as \WWqqtn\ are from the
analysis given in Sections \ref{sec-qqev-eid}--\ref{sec-qqlv-categ}, the
remaining events being selected as summarised below.  The \WWqqtn\ event
selection proceeds in a similar manner to the \WWqqen\ selection described
above, {\em i.e.} `lepton' identification, preselection and relative
likelihood selection. In this case no additional event categorisation is
performed.  The selection has been designed to be sensitive to the four main
tau decay classes: electron, muon, hadronic one prong and hadronic three
prong.  Consequently four separate selections are applied.  The lepton
identification of the \WWqqen\ selection is replaced by the identification of
the track most consistent with being from \Wten, \Wtmn\ and \Wtonep\ decays.
To be sensitive to three prong tau decays, the combination of three tracks
which is most consistent with a \Wtthreep\ decay is also identified. The
\WWqqtn\ selection then proceeds as four preselections, one for each of the
above cases, and four corresponding likelihood selections. The variables used
in the likelihood are similar to those used above but include more information
about the track (or tracks) identified as the tau decay {\em e.g.} the
invariant mass of all tracks and clusters within a 200 mrad cone around the
track. Figure~\ref{fig_wwqqtn} shows a sample of the variables used in the
likelihood selections.  An event with a relative likelihood greater than
0.75 for any one of the four tau likelihoods, is categorised as a \WWqqtn\ 
event. Events passing more than one of the \WWqqtn\ likelihood selections
enter the final event sample only once.



\section{\boldmath \WWqqqq\ Selection}
\label{app-qqqq} 

The selection of \WWqqqq\ events described below is divided into two parts,
namely a preselection and a likelihood based discriminant formed from the
combination of seven observables. A modified version of this selection,
which uses the same preselection and likelihood observables, is used to
obtain a weight for each event to be a \WWqqqq\ event. This variant,
described in Section~\ref{sec-event-weights}, leads to a small reduction in
the predicted uncertainty this channel contributes to the \WW\ cross-section
measurement and is therefore used in the analysis of cross-sections and
branching fractions.

\subsection{Preselection}
\label{sec-qqqq-pres}

Candidate events are required to be classified as hadronic \cite{bib:tkmh}
and not be selected by either the \WWlnln\ or \WWqqln\ selections.  Tracks
and calorimeter clusters are combined into four jets using the Durham
\cite{bib:durham} jet-finding algorithm, and the total momentum and energy
of each of the jets are corrected for double-counting of energy
\cite{bib:GCE}.  To remove events which are clearly inconsistent with
\WWqqqq\ events, predominantly radiative \Zgamma\ events, candidate events
are required to satisfy the following preselection criteria:
\begin{itemize}
\item $\sqrt{s^{'}}$, the fitted invariant mass of the hadronic system, must
  be greater than 140~GeV.
\item The energy of the most energetic isolated photon must be less than
  $0.3\roots$.
\item The visible energy of the event must be greater than $0.70\roots$.
\item $\Yc{34} > 0.003$, where \Yc{{ij}} is the value of the jet resolution
  parameter at which an event is reclassified from $i$ jets to $j$ jets.
\item Each jet is required to contain at least one charged track.
\item A kinematic fit, which imposes energy and momentum conservation and
  equality of the W masses, is performed on all three possible assignments
  of jets to W candidates in the event. At least one of these combinations
  must result in a convergent fit.
\end{itemize}
The efficiency of these preselection requirements for \WWqqqq\ events,
$\epsilon^{pre}_{sig}$, is 90.3\%, whilst rejecting 96.6\% of the \Zqq\ 
events. The total background cross-section after the preselection,
$\sigma^{pre}_{bgd}$, is estimated to be 4.5~pb.  The preselection accepts
99 events in data.

\subsection{Likelihood Variables}
\label{sec-qqqq-likevar}

Events satisyfing the preselection criteria are subjected to a likelihood
selection, which discriminates between signal and the remaining
four-jet-like QCD background.  The likelihoods are based on a set of seven
variables, ${y_i}$.  Probability density distributions are determined for
each $y_i$, using both simulated \WWqqqq\ and \Zqq\ events.  Each of these
$y_i$ distributions are used to construct probabilities corresponding to the
hypotheses that a given event in data is either \WWqqqq\ or \Zqq.  The
likelihoods, $L^{\qqqq}$ and $L^{\qq}$, are calculated as the product of
these probabilities.  The relative likelihood discriminant itself,
$\cal{L}^{\qqqq}$, is defined in terms of these two likelihoods as:
\begin{eqnarray*}
  \cal{L}^{\qqqq} &=& {{L^{\qqqq}}\over{L^{\qqqq}+L^{\qq}}}.
\end{eqnarray*} 
The value of $\cal{L}^{\qqqq}$, between zero and one, is used to
discriminate between signal and background events.

\begin{figure}[htb]  
 \centerline{\epsfig{file=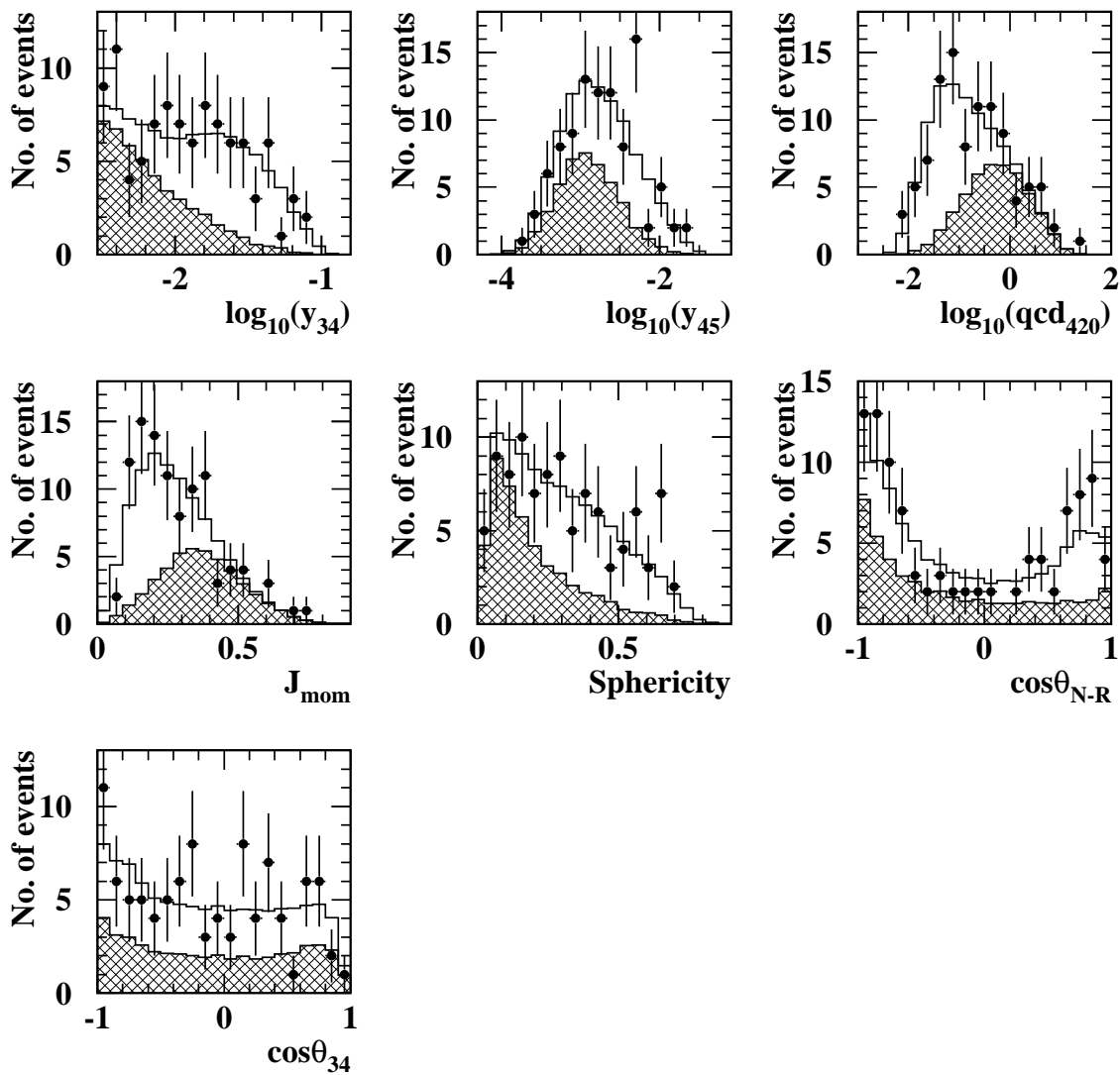,width=\textwidth}}
 \caption{Distributions of the seven variables that have been used in
   the \WWqqqq\ likelihood selection, after preselection. The points
   indicate the data, the open histogram shows the \WWqqqq\ Monte Carlo and
   the hatched histogram shows the total background expectation.}
 \label{qqqq_lvar1}
\end{figure}

The following seven variables, which use the characteristic four-jet-like
nature, momentum balance and angular structure of \WWqqqq\ to distinguish
events from the remaining background, are used to construct the
likelihoods:
\begin{itemize}
\item the logarithm of $y_{34}$,
\item the logarithm of $y_{45}$,
\item the sphericity of the event,
\item the quantity \Jmom, defined as
 \begin{eqnarray*}
   \Jmom & = & \frac{(p_1+p_2-p_3-p_4)}{\sqrt{s}},
 \end{eqnarray*}
 where $p_i$ are the jet momenta, and the jets are ordered by energy such
 that $p_1$ is the momentum of the most energetic jet,
\item the cosine of the modified Nachtmann-Reiter angle,
  $\cos\theta_{N-R}$ (see \cite{nr}),
\item the cosine of the angle between the two least energetic jets,
  $\cos\theta_{34}$,
\item the logarithm of the QCD event weight, $qcd_{420}$\cite{qcd}, which is
  calculated using the tree level matrix element for the processes
  $\epem\rightarrow \mathrm{q\bar{q}q\bar{q},q\bar{q}gg}$~\cite{qcd}.  This
  quantity should have large values for hadronic \Zgamma\ decays and smaller
  values for \WW\ events.
\end{itemize}
The distributions of data passing the preselection criteria for each of
these quantities is shown in Figure~\ref{qqqq_lvar1}, together with the
predictions of the \PYTHIA\ Monte Carlo for \WWqqqq\ and hadronic \Zgamma\ 
decays. Good agreement is seen. The resulting likelihood distribution is
given in Figure~\ref{qqqq_like}(a).  A cut value of $>0.2$ on the likelihood
is used for the measurements of the mass and event properties.

\subsection{\boldmath \WWqqqq\ Event Weights}
\label{sec-event-weights}

\begin{figure}
 \centerline{\epsfig{file=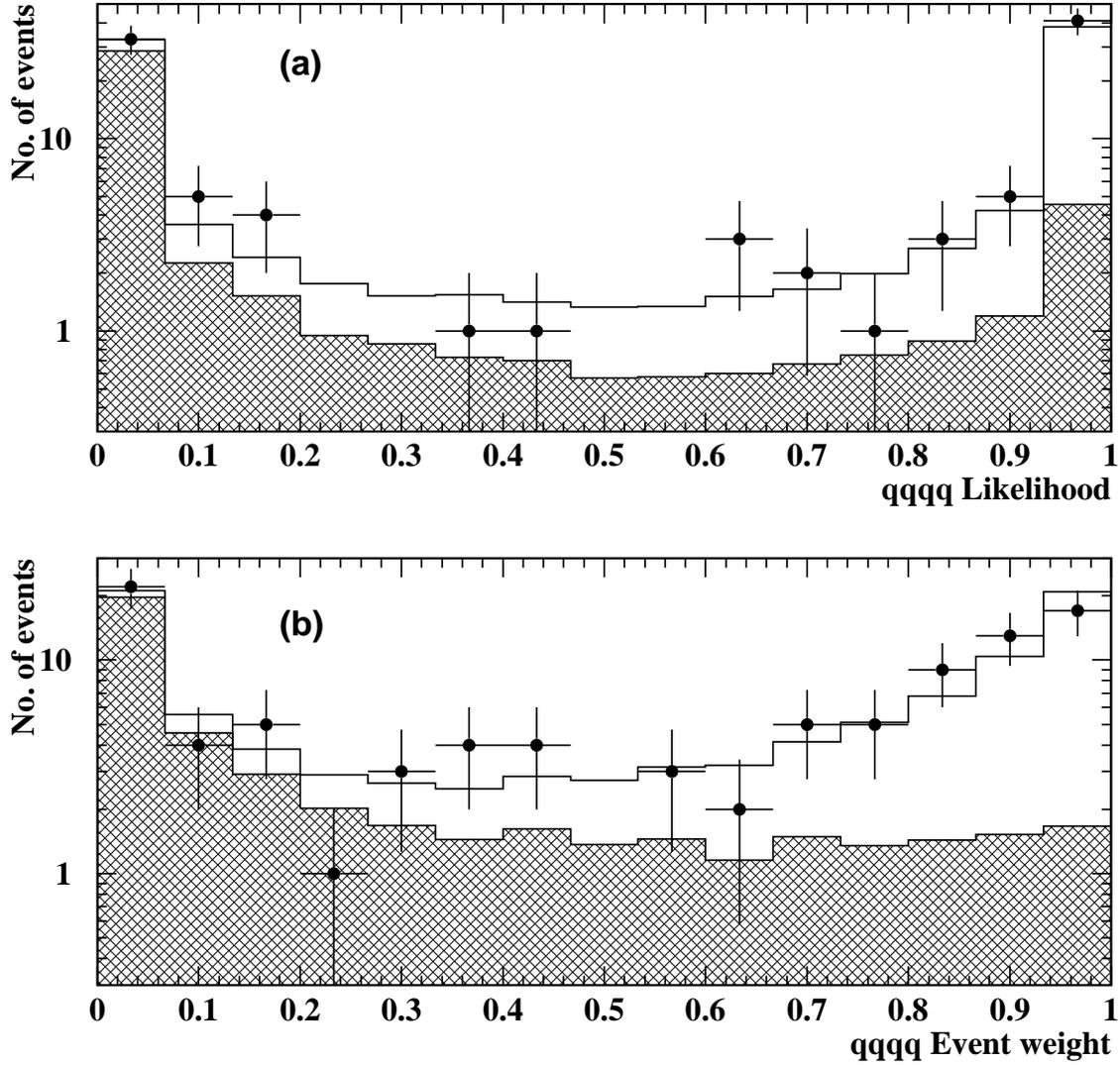,width=\textwidth}}
 \caption{Distributions of
   (a) the likelihood distribution, used to select events for the mass and
   event properties analyses and (b) the event weights, used for the
   cross-section measurement.  The points indicate the data, the open
   histogram shows the \WWqqqq\ Monte Carlo and the hatched histogram shows
   the total background expectation.}
 \label{qqqq_like}
\end{figure}

For the \WW\ cross-section and W branching fraction measurements the 99 events
passing the \WWqqqq\ preselection are assigned weights, between zero and
one, reflecting the probability that the event is from \WWqqqq\ rather than
background. The use of event weights results in a 5\% smaller statistical
uncertainty on these measurements\footnote{ In the case of the 
\WWqqln\ selection the discrimination between
  signal and background is much greater and event
  weights give a negligible improvement in statistical uncertainty and
  are therefore not used.}.  The weighting method of event classification is
discussed in \cite{evtweight}.  The optimal choice of event weight is the
probability that the event arose from signal rather than background.  In the
absence of correlations between the likelihood variables, the relative
likelihood gives the probability of an event being \WWqqqq.  The likelihood
variables described in Section~\ref{sec-qqqq-likevar} are not uncorrelated.  
For this reason a linear transformation is applied to the
likelihood variables, in order to obtain a new set of variables where the
off-diagonal elements of the covariance matrix are zero. These transformed
variables are then used to construct the relative likelihood function which
is used as an event weight.  The transformation is performed using the
orthogonal matrix which results in transformed variables with diagonal
covariance matrices for both the signal and background
separately.  The event weight, $w_i$, calculated using these
transformed variables is defined:
\begin{eqnarray*}
                w_i  & = & 
  {N_{\qqqq}{L^{\qqqq}}\over{N_{\qqqq} L^{\qqqq}+N_\qq L^{\qq}}} ,
\end{eqnarray*} 
where $N_{\qqqq}$ and $N_\qq$ are the expected numbers of preselected
\WWqqqq\ and background events respectively.  The weights for the
preselected events are compared to the Monte Carlo expectation in
Figure~\ref{qqqq_like}(b).  In calculating the event weights, the small
background contributions from \ZZqqqq\ and \WWqqlnu\ are included with the
dominant \Zqq\ background in the background covariance matrix and the
probability density distributions.

\clearpage

\clearpage
%
%
\begin{table}[htbp]
\centering
\begin{tabular}{|c|c|}
\hline
$\WW\rightarrow$ & Efficiency (\%) \\ \hline
$\enen$          & $82.4 \pm 1.1 \pm 3.2$ \\
$\enmn$          & $83.8 \pm 0.8 \pm 2.2$ \\
$\entn$          & $76.8 \pm 0.9 \pm 2.2$ \\
$\mnmn$          & $86.7 \pm 1.0 \pm 2.2$ \\
$\mntn$          & $77.0 \pm 0.9 \pm 2.2$ \\ 
$\tntn$          & $60.6 \pm 1.4 \pm 4.6$ \\ \hline
\end{tabular}
\caption{
Efficiency for selecting each \lnln\ final
state, evaluated using the \KORALW\ Monte Carlo program, after
correcting for detector occupancy. The errors are statistical and
 systematic respectively.}
\label{tab-lnln-eff}
\end{table}
\begin{table}[htbp]
\centering
\begin{tabular}{|l|r@{$\ \pm \ $}l|}
\hline
Source   & \multicolumn{2}{c|}{Cross-section (fb)} \\ \hline
$\ff$ ($\mathrm{f} = \mathrm{e},\mu,\tau,\nu, \mathrm{q} $)& \hspace{12mm}9 & 1 \\
$\epem\ff$ ($\mathrm{f}=\mathrm{e},\mu,\tau,\mathrm{q}$)   & 32& 19 \\
${\ell_1}^{+} {\ell_1}^{-} {\ell_2}^{+} {\ell_2}^{-} $
(${\ell_i}=\mu,\tau$)                                    & 1.2 & 0.5 \\
${\ell} {\overline{\ell}} \mathrm{q} {\overline{\mathrm{q}}}$
(${\ell}=\mu,\tau,\nu_{\mathrm{e}},\nu_{\mu},\nu_{\tau}$) & 0.6 & 0.3 \\
$\qq \enu$                         & 0.4 & 0.3 \\ 
${\ell_1}^{+} {\ell_1}^{-} \nu_{\ell_2} \overline{\nu}_{\ell_2}$
(${\ell_i}=\mathrm{e},\mu,\tau$) (${\ell_1} \ne {\ell_2}$) & 18 & 2 \\
${\ell_1}^{+} \nu_{\ell_1} {\ell_2}^{-} \overline{\nu}_{\ell_2}$
(${\ell_i}=\mathrm{e},\mu,\tau$) (${\ell_1}{\ell_2} \ne \mu \tau$) & 16 & 13 \\ 
\hline
Total                                        & 77 & 24 \\ \hline
\end{tabular}
\caption{Expected background cross-sections for different
  processes, assuming an average centre-of-mass energy of $\rroots$~GeV.
  The uncertainties include statistical and systematic components.}
\label{tab-lnln1}
\end{table}
\nopagebreak
\begin{table}[htbp]
 \begin{center}
 \begin{tabular}{|l|r|r|r|c|} \hline
  Selected as & Expected signal & Expected back. & Total & Observed \\ \hline
  \WWenen  & $ 1.3 \pm 0.1 $ & $ 0.1\pm0.0 $  & $ 1.4 \pm0.1$ &  2   \\ 
  \WWenmn  & $ 2.6 \pm 0.1 $ & $ 0.1\pm0.1 $  & $ 2.7 \pm0.1$ &  2   \\
  \WWentn  & $ 2.3 \pm 0.1 $ & $ 0.2\pm0.1 $  & $ 2.6 \pm0.1$ &  1   \\ 
  \WWmnmn  & $ 1.4 \pm 0.1 $ & $ 0.0\pm0.0 $  & $ 1.5 \pm0.1$ &  0   \\ 
  \WWmntn  & $ 2.2 \pm 0.1 $ & $ 0.1\pm0.1 $  & $ 2.3 \pm0.1$ &  1   \\
  \WWtntn  & $ 0.9 \pm 0.1 $ & $ 0.1\pm0.0 $  & $ 1.0 \pm0.1$ &  2   \\ \hline
  \WWqqen  & $16.8 \pm 0.4 $ & $ 1.2\pm0.4 $  & $18.0 \pm0.6$ & 19   \\
  \WWqqmn  & $17.4 \pm 0.4 $ & $ 0.6\pm0.1 $  & $17.9 \pm0.4$ & 16   \\
  \WWqqtn  & $13.5 \pm 0.4 $ & $ 2.7\pm0.7 $  & $16.2 \pm0.8$ & 20   \\ \hline
  \WWqqqq  & $41.3 \pm 1.5 $ & $13.1\pm2.0 $  & $54.4 \pm2.5$ & 54.1 \\ \hline
  Combined & $99.7 \pm 2.6 $ & $18.3\pm2.2 $  & $117.9\pm3.4$ & 117.1\\ \hline
 \end{tabular}
 \end{center}
\caption{Observed numbers of candidate events in each \WW\ decay
  channel for an integrated luminosity of 
 $\intLdt\pm\dLtot$~pb$^{-1}$ at $\rroots\pm0.06$~GeV,
  together with expected numbers of signal and background events,
  assuming $\Mw=80.33\pm0.15$~\GeVcc. The numbers for the \WWqqqq\ channel
  are the sums of the event weights for the 99 events passing the 
  preselection (see text). The predicted numbers of
  signal events include systematic uncertainties from the efficiency,
  luminosity, beam energy, \WW\ cross-section 
  and \Mw , while the background
  estimates include selection and luminosity uncertainties. The errors
  on the combined numbers account for correlations. }
\label{tab-sel_summary}
\end{table}
\begin{table}[htbp]
\begin{center}
\begin{tabular}{|l|r|r|r|} \hline
 \multicolumn{1}{|l|}{} &
 \multicolumn{3}{c|}{Generated as} \\
   Selected as & \WWqqen    & \WWqqmn & \WWqqtn                   \\ \hline
   \WWqqen & $85.1\pm0.9\%$ & $ 0.1\pm0.1\%$ & $ 3.9\pm0.3\%$\\
   \WWqqmn & $ 0.2\pm0.1\%$ & $87.6\pm0.8\%$ & $ 4.4\pm0.3\%$\\
   \WWqqtn & $ 4.7\pm0.5\%$ & $ 5.2\pm0.5\%$ & $61.4\pm1.2\%$\\ \hline
 \end{tabular}
 \end{center}
 \caption{Selection efficiencies for the  
  different \WWqqln\ channels after event categorisation showing the 
  cross contamination in each selection. The efficiencies, based on
  \KORALW, have been 
  corrected for differences between data and Monte Carlo. 
  The errors include both statistical and systematic uncertainties.}
 \label{qqln-xtalk}
\end{table}
\begin{table}[htbp]
\begin{center}
\begin{tabular}{|l|c|c|c|} \hline
 \multicolumn{1}{|l|}{} &
 \multicolumn{3}{c|}{Signal efficiency  error (\%) }              \\
Source of uncertainty          & \WWqqen\  & \WWqqmn\  & \WWqqtn\  \\ \hline
Statistical                    & 0.3 & 0.3 & 0.5 \\ 
Comparison of MC models        & 0.4 & 0.2 & 0.2 \\
Data/Monte Carlo               & 0.6 & 0.4 & 0.7 \\ 
\Mw\ dependence ($\pm150$~MeV) & 0.1 & 0.3 & 0.1 \\ 
Beam energy dependence         & 0.5 & 0.5 & 0.8 \\ \hline
Total                          & 0.9 & 0.8 & 1.2 \\ \hline
 \end{tabular}
 \end{center}
 \caption{ Sources of uncertainty on the \WWqqln\ selection efficiencies.}
%
 \label{qqln-sys}
\end{table}
\begin{table}[htbp]
 \begin{center}
 \begin{tabular}{|l|r@{$\ \pm \ $}l|r@{\ $\pm$ \ }l|r@{$\ \pm \ $}l|} \hline
 \multicolumn{1}{|l|}{} &
 \multicolumn{6}{c|}{Background cross-sections (fb) }                 \\
 \multicolumn{1}{|l|}{Source} & 
 \multicolumn{2}{c|}{\WWqqen} & \multicolumn{2}{c|}{\WWqqmn}& 
 \multicolumn{2}{c|}{\WWqqtn} \\ \hline
 \qqen    & \hspace{10mm}  0&25   & \hspace{10mm} 1 &2  & \hspace{8mm} 42&19  \\
 \qqff    &   1&1    & 24&5  &  40&4   \\
 \eeff    &  82&27   & 6 &2  &  31&10  \\
 \Zqq     &  29&13   & 23&10 & 143&61 \\
 \Ztt     &  6 &1    & 1 &1  &   1&1    \\ \hline
 Combined & 118&39   & 55&12 & 257&65 \\ \hline
 \end{tabular}
\end{center}
\caption{Background  cross-sections for the \WWqqln\ selections in fb.
  The four-fermion backgrounds  have been split into final states 
  containing no electron (\qqff),  where f refers to
  a fermion other than an electron, one electron  (\qqen)  
  and  two electrons (\eeff). 
  The background from \eeff\ final states includes two-photon processes.
  The \Zqq\ background includes the uncertainty on the background
  correction factor of $1.2\pm0.5$.
  All errors include both statistical and systematic contributions. }
\label{qqln-back}
\end{table}
\begin{table}[htbp]
\begin{center}
\begin{tabular}{|l|r@{$\ \pm \ $}l|r@{\ $\pm$ \ }l|} \hline
                      & \multicolumn{4}{c|}{Background cross-section (fb)} \\ 
                      & \multicolumn{2}{c|}{Likelihood cut} & 
                        \multicolumn{2}{c|}{Event weight}  \\ 
                      & \multicolumn{2}{c|}{$\sigma_{\mathrm{bgd}}$} &   
                        \multicolumn{2}{c|}
       {$\sigma^{\mathrm{pre}}_{\mathrm{bgd}}\overline{w_b}$} \\ \hline
\qqln            &\hspace{9mm}1&   1    &    \hspace{8mm}1 &   1 \\
\qq              &1237& 281 & 1142 & 181 \\
\qqff (f$\neq$e) &127&  46  &  108 &  47 \\
\qqee            &18&   3   &   14 &   3 \\ \hline
Combined         &1380& 280 & 1260 & 190 \\ \hline
\end{tabular}
\end{center}
\caption{Background cross-sections for the \WWqqqq\ channel
  in fb, assuming an average centre-of-mass energy of $\rroots$~GeV.
  The uncertainties include systematic contributions.}
\label{qqqq-back}
\end{table}

\begin{table}[htbp]
\begin{center}
\begin{tabular}{|l|c|c|c|c|} \hline
                      & \multicolumn{4}{c|}{Selection} \\ 
                      & \multicolumn{2}{c|}{Likelihood cut} 
                      & \multicolumn{2}{c|}{Event weight}  \\
Source of uncertainty & $\Delta \epsilon_{\mathrm{sig}}$  (\%) 
& $\Delta \sigma_{\mathrm{bgd}}$ (pb)  & 
$\Delta(\epsilon^{\mathrm{pre}}_{\mathrm{sig}}\overline{w_s})$ (\%) &
 $\Delta(\sigma^{\mathrm{pre}}_{\mathrm{bgd}}\overline{w_b})$ (pb) \\ \hline
Monte Carlo models                   & 0.30 & 0.08 & 0.21 & 0.07 \\
QCD parameter variation              & 0.93 & 0.12 & 1.14 & 0.12 \\
Data/Monte Carlo $\sqrt{s}=172$~GeV  & 0.70 & 0.18 & 1.59 & 0.02 \\
Data/Monte Carlo $\sqrt{s}=91.2$~GeV & ---  & 0.11 & ---  & 0.09 \\
Beam energy dependence               & 0.25 & 0.11 & 0.34 & 0.06 \\
\Mw\ dependence ($\pm 150$ MeV)      & 0.03 & 0.00 & 0.17 & 0.00 \\
Binning effects                      & 0.15 & 0.04 & 0.32 & 0.02 \\ \hline
{Total}                              & 1.24 & 0.28 & 2.03 & 0.18  \\ \hline
\end{tabular}
\end{center}
\caption{Sources of systematic error on the signal efficiency and
expected background cross-section for the \WWqqqq\ likelihood
selection, which is used in the W mass and event properties analyses.
The systematic errors on the event weight based
quantities used for the cross-section measurement are also given.
The total is the quadrature sum of all uncertainties.}
\label{qqqq-sys}
\end{table}
\begin{table}[htbp]
 \begin{center}
 \begin{tabular}{|l|c|c|c|r|} \hline
 \WWqqqq        & Expected signal & Expected back. & Total & Observed \\ 
 Selection      &                 &                &       &              \\ \hline
 Preselection   & $53.0 \pm 1.5 $ & $46.7\pm2.1 $  & $99.7 \pm2.6 $ & 99 \\ 
 Likelihood cut & $46.8 \pm 1.2 $ & $14.3\pm2.9 $  & $61.0 \pm3.1 $ & 57 \\
 Event weight   & $41.3 \pm 1.5 $ & $13.1\pm2.0 $  & $54.4 \pm2.5 $ & 54.1$\pm$6.6 \\
\hline
 \end{tabular}
 \end{center}
 \caption{ Comparison of expected and observed numbers of events in
           the \WWqqqq\ channel, 
           based on an integrated luminosity of \intLdt~pb$^{-1}$,
           $\Mw\ = 80.33~\GeVcc$ and
           a total \WW\ production cross-section of \GENTxs~pb. 
           Results are given for the preselection cuts, for the likelihood
           cut selection used in the mass and event properties analyses
           and for the event weight analysis used in the cross-section
           measurement. The 
           errors on the expected numbers of events 
           include contributions arising 
           from luminosity, \sigccthree\ and \Mw\ uncertainties.
           The error on the observed event weight is calculated
           as the square root of the sum of the event weights 
           squared for the 99 events passing the preselection.}
 \label{qqqq-events}
\end{table}
\renewcommand{\arraystretch}{1.2}
\begin{table}[htbp]
 \begin{center}
 \begin{tabular}{|l|cccccc|ccc|c|} \hline
 Event    &     \multicolumn{10}{c|}{Efficiencies[\%] for $\WW\rightarrow$}    \\    
 selection&\senen$\!\!$&\senmn$\!\!$&\sentn$\!\!$&\smnmn$\!\!$&\smntn$\!\!$&
    \stntn$\!\!$&\qqen$\!\!\!$&\qqmn$\!\!\!$& \qqtn$\!\!\!$&\qqqq$\!\!\!$\\ \hline
 \senen    & 72.3&  0.1 &  5.9 &  0.0 &  0.0 &  0.2 &  0.0 &  0.0 &  0.0 &  0.0 \\
 \senmn    & 1.9 & 74.1 &  4.6 &  0.3 &  5.3 &  1.0 &  0.0 &  0.0 &  0.0 &  0.0 \\
 \sentn    & 7.6 &  4.3 & 62.8 &  0.0 &  1.3 &  9.0 &  0.0 &  0.0 &  0.0 &  0.0 \\
 \smnmn    & 0.0 &  1.0 &  0.1 & 78.3 &  6.4 &  0.3 &  0.0 &  0.0 &  0.0 &  0.0 \\
 \smntn    & 0.1 &  3.8 &  0.4 &  7.9 & 60.3 &  5.9 &  0.0 &  0.0 &  0.0 &  0.0 \\
 \stntn    & 0.3 &  0.3 &  2.9 &  0.2 &  3.7 & 44.3 &  0.0 &  0.0 &  0.0 &  0.0 \\ \hline
 \qqen    & 0.0 &  0.0 &  0.3 &  0.0 &  0.0 &  0.0 & 85.1 &  0.1 &  3.9 &  0.0 \\ 
 \qqmn    & 0.0 &  0.0 &  0.0 &  0.0 &  0.0 &  0.0 &  0.2 & 87.6 &  4.4 &  0.1 \\
 \qqtn    & 0.0 &  0.0 &  0.0 &  0.0 &  0.0 &  0.3 &  4.7 &  5.2 & 61.4 &  0.2 \\ \hline
 \qqqq    & 0.0 &  0.0 &  0.0 &  0.0 &  0.0 &  0.0 &  0.0 &  0.0 &  0.4 & 70.3 \\ \hline
 \end{tabular}
\end{center}
\caption{ Efficiency matrix, $\epsilon_{ij}$, for the 172~GeV event selections
          determined using \KORALW\ (\CC) 
          Monte Carlo events. Each entry represents the percentage
          of generated  events in decay channel $i$ which are accepted by the 
          selection for channel $j$.
          The numbers for the \qqln\ selections have been
          corrected for differences between data and Monte Carlo.
           The \qqqq\ numbers are calculated as the preselection
          efficiency multiplied by the average event weight for each
          event type. The \lnln\ efficiencies have been corrected by 
          a factor 0.99 to account for detector occupancy.}
\label{tab:eff_matrix}
\end{table}
\renewcommand{\arraystretch}{1.0}
\renewcommand{\arraystretch}{1.2}
\begin{table}[htbp]
 \begin{center}
 \begin{tabular}{|clc|r|r|r|} \hline
& Fitted parameter&   &  \multicolumn{3}{c|}{Fit assumptions : }    \\   
& &      &   & Lepton universality & SM branching fractions   \\ \hline
& $\Br(\Wev)$ &  & $ 0.098^{+0.022}_{-0.020} \pm 0.003$ &  &  \\ 
& $\Br(\Wmv)$ &  & $ 0.073^{+0.019}_{-0.017} \pm 0.002$ &  &  \\ 
& $\Br(\Wtv)$ &  & $ 0.140^{+0.030}_{-0.028} \pm 0.005$ &  &  \\ 
& $\Br(\Wlv)$ &  &     &  $0.101^{+0.011}_{-0.010}\pm 0.002$ &  \\ 
& $\Br(\Wqq)$ &  & $0.690^{+0.030}_{-0.032} \pm 0.007$ & 
                              $0.698^{+0.030}_{-0.032}\pm0.007$ &  \\ \hline 
& \sigccthree(161 GeV) [pb]& & $ 3.9^{+1.0}_{-0.9} \pm 0.2$ & 
       $ 3.7^{+0.9}_{-0.8} \pm 0.2$ &  $3.6^{+0.9}_{-0.8} \pm 0.2$\\
& \sigccthree(172~\GeV) [pb]& & $12.6\pm 1.3 \pm 0.4$ & 
                        $12.3\pm 1.3 \pm 0.4$ & $12.3 \pm 1.3 \pm 0.3$\\ \hline
 \end{tabular}
\end{center}
\caption{ Summary of cross-section and branching fraction results from 
          the combined 161.3~GeV and 172.12~GeV data. The results from three 
          different fits described in the text are shown. 
          The correlations between the branching fraction measurements
          from the fit without the assumption of 
          lepton universality are less than 30\%.}
\label{tab:xsecbr_results}
\end{table}
\renewcommand{\arraystretch}{1.0}
\begin{table}[htbp]
\begin{center}
\begin{tabular}{|l|cc|}
\hline
   & \multicolumn{2}{c|}{Selected events in fit range }\\ 
Channel &  BW method  & RW method  \\ 
\hline
 \WWqqqq\ 
   (1 jet-jet pairing per event)  & 30 & 22 \\
 \WWqqqq\ 
   (2 jet-jet pairings per event)  & 12 & 21 \\
 \WWqqenu\ + \WWqqmnu\                                & 28 & 31 \\
 \WWqqtnu\                                            & 14 & 17 \\ \hline
 Total                                                & 84 & 91 \\ \hline
\end{tabular}\end{center}
\caption[foo]{\label{tab:evno}
 Numbers of selected events in each channel for use in the W mass
 determination. For the \WWqqqq\ process, up to two jet-jet pairings
 are used in each event.
 For the BW method, only events with a reconstructed mass of 
 $40~\GeV < \mrec < 84~\GeV$ are included, while for the RW method, 
 reconstructed masses from $65~\GeV$ up to the kinematic limit are used.
}
\end{table} 
\begin{table}[htbp]
\begin{center}
\begin{tabular}{|l|r|r|r|} \hline
                      & \WWqqqq\         & \WWqqlnu\        & combined 
                      \\ \hline
 Fitted mass (GeV)    & $80.19 \pm 0.39$ & $80.80 \pm 0.37$ & $80.50 \pm 0.27$ 
                      \\
 MC correction (GeV)  & $-0.19 \pm 0.01$ & $-0.32 \pm 0.01$ & $-0.26 \pm 0.01$ 
                      \\
 Corrected mass (GeV) & $80.01\pm0.43\pm0.14$ 
                      & $80.48\pm0.40\pm0.12$ 
                      & $80.24\pm0.30\pm0.10$ 
                      \\ \hline
\end{tabular}\end{center}
\caption[foo]{\label{tab:simplefit}
 Summary of fit results and Monte Carlo corrections to \Mw\ for the BW fit
 method. The errors on the fitted mass and corrections are statistical only,
 while
 for the corrected mass the statistical and systematic errors are both listed 
 (see Section~\ref{sec:mwsyserr} for details).
}
\end{table} 
\begin{table}[htbp]
\begin{center}\begin{tabular}{|l|c|c|c|c|c|c|c|c|} \hline
    & \multicolumn{3}{|c|}{BW-fit to \Mw}  &
      \multicolumn{3}{|c|}{RW-fit to \Mw}  &
      \multicolumn{2}{|c|}{RW-fit to \Mw\ and \Gw}  
    \\
  Systematic errors
    & \multicolumn{3}{|c|}{fixed width}  &
      \multicolumn{3}{|c|}{SM constrained width}  &
      \multicolumn{2}{|c|}{ }  
    \\
  ~~~~~~~~(MeV) 
    & \qqqq & \qqln & comb. & \qqqq & \qqln & comb. & ~~~$\Mw$~~~ & $\Gw$ 
    \\ \hline
 Beam Energy      &27 &  27 & 27 & 30 & 30 & 30 & 30 & 15  \\
 ISR              &22 &  22 & 22 & 20 & 20 & 20 & 20 & 61  \\
 Hadronisation    &16 &  16 & 12 & 52 & 26 & 20 & 21 & 52  \\
 Four-fermion     &42 &  37 & 37 & 42 & 48 & 42 & 35 & 70  \\
 Detector Effects &67 &  87 & 54 & 62 & 76 & 48 & 48 & 98 \\
 Fit procedure    & & & & & & & & \\
 ~~~~and Background &33 &  65 & 36 & 32 & 28 & 25 & 25 & 85  \\
 Colour Reconnection & & & & & & & & \\
 ~~~~and Bose-Einstein    
                  & 100 & 0 & 50 & 100 & 0 & 50 & 50 & 50  \\ \hline
 Total systematic error   & 137 & 121 & 96 & 144 & 104 & 94 & 92 & 175
    \\ \hline
\end{tabular}\end{center}
\caption[foo]{\label{tab:sys}
 Summary of the systematic errors for the various fit methods.  For the 
 one-parameter fits to determine \Mw,
 uncertainties are given for the fits to
 \WWqqqq, \WWqqln\ and for the combined sample. For the 
 two-parameter fit, the uncertainties for \Mw\ and \Gw\ are listed
 separately only for the fits to the combined sample.}
\end{table} 
\begin{table}[htbp]
 \begin{center}
 \begin{tabular}{|l|c|c|c|c|c|c|c|c|} \hline
 Systematic variation & 
 \nchQQQQ & \nchQQLV & \Dnch & \xpQQQQ
                                    & \xpQQLV & \Dxp
                                              & \thrQQQQ & \yQQQQ \\ 
          &          &       & & $\times 10^{2}$
                                    & $\times 10^{2}$ & $\times 10^{2}$ 
                                              &          &        \\ 
\hline
 \WW\ model dependence & 
 $0.30$ & $0.16$ & $0.19$ & $0.018$ & $0.035$ & $0.027$ & $0.005$ & $0.006$  \\
 Track quality cuts &                                  
 $0.28$ & $0.18$ & $0.26$ & $0.043$ & $0.055$ & $0.065$ & ---      & $0.011$  \\
 Event selection &                                    
 $0.15$ & $0.25$ & $0.52$ & $0.047$ & $0.067$ & $0.020$ & $0.008$ & $0.017$  \\
 Background scaling &                                 
 $0.05$ & $0.06$ & $0.15$ & $0.018$ & $0.005$ & $0.054$ & $0.008$ & $0.011$  \\
 Z/$\gamma$ background &                         
 $0.27$ & $0.03$ & $0.22$ & $0.012$ & $0.005$ & $0.017$ & $0.007$ & $0.005$  \\
 Other backgrounds &                                  
 $0.06$ & $0.07$ & $0.11$ & $0.017$ & $0.009$ & $0.022$ & $0.001$ & $0.002$  \\
 Beam energy dependence &                             
 $0.15$ & $0.01$ & $0.18$ & $0.001$ & $0.015$ & $0.016$ & $0.004$ & $0.002$  \\
 Unfolding procedure &                                
 $0.15$ & $0.22$ & $0.29$ & $0.025$ & $0.051$ & $0.026$ & $0.015$ & $0.001$ \\
\hline                                                
 Total &                                              
 $0.56$ & $0.42$ & $0.75$ & $0.076$ & $0.11$  & $0.10$  & $0.021$  & $0.025$  \\
\hline
 \end{tabular}
 \end{center}
\caption{Individual contributions to the systematic
uncertainties on the average event properties.}
\label{tab-wwprop-syst}
\end{table}

\clearpage


\end{document}